\documentclass[]{emulateapj}
\usepackage{rotating}
\newcommand{\comment}[1]{}

\newcommand*{\savedfootnotes}{}
\newcommand*{\resetsavedfootnotes}{\global\let\savedfootnotes\empty}
\newcommand{\tablefootnote}[1]%
  {%
    \footnotemark
    \xdef\savedfootnotes%
      {\unexpanded\expandafter{\savedfootnotes}\noexpand\footnotetext{#1}}%
  }
\edef\endtable%
  {%
    \aftergroup\noexpand\savedfootnotes
    \aftergroup\noexpand\resetsavedfootnotes
    \unexpanded\expandafter{\endtable}%
  }
\usepackage{lipsum}

\newlength{\colwidth}

\newcommand{\solar}{_{\mathord\odot}}
\newcommand{\hmpc}{\ifmmode{h^{-1}{\rm Mpc}}\;\else${h^{-1}}${\rm Mpc}\fi}
\newcommand{\hMpc}{\ifmmode{h^{-1}{\rm Mpc}}\;\else${h^{-1}}${\rm Mpc}\fi}
\newcommand{\hGpc}{\ifmmode{h^{-1}{\rm Gpc}}\;\else${h^{-1}}${\rm Gpc}\fi}
\newcommand{\hkpc}{\ifmmode{h^{-1}{\rm kpc}}\;\else${h^{-1}}${\rm kpc}\fi}

\newcommand{\msun}{{\rm M}_{\solar}}

\newcommand{\beq}{\begin{equation}}
\newcommand{\eeq}{\end{equation}}

\newcommand{\mr}{\ifmmode{M_r}\;\else$M_r$\fi}
\newcommand{\rd}{\ifmmode{R_\delta}\;\else$R_\delta$\fi}
\newcommand{\ngals}{\ifmmode{N_{\rm gals}}\;\else$N_{\rm gals}$\fi}

\newcommand{\mcooldot}{\dot{m}_{\rm cool}}
\newcommand{\vvir}{V_{\rm vir}}
\newcommand{\rvir}{R_{\rm vir}}

\newcommand{\be}{\begin{equation}}
\newcommand{\ee}{\end{equation}}

\usepackage{graphicx}
\usepackage{amsmath}
\usepackage[usenames]{color}

\begin{document}

\title{A Comparison between Semi-Analytic Model Predictions for the CANDELS Survey}

\author{Yu Lu\altaffilmark{1}, Risa H. Wechsler\altaffilmark{1}, Rachel S. Somerville\altaffilmark{2}, Darren Croton\altaffilmark{3}, 
Lauren Porter\altaffilmark{4}, Joel Primack\altaffilmark{4}, 
Peter S. Behroozi\altaffilmark{5}, Henry C. Ferguson\altaffilmark{5}, 
David C. Koo\altaffilmark{6},
Yicheng Guo\altaffilmark{6},
Mohammadtaher Safarzadeh \altaffilmark{7},
Kristian Finlator\altaffilmark{8},
Marco Castellano\altaffilmark{9},
Catherine E. White \altaffilmark{7},
Veronica Sommariva\altaffilmark{9},
Chris Moody\altaffilmark{4}
}
\altaffiltext{1}{Kavli Institute for Particle Astrophysics \& Cosmology,
  Physics Department, and SLAC National Accelerator Laboratory,
  Stanford University, Stanford, CA 94305 {\tt luyu@stanford.edu, rwechsler@stanford.edu}}
\altaffiltext{2}{Department of Physics and Astronomy, Rutgers University, 136 Frelinghuysen Road, Piscataway, NJ 08854, USA}
\altaffiltext{3}{Centre for Astrophysics \& Supercomputing, Swinburne University of Technology, PO Box 218, Hawthorn, VIC 3122, Australia}
\altaffiltext{4}{Department of Physics, University of California at Santa Cruz, Santa Cruz, CA 95064, USA}
\altaffiltext{5}{Space Telescope Science Institute, 3700 San Martin Drive, Baltimore, MD 21218, USA}
\altaffiltext{6}{UCO/Lick Observatory, Department of Astronomy and Astrophysics, University of California, Santa Cruz, CA 95064, USA}
\altaffiltext{7}{Department of Physics \& Astronomy, The Johns Hopkins University, 3400 North Charles Street, Baltimore, MD 21218, USA}
\altaffiltext{8}{Dark Cosmology Centre, Niels Bohr Institute, University of Copenhagen, Copenhagen, Denmark}
\altaffiltext{9}{INAF - Osservatorio Astronomico di Roma, via Frascati 33, 00040, Monteporzio, Italy}

\begin{abstract}
  We compare the predictions of three independently developed
  semi-analytic galaxy formation models (SAMs) that are being used to
  aid in the interpretation of results from the CANDELS survey.  These
  models are each applied to the same set of halo merger trees
  extracted from the ``Bolshoi'' high-resolution cosmological $N$-body
  simulation and are carefully tuned to match the local galaxy stellar
  mass function using the powerful method of Bayesian Inference
  coupled with Markov Chain Monte Carlo (MCMC) or by hand.  
The comparisons reveal that in spite of the significantly different
parameterizations for star formation and feedback processes, the three
models yield qualitatively similar predictions for the assembly
histories of galaxy stellar mass and star formation over cosmic time.  
Comparing SAM predictions with existing estimates of the stellar mass
function from $z=0-8$, we show that the SAMs generally require strong
outflows to suppress star formation in low-mass halos to match the
present day stellar mass function, as is the present common wisdom.
However, all of the models considered produce predictions for the
star formation rates
and metallicities of low-mass galaxies that are
inconsistent with existing data.
The predictions for metallicity-stellar mass relations and their
evolution clearly diverge between the models.  We suggest that
large differences in the metallicity relations and small differences in the
stellar mass assembly histories of model galaxies stem from different
assumptions for the outflow mass-loading factor produced by feedback.
Importantly, while more accurate observational measurements for
stellar mass, star formation rate and metallicity of galaxies at
$1<z<5$ will discriminate between models, the discrepancies between
the constrained models and existing data of these observables have
already revealed challenging problems in understanding star formation
and its feedback in galaxy formation.
The three sets of models are being used to construct catalogs of mock 
galaxies on light cones that have the same geometry as the CANDELS 
survey, which should be particularly useful for quantifying the biases and 
uncertainties on measurements and inferences from the real observations.
\end{abstract}

\section{Introduction}\label{sec:introduction}
Over the past three decades, a number of theoretical methods have been
developed to study the formation and evolution of galaxies.  Among
these, semi-analytic models (SAMs) have become a widely used tool to
study the formation and evolution of galaxies in a cosmological
context, owing to their relatively light computational demands and
great flexibility for exploring the effects of different physical
assumptions.  SAMs follow astrophysical processes affecting the
baryonic components of galaxies using simple, yet physically, and/or
empirically motivated recipes. In those recipes, the evolution of
each baryonic component is followed by a set of differential equations
that track the mass of each component 
as a function of time, depending on other properties (e.g. halo mass
and angular momentum, gas mass, metallicity, etc).  The models are
embedded either in Monte Carlo realizations of an analytic
representation of dark matter halo formation histories
\citep{White1991, Kauffmann1993, Cole1994, Somerville1999} or in a
direct numerical simulation \citep{Kauffmann1999, Springel2001,
  Springel2005} of the evolution of the underlying dark matter
distribution. The pioneering work by \citet{White1978} first laid the
theoretical groundwork for this type of model and more detailed
processes started to be included in the 1990s \citep{Cole1991,
  White1991, Kauffmann1993, Somerville1999a}. 
Over the past ten years, such models have been substantially extended
and refined by a number of different groups \citep[for reviews,
  see][]{Baugh2006, Benson2012}.  SAMs can yield a rich set of
predictions for the evolution of various galaxy properties, such as
the distributions of stellar mass, luminosity, star formation rate,
size, rotation velocity, morphology, gas content, and metallicity, as
well as the scaling relations linking these properties.

Because the physical processes of galaxy formation are not well
understood, SAMs generally have large uncertainties in their model
parameterization and their choice of parameter values. A key goal of
any semi-analytic model is to constrain these uncertainties using
observational data and to better understand the processes through
detailed calculations or simulations. Currently, such uncertainties
are represented by adjustable parameters governing the efficiencies
and scalings of these processes (note that this is true also for
cosmological hydrodynamical simulations). Different groups typically
adopt different parameterizations for modeling the processes and
different values for those parameters. It is important to understand
the impact of these model recipes and their parameters on the
predicted galaxy properties.

After a SAM is developed or significantly updated, a new paper is
generally written to report on how the model is tuned to match
observational data, and to make new predictions for various
observables \citep[e.g.][]{Bower2006, Cattaneo2006, Croton2006,
  Menci2006, DeLucia2007, Monaco2007, Somerville2008,
  Guo2011}. However, to date there have been only a few direct
comparisons between different semi-analytic model codes. 
One approach is to compare the recipes adopted by various groups by
implementing them as modules within a single code, as in an early
study by \citet{Somerville1999a} and more recently in \citet{Lu2011}.
Other studies have compared models
that are based on different merger trees, and for which no attempt has
been made to normalize the models in the same way
\citep{Fontanot2009,Kimm2009}. Similarly, \citet{Wang2012} and
\citet{Weinmann2012} took models by \citet{Guo2011} and
\citet{Neistein2010} with varied parameters to understand the effect
of the star formation and SN feedback. \citet{Maccio2010} compared
three different SAMs run in the same merger trees, but focused only on
Milky Way mass halos. \citet{DeLucia2010} presented a comparison of
three different
SAMs run within the same merger trees, but for ``stripped down''
versions of the SAMs that contained only cooling and merging.

The above studies have shown that the model predictions are sensitive
to some of the model parameters, while other parameters are
degenerate. Furthermore, models based on very different
parameterizations can make similar predictions for some observables,
which suggests that those predictions are insensitive to the choice of
model recipes (i.e. how the physics itself is implemented). In this
paper, we compare model predictions from three independently developed
semi-analytic models, namely the Croton model \citep[][Croton et
  al. in prep]{Croton2006} the Somerville model \citep{Somerville2008,
  Somerville2012} and the Lu model \citep{Lu2011a}. Our goal is to
understand the impact of the different model assumptions on the
predicted observables when the different models are applied to the
same underlying merger trees and are carefully tuned to match some
basic observational constraints. The present paper focuses on the most
fundamental physical properties, e.g. stellar mass, star formation
rate (SFR), stellar and cold gas metallicities, cold gas mass, etc,
predicted by the models in this large comparison.

All the models presented in this paper are being used to produce mock
catalogs for the Cosmic Assembly Near-IR Deep Extragalactic Legacy
Survey \citep[CANDELS][]{Koekemoer2011, Grogin2011}. CANDELS is an
observational project that has just completed a three year run of data
taking on the Hubble Space Telescope. The survey is designed to reveal
the evolution of galaxies from their infancy to the present day via
deep imaging of more than 250,000 galaxies with the new wide-field
near-infrared camera (WFC3) installed on Hubble in 2009. A number of
light cone realizations that mimic the geometry of the CANDELS fields
are populated with galaxies using these three SAMs. The mock catalogs
contain the most important physical properties of galaxies, such as
stellar mass and star formation rate, as well as observables such as
luminosities in all CANDELS bands. This paper thus also serves to
present the basic properties predicted by the three models.  Note that
only a subset of the quantities in the present comparison can be
directly measured by CANDELS.  Those quantities are the most important
ones for us to understand the similarities and differences between the
models, i.e., the rate at which galaxies assemble their stellar mass,
form their stars, and build up their metals.  In a series of
follow-up papers,
we will investigate more
directly observable quantities, i.e. magnitudes and colors, which
depend on these underlying quantities and also on additional model
assumptions, for example, the stellar population synthesis model and
dust model.  The comparisons of the most basic quantities in this
paper will guide us in interpreting the results presented in future
papers, as well as provide insights into how robust these predictions
are to variations in the parameterization of physical recipes and
their detailed implementation in different codes.

The paper is organized as follows. In Section \ref{sec:models}, we
describe the SAMs that are studied in the comparison. In Section
\ref{sec:method}, we summarize the merger trees on which all three
SAMs are built, and the method we use to calibrate the models. We show
the model predictions and comparisons in Section \ref{sec:results},
and conclude our study in Section \ref{sec:conclusion} with an
overview of our key results.  Throughout the paper, we use a
$\Lambda$CDM cosmology with $\Omega_{\rm M,0} = 0.27$,
$\Omega_{\Lambda,0} = 0.73$, $\Omega_{\rm b,0} = 0.044$, $h = 0.70$,
$n = 0.95$, and $\sigma_8 = 0.82$. These values are adopted in the
Bolshoi simulation \citep{Klypin2011}, and are consistent with the
7-year Wilkinson Microwave Anisotropy Probe (WMAP7) data
\citep{Jarosik2011} and the WMAP5 data \citep{Dunkley2009,
  Komatsu2009}\footnote{
The values used here are not fully consistent with the recent results
from the Planck collaboration \citep{Planck2013}, but our key results
are largely insensitive to changes in the cosmological parameters
at this level.}.  To focus our comparison on the different treatments
of galaxy formation processes rather than stellar evolution models, we
fix the initial mass function (IMF) to the Chabrier IMF
\citep{Chabrier2003} with the stellar mass range of
$0.1\msun<M<100\msun$. In the comparisons between model predictions
and observational data, the 
observationally derived stellar masses are converted by adopting the
same IMF and are based on the BC03 stellar population synthesis model
\citep{Bruzual2003}.

\begin{table*}[htb]
\begin{center}
\caption{Summary of model recipes and parameter values} 
\centering
\begin{tabular}{p{2cm} p{4.5cm} p{4.5cm} p{4.5cm}}
\hline\hline
process & Croton & Somerville & Lu$\, ^{\rm a}$ 
\\ [0.5ex]
\hline
re-ionization & Gnedin et al. 2000; Kravtsov et al. 2004 & same & same \\
equation & Eq.\ref{equ:reion} & Eq.\ref{equ:reion} & Eq.\ref{equ:reion} \\
parameters & $z_{\rm reion}=7$, $z_{\rm overlap}=8$ & 
$z_{\rm reion}=11$, $z_{\rm overlap}=12$ & 
$z_{\rm reion}=10$, $z_{\rm overlap}=11$ \\
\hline
cooling & Croton et al. 2006 & Croton et al. 2006, but assumes half of the hot gas accretes onto a central galaxy in a dynamical time when the cooling radius is larger than the virial radius to avoid the discontinuity in the cooling rate when $r_{\rm cool}=R_{\rm vir}$ & Croton et al. 2006 \\
\hline
star formation & 
only cold gas above a critical mass forms stars; constant star formation efficiency;     & 
Schmidt-Kennicutt law; only cold gas above a threshold surface density forms stars;  & 
only cold gas above a threshold surface density forms stars; efficiency depends on $\vvir$;\\
equations &
Eq. \ref{equ:sf_croton}  &
Eq. \ref{equ:sf_sd_somerville}, \ref{equ:sf_somerville} &
Eq. \ref{equ:sf_msf_lu}, \ref{equ:sf_ef_lu}, \ref{equ:sf_lu}\\
parameters &
$\alpha_{\rm SF}=0.05$  &
$A_{\rm Kenn}=8.33\times10^{-5}\msun{\rm yr}^{-1}{\rm kpc}^{-2}$, $N_{\rm K}=1.4$,  $\Sigma_{\rm crit}=6\msun\,{\rm pc}^{-2}$ &
$\alpha_{\rm SF}\in[0.0056,0.41]$, $\beta_{\rm SF}\in[0.05, 4.7]$, $V_{\rm SF}\in[32.4,  281]$km/s, $f_{\rm SF}\in[0.11, 3.9]$\\
\hline
SN feedback & 
reheat cold gas, eject hot gas  & 
reheat cold gas, eject cold gas  & 
reheat cold gas, eject cold gas and hot gas\\
equations &
Eq. \ref{equ:rh_croton}, \ref{equ:ej_croton} &
Eq. \ref{equ:rh_somerville}; \ref{equ:ej_somerville}&
Eq. \ref{equ:rh_lu}, \ref{equ:ej_somerville}, \ref{equ:wind_lu}\\
parameters &
$\alpha_{\rm LD}=3.5$, $\alpha_{\rm SN}=0.35$, $\chi_{\rm ri}=0.5$ &
$\alpha_{\rm LD}=1.5$, $\beta_{\rm LD}=2.25$, $V_{\rm EJ}=130{\rm km/s}$, $\beta_{\rm EJ}=6$&
$\alpha_{\rm LD}\in[0.011, 1.1]$, $\beta_{\rm LD}\in[3.3, 9.9]$, $V_{\rm EJ}\in[35, 970]$km/s, $\beta_{\rm EJ}\in[0.08, 7.7]$, $\alpha_{\rm SN}\in[0.016, 9.3]$, $\epsilon_{\rm W}\in[0.0011, 0.71]$, $\chi_{\rm ri}\in[0.01, 0.66]$\\
\hline
quenching model & 
radio mode and quasar mode & 
radio mode and quasar mode & 
halo quenching \\
equations &
Eq. \ref{equ:mbh_radio_croton}, \ref{equ:mbh_quasar_croton}, \ref{equ:agn_radio_croton} &
Eq. \ref{equ:mbh_radio_somerville}, \ref{equ:agn_radio_somerville}, \ref{equ:agn_hr_somerville} &
\\
parameters &
$f_{\rm BH}=0.015$, $\kappa_{\rm AGN}=1\times10^{-3}\msun/{\rm yr}$, $\eta_{\rm AGN}=0.005$&
$\sigma_{\rm BH}=0.3$, $f_{\rm BH, FINAL}=2$, $\kappa_{\rm radio}=4.3\times 10^{-3}$, $\kappa_{\rm heat}=1$, $\eta_{\rm rad}=0.1$, $\epsilon_{\rm wind}=0.5$&
$M_{\rm cc}\in[10^{11}, 10^{12.1}]\msun/h$, $\sigma_{\rm CC}\in[0.064, 1.1]$ \\
\hline
starburst &  
when two galaxies merge; efficiency depends on the baryonic mass ratio &  
when two galaxies merge; efficiency depends on the mass ratio and the 
gas fraction
& 
when two galaxies merge; efficiency depends on the mass ratio \\
equations &
Eq. \ref{equ:sb}
&
Eq. 7-10 in Somerville et al. (2008)
&
Eq. \ref{equ:sb}
\\
parameters &
$\alpha_{\rm burst}=0.56$, $\beta_{\rm burst}=0.73$
&
$e_{\rm burst}$, $\tau_{\rm burst}$
&
$\alpha_{\rm burst}=0.56$, $\beta_{\rm burst}=0.73$
\\
\hline
satellite kinematics & 
subhalo info from simulation; dynamical friction model is used to determine whether a satellite is stripped or merged with a central galaxy   & 
dynamical friction model \citep{Boylan-Kolchin2008} & 
subhalo info from simulation and dynamical friction when subhalo is unresolved \\
equation &
Eq. \ref{equ:df_croton}
&
Eq. \ref{equ:df_somerville}
&
Eq. \ref{equ:df_lu}
\\
parameters &
&
&
$f_{\rm DF}\in[1.6, 91.6]$
\\
\hline
satellite stripping &
hot gas is stripped in proportion to the subhalo dark matter mass stripping; 
the entire satellite galaxy is stripped or merged with a central galaxy when a subhalo mass is reduced below a certain halo mass to galaxy mass ratio 
&
hot gas is instantaneously stripped when a halo enters the virial radius of a bigger halo;
the entire stellar and cold gas mass is instantaneously stripped when the galaxy is considered tidally disrupted
&
hot gas is instantaneously stripped when a halo becomes a subhalo; the cold gas is instantaneously stripped when a subhalo loses its identity in the merger tree; 
a fraction of stellar mass is stripped in every orbital timescale after the subhalo is lost \\
equation &
&
&
\\
parameters &
$f_{\rm ST,crit}=2$
&
$f_{\rm strip}= 1$, $f_{\rm dis}=1$ 
&
$f_{\rm ST}=[0.11, 0.98]$
\\

\hline 
\hline
\end{tabular}
\label{tab:model}
\end{center}
\footnotetext{For the Lu model, numbers in the parentheses encompass 95\% posterior range of the free parameters in the model.}
\end{table*}

\section{Semi-Analytic models}\label{sec:models}
Throughout this paper we use the Croton model \citep[][Croton et
  al. in prep]{Croton2006} (also known as the Semi-Analytic Galaxy
Evolution, or SAGE, model), the Somerville model
\citep{Somerville2008}, and the Lu model \citep{Lu2011a}. Below we
review the implementations and parameterizations of the most important
processes included in each model. These processes include
re-ionizatiion (\S \ref{sec:reionization}), radiative cooling (\S
\ref{sec:cooling}), `normal' star formation in galaxy disks (\S
\ref{sec:starformation}), supernova (SN) feedback (\S
\ref{sec:feedback}), AGN feedback and halo quenching (\S
\ref{sec:AGN}), galaxy-galaxy mergers (\S \ref{sec:merger}), and
chemical evolution (\S \ref{sec:chemical}).

\subsection{Baryonic matter budget}
In a SAM, every dark matter halo hosts a number of baryonic matter
components. A typical SAM contains hot gas, cold gas, stellar mass in
a bulge and disk component, ejected 
gas, and black hole mass. The basic task of a SAM is to follow the
evolution of the mass in these components.

It is normally assumed that every dark matter halo can acquire a
baryonic mass totaling $f_b M_{\rm vir}$, where $f_{\rm b}=\Omega_{\rm
  b}/\Omega_0$ is the universal baryon fraction, and $M_{\rm vir}$ is
the virial mass of the dark matter halo. The physical processes
assumed in the model can, however, change the total amount of baryonic
matter by preventing baryons from accreting onto the halo or ejecting the
accreted baryons out of the halo. In the models adopted here, we
assume that at every timestep as we follow the assembly of a dark
matter halo, a halo accretes some stellar mass, cold gas and/or black
hole mass, which has formed in its progenitor halos. 
Sometimes a galaxy merges with another galaxy, at which time the
corresponding components in merger progenitors are combined together
and may trigger a starburst.

In addition to the cold gas in a galaxy disk, the models also assume
that a certain baryonic mass is accreted into the halo as a diffuse
hot gas component. The total mass of hot halo gas is
\begin{equation}
M_{\rm hot}=f_{\rm b,coll}f_{\rm b} M_{\rm vir} - \sum_i [M_{*,i} + M_{{\rm cold}, i} + M_{{\rm BH},i}+M_{{\rm ej}, i}] ~, 
\end{equation}
where $f_{\rm b,coll}$ is the baryon fraction that can accrete onto a
halo limited by some physical processes (for example, re-ionization),
$M_*$, $M_{\rm cold}$, $M_{\rm BH}$ and $M_{\rm ej}$ are the masses in
stars, cold gas, the supermassive black hole and ejected gas,
respectively, and the summation is over all galaxies in the halo.  In
the rest of this section we describe the physical processes that
transfer baryonic mass from one component to another and the
parameterizations for these processes in the three different models.

\subsection{Re-ionization}\label{sec:reionization}
\citet{Gnedin2000} showed that the fraction of baryons that can
collapse into halos of a given mass in the presence of a photoionizing
background can be described in terms of the `filtering mass', $M_{\rm
  F}$. Haloes less massive than $M_{\rm F}$ can confine less baryonic
mass than the universal average. \citet{Gnedin2000} parametrized the
collapsed baryon fraction as a function of redshift and halo mass with
the expression
\begin{equation}\label{equ:reion}
f_{\rm b, coll}(z, M_{\rm vir})= { f_{\rm b} \over [1+0.26 M_{\rm F}(z)/M_{\rm vir}]^3}~,
\end{equation}
where $f_{\rm b}$ is the universal baryon fraction and $M_{\rm vir}$
is the halo virial mass. The filtering mass is a function of redshift,
and this function depends on the re-ionization history of the
Universe.  \citet{Kravtsov2004} provide a fitting formula for the
filtering mass as a function of the redshift at which the first HII
regions begin to overlap ($z_{\rm overlap}$) and the redshift at which
most of the medium is re-ionized ($z_{\rm reion}$). Recent results
from the WMAP experiment suggest an earlier epoch of re-ionization,
$z_{\rm reion}>10$ \citep{Spergel2007}. All three of our models make
use of the fitting functions (B2) and (B3) from appendix B of
\citet{Kravtsov2004} to compute the initial fraction of baryons that
can collapse as a function of halo mass and redshift, $f_{\rm b,
  coll}$, with two parameters $z_{\rm overlap}$ and $z_{\rm reion}$.

\subsection{Radiative cooling: hot gas to cold gas}\label{sec:cooling}
All three models follow the treatment of \citet{Croton2006}, first introduced 
by \citet{Springel2001}, to predict accretion and radiative cooling of hot gas.
In the model, the halo hot gas is redistributed in every time-step,
and the density of the hot gas is assumed to have a singular
isothermal profile,
\begin{equation}
\rho_{\rm hot}={M_{\rm hot} \over 4\pi R_{\rm vir}} r^{-2}~,
\end{equation}
where $R_{\rm vir}$ is the virial radius of the halo. The temperature
of the hot gas, $T_{\rm hot}$, is assumed to be a constant for each
halo with $T_{\rm hot} = 35.9(\frac{\vvir}{\rm km s^{-1}})^2$K,
where $\vvir$ is the circular velocity of the halo at the virial
radius. The cooling timescale of the gas at radius $r$ is then
estimated by
\begin{equation}
\tau_{\rm cool}(r)=\frac{3}{2}{\mu m_{\rm H} kT_{\rm hot} \over
  \rho_{\rm hot}(r)\Lambda(T_{\rm hot},Z_{\rm hot})}~,
\end{equation}
where $\mu$ is the mean molecular weight in units of the mass of
hydrogen atom, $\Lambda$ is the cooling function from
\citet{Sutherland1993}, 
and $Z_{\rm hot}$ is the metallicity of the
hot gas.  At each timestep, each model calculates the cooling radius
$r_{\rm cool}$ by equating the cooling timescale with the dynamical
timescale of the host halo, $\tau_{\rm cool} =\tau_{\rm dyn} \equiv
{R_{\rm vir}}/{\vvir}$. If the cooling radius is equal to or smaller
than the virial radius, the estimated cooling rate is
\begin{equation}
\mcooldot
=0.5M_{\rm hot}{r_{\rm cool}\vvir \over r_{\rm vir}^2}~.
\end{equation}
In other words, half of the hot gas mass enclosed by the cooling
radius cools and accretes onto the central galaxy of the halo in a
dynamical time. If the cooling radius is larger than the virial
radius, the Croton and Lu models set the cooling rate to be equal to
the total hot gas mass in the halo divided by the dynamical time, but
the Somerville model sets the cooling rate to be half of the hot gas
mass, rather than the total amount, divided by the dynamical time. The
Lu model and the Somerville model implicitly assume that all hot gas
is associated with the primary halo and only the central galaxy of the
primary halo can accrete cooling gas; satellite galaxies, even if they
have associated subhalos, do not accrete any hot gas.

Although this radiative cooling model does not have any explicit free
parameters, this does not mean the theory for radiative cooling of
halo gas is certain. \citet{Lu2011} have explored various model
recipes adopted in different SAMs and found that they predict cooling
rates that differ from each other by a factor of 2 for low mass halos
($<10^{12}\msun$), and even larger for more massive halos \citep[also
  see][]{Saro2010}. In this paper, we fix the cooling recipe for all
the models and focus our comparisons on the impact of other parts of
the model related to the star formation and feedback processes.

\subsection{Star formation: cold gas to stars}\label{sec:starformation}
All three models assume that cooled gas settles into a disk in the
central galaxy of a halo, but they adopt different prescriptions for
the rate at which this gas turns into stars. In addition, each
distinguishes between `normal' star formation in the disk and
merger-driven starbursts. We first describe disk star formation.

\subsubsection{The Croton model}
The Croton model assumes that cold gas is distributed in an
exponential disk, and star formation is regulated by a critical
surface density \citep{Kennicutt1998}, below which stars do not
form. Practically, star formation occurs when the cold gas mass of the
galaxy, $M_{\rm cold}$, exceeds a critical mass, $m_{\rm crit}$,
suggested by the empirical relation of \citet{Kauffmann1996}
\begin{equation}
m_{\rm crit}=3.8\times 10^9 \left( {\vvir \over 200\, {\rm km\,s}^{-1}}\right) \left({ r_{\rm gas} \over 10\,{\rm kpc}}\right) \msun~,
\end{equation}
where $r_{\rm gas}$ is the characteristic radius of the cold gas disk
which is set to be three times larger than the stellar disk
scale-length, $r_{\rm s}$. 
Following the simplified model of \citet{Mo1998}, where the density
profile of the host halo is assumed to be a singular isothermal sphere
and halo contraction is ignored, the model sets $r_{\rm s} =
(\lambda/\sqrt{2})\rvir$, where $\lambda$ is the spin parameter of the
host halo.
The star formation rate of the galaxy, $SFR$, is
then given by
\begin{equation}\label{equ:sf_croton}
SFR=\alpha_{\rm SF} { M_{\rm cold} - m_{\rm crit} \over \tau_{\rm disk}},
\end{equation}
where $\tau_{\rm disk} = r_{\rm gas} / \vvir$ is the dynamical
time of the gaseous disk and $\alpha_{\rm SF}$ is a free parameter
controlling the efficiency with which the cold gas is converted
into stars over this timescale.

\subsubsection{The Somerville model}
The Somerville model also adopts a star formation recipe based on the
empirical Schmidt-Kennicutt relation \citep{Kennicutt1989,
  Kennicutt1998}, but implemented in a somewhat different
manner. Here, the star formation rate surface density of the disk is
calculated according to
\begin{equation}\label{equ:sf_sd_somerville}
\Sigma_{\rm SFR} = \left\{ \begin{array}{ll}
	A_{\rm Kenn}\Sigma_{\rm gas}^{N_{\rm K}}, & \mbox{$\Sigma_{\rm gas} > \Sigma_{\rm crit}$}; \\
	0, & \mbox{otherwise}. \end{array} \right.
\end{equation}
Here, $A_{\rm Kenn} = 1.67 \times 10^{-4}\msun\,{\rm yr}^{-1}{\rm
  kpc}^{-2}$, $N_{\rm K} = 1.4$ and $\Sigma_{\rm gas}$ is the surface
density of cold gas in the disk in units of $\msun\,{\rm pc}^{-2}$,
which are taken from the observationally derived values of
\citet{Kennicutt1998}, converted to a Chabrier IMF. The gas
distribution follows an exponential profile, with a characteristic
radius $r_{\rm gas}=1.5 r_{\rm s}$, where $r_{\rm s}$ is calculated
using the full model of \citet{Mo1998} with the assumption that the
host halo has an NFW density profile and the halo mass distribution is
contracted in response to the formation of a central galaxy in the
halo center:
\begin{equation}
r_{\rm gas}={1 \over \sqrt{2}} f_j \lambda R_{\rm vir} f_c^{-1/2} f_{\rm R}(\lambda, c, f_{\rm d})~,
\end{equation}
where $f_{j}\equiv(J_{\rm d} /m_{\rm d})/(J_{\rm h}/M_{\rm vir})$ is
the ratio of the specific angular momentum of the disk and the halo,
$c$ is the concentration of the halo, and $f_{\rm d}$ is the disk mass
to the halo mass ratio. Comparing with the Croton model, $f_c^{-1/2}$
reflects the difference in energy of a singular isothermal profile
versus the NFW profile, and $f_{\rm R}$ reflects the adiabatic
contraction of a NFW halo (see \citet{Mo1998} for expressions governing
$f_{\rm R}$ and $f_c$).

The model assumes that only the gas above a critical surface density
threshold $\Sigma_{\rm crit}$ ($= 6 \msun {\rm pc}^{-2}$) is available
for star formation. The critical surface density gives a critical
radius,
\begin{equation}
r_{\rm crit}=-\ln \left[ {\Sigma_{\rm crit} \over \Sigma_0}\right] r_{\rm gas},~
\end{equation}
where $\Sigma_0=M_{\rm cold}/(2\pi r_{\rm gas}^2)$. 
The total star formation rate is then
\begin{equation}\label{equ:sf_somerville}
\begin{split}
SFR=2\pi \int_0^{r_{\rm crit}} \Sigma_{\rm SFR}(r) r {\rm d}r = { 2\pi A_{\rm Kenn} \Sigma_0^{N_{\rm K}} r_{\rm gas}^2 \over N_{\rm K}^2} \\
\times \left[ 1-\left(1+ {N_{\rm K} r_{\rm crit} \over r_{\rm gas}}\right) \exp(-N_{\rm K} r_{\rm crit} /r_{\rm gas})\right]~.
\end{split}
\end{equation}

\subsubsection{The Lu model}
The Lu model also assumes that cold gas is distributed in an
exponential disk with scale radius $r_{\rm gas}$, and only gas mass
with a surface density higher than a certain threshold, $\Sigma_{\rm
  crit}$, can form stars \citep[e.g.][]{Kennicutt1998, Kennicutt2007,
  Bigiel2008}. The model assumes a fiducial value of the parameter
for the threshold surface density of $\Sigma_{\rm crit,0}=10\msun {\rm
  pc}^{-2}$ but allows it to change in a range covering the observational uncertainty, $\Sigma_{\rm crit}\sim 3-10 \msun {\rm pc}^{-2}$ 
  \citep[][and references therein]{Schaye2004}, by introducing a free parameter $f_{\rm SF}$ 
  defined below.  
For simplicity, when calculating $r_{\rm gas}$,
the model adopts a single value $\lambda=0.035$ for all halo
spins. The fiducial size of the gaseous disk is then $r_{\rm
  gas,0}={0.035 \over \sqrt{2}} \rvir$. The cold gas mass that has a
surface density above a threshold surface density $\Sigma_{\rm crit}$
and is available for star formation is then
\begin{equation}\label{equ:sf_msf_lu}
m_{\rm sf}=M_{\rm cold} 
	\left[1-\left[1+\ln \left({\Sigma_{\rm cold,0} \over f_{\rm SF}
       \Sigma_{\rm crit,0}}\right)\right]
       {f_{\rm SF} \Sigma_{\rm crit,0} \over \Sigma_{\rm cold,0}}\right]~,
\end{equation}
where $\Sigma_{\rm cold, 0}=M_{\rm cold}/(2\pi r_{\rm gas, 0}^2)$, and the model parameter $f_{\rm SF}$ is defined as
\begin{equation}
f_{\rm SF}=\left({r_{\rm gas} \over r_{\rm gas,0}}\right)^2 \left({\Sigma_{\rm crit} \over \Sigma_{\rm crit,0}}\right).
\end{equation}
The model parameter $f_{\rm SF}$ absorbs the uncertainties in both the
size of the gaseous disk and the threshold surface density of the cold
gas for star formation. As before, we take the star formation rate
to be proportional to the mass of star forming gas and inversely
proportional to the dynamical timescale of the disk, $\tau_{\rm
  disk}={r_{\rm gas,0} \over \vvir}$, yielding
\begin{equation}\label{equ:sf_lu}
SFR=\epsilon_{\rm sf}{m_{\rm sf} \over \tau_{\rm disk}}~,
\end{equation}
where $\epsilon_{\rm sf}$ governs star formation efficiency. 

A variety of factors, such as small scale stellar feedback,
turbulence, etc, could affect the timescale for the gas to be depleted
by star formation, so to generalize this model for the star formation
efficiency, the Lu model further assumes a halo circular velocity
dependence for the star formation efficiency such that $\epsilon_{\rm
  sf}$ has a broken power-law dependence on the circular velocity of
the host halo \citep[e.g.][]{Cole1994, Kang2005}:
\begin{equation}\label{equ:sf_ef_lu}
  \epsilon_{\rm sf} = \left\{ \begin{array}{ll}
      \alpha_{\rm SF} & \mbox{$\vvir \geq V_{\rm SF}$}; \\
      \alpha_{\rm SF}\left({\vvir \over V_{\rm SF}}\right)^{\beta_{\rm SF}}
      & \mbox{$\vvir < V_{\rm SF}$}, \end{array} \right. 
\end{equation}
where $\alpha_{\rm SF}$, $\beta_{\rm SF}$ and $V_{\rm SF}$ are model parameters in addition to $f_{\rm SF}$.

\subsection{SN feedback model: outflow and re-infall}\label{sec:feedback}
With each new star formation episode, the high mass stars
($\gtrsim8\msun$) will rapidly evolve and end their lives as energetic
supernovae, on timescales much shorter than the time
resolution of the models. The injection of this energy into the galaxy
interstellar medium will heat up a fraction of the cold gas, expelling
it from the disk. SAMs generally assume that supernova (SN) feedback
can affect the interstellar medium (ISM) and hot halo gas in three
distinct ways: (i) a fraction of the disk ISM can be ``re-heated''
from the cold phase to the hot phase, where the re-heated gas is mixed
with the hot halo gas; (ii) a fraction or all of the re-heated gas can
be directly ejected from the host halo without mixing with the hot
halo gas; and (iii) if the SN energy from all galaxies in a halo is
sufficiently large, the hot gas in the host halo can be heated,
causing a fraction of the halo hot gas to be ejected from the halo
itself. Gas that is ejected from the halo is stored in a separate
reservoir and may be re-accreted later --- we return to this below.

In all three models, the mass flux that is re-heated from the cold gas
is proportional to star formation rate,
\begin{equation}\label{equ:fb_rh}
\dot{m}_{\rm out}=f_{\rm ld} SFR~,
\end{equation}
where the coefficient $f_{\rm ld}$ is frequently referred to as the `mass-loading'
factor. Different models have different loading factors and treatments
for the ejection, which are described below
for our three models.

In addition, all three models assume that the ejected gas can
re-collapse into the halo at later times as hot halo gas. 
As in \citet{Springel2001} and \citet{DeLucia2007}, the rate of
re-infall of rejected gas is given by
\begin{equation}\label{equ:fb_ri}
\dot{m}_{\rm re-infall}=\chi_{\rm ri} \left( { M_{\rm ej} \over \tau_{\rm dyn}} \right).
\end{equation}
Here, $\chi_{\rm ri}$ is a free parameter, $M_{\rm ej}$ is the total
mass of gas in the `ejected' reservoir and $\tau_{\rm dyn} =
\rvir/\vvir$ is the dynamical time of the halo.

\subsubsection{The Croton model}
This model assumes that the amount of cold gas re-heated from the disk
depends only on the star formation rate, e.g.
\begin{equation}\label{equ:rh_croton}
\dot{m}_{\rm rh}=\alpha_{\rm LD} SFR~, 
\end{equation}
where $\alpha_{\rm LD}=3.5$, consistent with observations of
\citet{Martin1999}. The total amount of energy released by SN
explosions and coupled into feedback over the relevant time interval
during which an amount 
$SFR\Delta t$  of stellar mass formed is written as
\begin{equation}
\Delta E_{\rm SN}=0.5 \alpha_{\rm SN} V_{\rm SN}^2 SFR \Delta t~,
\end{equation} 
where $0.5V_{\rm SN}^2$ is the mean energy in kinetic form injected by
supernovae per unit mass of star formation, which is taken to be $V_{\rm SN}=630\ {\rm km\,s^{-1}}$,  
and the parameter
$\alpha_{\rm SN}$ controls the efficiency with which this energy can
actually power feedback. The amount of energy required to
adiabatically reheat $\Delta m_{\rm rh}$ of cold gas and add it to the
hot halo reservoir is
\begin{equation}
\Delta E_{\rm rh}=0.5 \Delta m_{\rm rh} \vvir^2~.
\end{equation}	
If $\Delta E_{\rm excess} = \Delta E_{\rm SN} - \Delta E_{\rm rh}$ is
positive then enough energy is provided to physically unbind some
fraction of the hot gas from the halo. The mass of the ejected hot gas
is then
\begin{equation}\label{equ:ej_croton}
\Delta M_{\rm ej} = { \Delta E_{\rm excess} \over E_{\rm hot} } M_{\rm hot} = \left( \alpha_{\rm SN} { V_{\rm SN}^2 \over \vvir^2} - \alpha_{\rm LD} \right) SFR  \Delta t~,
\end{equation}
where $E_{\rm hot}$ represents the binding energy of the hot halo gas,
$E_{\rm hot}=0.5M_{\rm hot}\vvir^2$.  This ejected gas is added to
an external reservoir of material, $M_{\rm ej}$. Ejected material may
fall back into the potential well of the halo at a later time and be
added back to the hot component, as described above.  The model
parameters are $\alpha_{\rm LD}=3.5$, $\alpha_{\rm SN}=0.35$, and
$\chi_{\rm ri}=0.5$.

\subsubsection{The Somerville model}
The mass reheating rate of the cold gas in the Somerville model is
given by
\begin{equation}\label{equ:rh_somerville}
\dot{m}_{\rm rh} = \alpha_{\rm LD} \left( {V_{\rm disk} \over 200 {\rm km\,s^{-1}}} \right)^{-\beta_{\rm LD}} SFR~,
\end{equation}
where $\alpha_{\rm LD}$ and $\beta_{\rm LD}$ are free parameters. The
circular velocity of the disk, $V_{\rm disk}$, is taken to be the
maximum rotation velocity of the (uncontracted) dark matter halo,
$V_{\rm max}$. After the gas is removed from the galaxy it can either
be trapped within the potential well of the dark matter halo and
deposited into the hot gas reservoir, or ejected from the halo into
the intergalactic medium (IGM). This model assumes that a fraction of
the re-heated gas can be blown out of the halo, governed by
\begin{equation}\label{equ:ej_somerville}
f_{\rm ej}(\vvir) = \left[ 1+ \left( { \vvir \over V_{\rm EJ}} \right)^{\beta_{\rm EJ}} \right]^{-1}~,
\end{equation}
with $\beta_{\rm EJ}$ and $V_{\rm EJ}$ as free parameters. This
function behaves as a smoothed step function. For halos with $V_{\rm
  c}$ much higher than $V_{\rm EJ}$, a negligible fraction of re-heated
gas is expelled out of the halo, and for halos with $\vvir$ much
lower than $V_{\rm EJ}$ all re-heated gas leaves the halo and is
deposited into $M_{\rm ej}$, where it is again subject to re-infall at
late times.

\subsubsection{The Lu model}

The Lu model assumes the loading factor for SN feedback is a power-law
function of halo circular velocity, similar to the Somerville model:
\begin{equation}\label{equ:rh_lu}
f_{\rm ld}=\alpha_{\rm LD} \left({\vvir \over V_0} \right)^{-\beta_{\rm LD}}~,
\end{equation}
where the power index $\beta_{\rm LD}$ and the normalization
$\alpha_{\rm LD}$ are model parameters, and $V_0$ is an arbitrary
scale, which is fixed at $220$ km/s.  To compute the fraction of gas
that is ejected from the halo, the model follows the same
parameterization as the Somerville model
(Eq. \ref{equ:ej_somerville}). Again, for a halo with a circular
velocity lower than $V_{\rm EJ}$, most of the outflow gas is ejected
out of the halo, while for halos with circular velocities much larger
than $V_{\rm EJ}$, the ejected fraction follows a power-law function
of the halo circular velocity.

In addition to reheating and ejecting the cold gas, the Lu model also 
expels the hot halo gas in a similar way as the Croton model. 
If there is still extra SN energy left after reheating and ejection, 
the surplus is assumed to power a wind, and the mass of the wind can be written as
\begin{equation}\label{equ:wind_lu}
\Delta m_{\rm wind} = \epsilon_{\rm W}   
  \left\{ \alpha_{\rm SN}\frac{V_{\rm SN}^2}{V_{\rm esc}^2} -
  f_{\rm ld} \left[\left(\frac{\vvir}{V_{\rm esc}}\right)^2+f_{\rm ej}\right]\right\}
   SFR  \Delta t~,
\end{equation}
where $V_{\rm esc}$ is the escape velocity of the halo. For a NFW halo
with a concentration $c$ \citep{Navarro1996}, $V_{\rm esc}=\vvir
\times \sqrt{ 2c \over \ln (1+c) -{c \over 1+c}}$. For halos with
$\vvir=100$km/s at $z=0$, the concentration is typically $\ga 10$ in
the current $\Lambda$CDM model \citep[e.g.][]{Zhao2009, Prada2012},
giving $V_{\rm esc}^2\approx 13 \vvir^2$. In total, there are 5
parameters -- $\alpha_{\rm LD}$, $\beta_{\rm LD}$, $\beta_{\rm EJ}$,
$V_{\rm EJ}$, and $\epsilon_{\rm W}$ -- governing SN feedback in this
model. Note that when $\beta_{\rm LD}$ and $\beta_{\rm EJ}$ in the Lu
model is set to be 0, and $V_{\rm esc}$ and $\vvir$ are not
distinguished, the Lu model is reduced to the Croton model. Similarly,
when $\epsilon_{\rm W}$ in Eq. \ref{equ:wind_lu} is 0, the Lu model
reduces to the Somerville model.

\subsection{AGN feedback and Halo quenching model}\label{sec:AGN}
The sharp decline of the number density of galaxies at the high
mass/luminosity end of the galaxy mass/luminosity function suggests a
mechanism (or mechanisms) at work in high mass halos to effectively
suppress star formation. 
The large reservoirs of hot gas that are detected observationally in
massive halos further suggest that cooling is also suppressed. Recent
models implementing the feedback from supermassive black holes have
demonstrated that AGN activity can shut off radiative cooling of the
hot halo gas and quench star formation in high-mass galaxies
\citep{Croton2006, Bower2006, Cattaneo2007, Somerville2008}. Although
both the Croton model and Somerville model make different assumptions
about the details of the AGN feedback processes, they share a common
conceptual picture of the growth of supermassive black hole mass and
its feedback. In these two models, the black holes grow their mass and
affect galaxy formation through two different modes: quasar mode and
radio mode. The quasar mode is the bright mode of black hole growth
observed as optical or X-ray bright AGN radiating at a significant
fraction of their Eddington limit ($L\approx(0.1-1)L_{\rm Edd}$)
\citep[][]{Vestergaard2004, Kollmeier2006}. Such bright AGNs are
believed to be fed by optically thick, geometrically thin accretion
disks \citep{Shakura1973}. In these models, the quasar mode is assumed
to be triggered by merger events, which also increase the mass of the
galactic bulge.  
A large fraction of massive galaxies are detected at radio wavelengths
\citep{Best2007} without showing the characteristic emission lines of
classical optical or X-ray bright quasars \citep{Kauffmann2008}. Their
accretion rates are believed to be a small fraction of the Eddington
rate and they are radiatively extremely inefficient. Although AGN
spend most of their time in the radio mode, they gain most of their
mass during the short and Eddington-limited quasar mode.

For the Croton model and the Somerville model in this paper, the
quasar phase is triggered solely by galaxy merger events. Whenever the
two progenitor galaxies merge, their black holes are assumed to also
merge and form a single black hole whose mass is the sum of the
progenitor black holes' masses. After that, rapid gas accretion onto
the black hole occurs. This is motived by the gas inflow into the
nuclear regions of galaxies seen in disk merger simulations
\citep{Springel2005, Cox2006a, Hopkins2006a, Hopkins2007a,
  Robertson2006}. During the merger, the black hole is assumed to grow
rapidly with accretion rates near the Eddington limit. This rapid
accretion continues until the energy being deposited into the ISM in
the central region of the galaxy is sufficient to significantly offset
and eventually halt accretion via a pressure-driven outflow. We now
describe the models in more detail for both quasar and radio mode, and
their effect on galaxy formation.

\subsubsection{The Croton model}
In the Croton model, during a quasar mode event progenitor black holes
are assumed to coalesce with no loss of mass due to dissipative
processes. A fraction of the cold gas of the progenitor galaxies is
also accreted by the central black hole, causing the black hole mass to increase by:
\begin{equation}\label{equ:mbh_quasar_croton}
\Delta m_{\rm BH, quasar} = { f_{\rm BH} M_{\rm cold} M_{\rm sat} / M_{\rm central} \over 1 + (280 {\rm km\,s^{-1} /\vvir})^2}~,
\end{equation}
where $M_{\rm central}$ and $M_{\rm sat}$ are the cold baryon mass
(stars and cold gas) of the central galaxy and the merging satellite,
and $f_{\rm BH}$ is a free parameter governing the efficiency of the
accretion event.  As described above, the quasar mode is the dominant
growth mechanism for black holes in the model. 
The version of the Croton model we adopt in this paper calculates the total energy in a quasar wind
multiplied by a coupling parameter, $E_{\rm quasar}=0.1\eta_{\rm AGN}
\Delta m_{\rm BH, quasar} c^2$, and the binding energy of the cold gas
in the galaxy, $1/2 M_{\rm cold} \vvir^2$, and the binding energy
of the hot gas in the halo, $1/2 M_{\rm hot} \vvir^2$.  The model
then compares the energy radiated by the quasar and the binding energy
of the cold gas. If the quasar energy is larger than the cold gas
binding energy, all the cold gas is ejected from the disk and its mass
is added into the ejected reservoir. If the quasar wind energy is
larger than the combined cold and hot gas binding energy, then the
model ejects both the cold and hot gas.
It is worth noting that the Croton model does not explicitly impose an Eddington limit 
for the accretion rate in the quasar mode and, therefore, does
not make assumptions about black hole seed mass. 

The second mode of AGN, the radio mode, also enables the black hole to
gain mass, but generally at a much slower rate. In contrast to the
quasar mode, the radio mode has low Eddington ratio accretion rates,
is radiatively inefficient and is associated with efficient production
of radio jets and buoyant bubbles that can heat gas in a
quasi-hydrostatic hot halo. Here, the accretion is from the
surrounding hot gas and is assumed to be continuous as long as a hot
halo is present. In the Croton model, it is characterized by the
following simple relation:
\begin{equation}\label{equ:mbh_radio_croton}
\dot{m}_{\rm BH,radio} = \kappa_{\rm AGN} \left( {M_{\rm BH} \over 10^8 \msun}\right) \left({f_{\rm hot} \over 0.1}\right) \left({\vvir \over 200 {\rm km \, s^{-1}}}\right)^3~,
\end{equation}
where $\kappa_{\rm AGN}$ is a free parameter governing the rate of the
black hole mass growth in the radio mode, and $f_{\rm hot}$ is the
fraction of the mass of the halo in the hot gas component. In contrast
to the quasar mode, the radio mode results in the injection of
feedback energy directly into the hot halo gas. The injected energy is
\begin{equation}\label{equ:agn_radio_croton}
L_{\rm heat} = \eta_{\rm AGN} \dot{M}_{\rm BH}c^2 ~,	
\end{equation}
where $\eta_{\rm AGN}$ is a free parameter governing the efficiency of
radio mode heating.  Assuming that gas is thermalized to the virial
temperature of the halo, the heating rate is
\begin{equation}\label{equ:agn_hr_croton}
\dot{m}_{\rm heat} = { L_{\rm heat} \over 1/2 \vvir^2}~.
\end{equation}
The net cooling rate is then the usual cooling rate minus the heating
rate.  Radio mode feedback can result in a reduction, or even complete
cessation, of cooling onto the disk, depending on how this heating
rate compares with the cooling rate of the halo gas. This can have a
dramatic effect on the production of new stars, especially in massive
galaxies.

\subsubsection{The Somerville model}
The Somerville model seeds a black hole with mass $\sim 100\msun$ for
every top-level progenitor halo in the merger tree. When galaxy-galaxy
mergers occur, and hence a quasar mode event is triggered, the final
black hole mass 
$m_{\rm BH,final}$  at the end of the blow-out phase
is assumed to be related to the mass of the spheroidal component after
the merger. The limiting black hole mass is given by the following
black hole mass to spheroid mass ratio scaling relation from
simulations \citep{Hopkins2007a},
\begin{equation}\label{eqn:bhm}
\log \left({M_{\rm BH,0} \over M_{\rm sph}}\right) = -3.27 + 0.36 \cdot  {\rm erf} \left[(f_{\rm cold} - 0.4)/0.28\right]\,,
\end{equation}
with a scatter around the relationship of $\sigma_{\rm BH}\sim
0.2-0.3$dex, where $\sigma_{\rm BH}$ is a free parameter. For the
model adopted in this paper, $\sigma_{\rm BH}=0.3$dex. In the above
scaling relation, $f_{\rm cold}$ is simply the cold gas fraction of
the combined merging galaxy pair, $f_{\rm cold}=M_{\rm cold} / (M_{\rm
  cold}+M_*)$.  The final black halo mass is the value of the black
hole mass computed from Eq.\ref{eqn:bhm} times a free parameter,
$f_{\rm BH,FINAL}$, e.g.
\begin{equation}
M_{\rm BH,final}= f_{\rm BH,FINAL} M_{\rm BH,0}~.
\end{equation}
The value of $f_{\rm BH,FINAL}$ is taken from numerical hydro
simulations of galaxy mergers.

During the quasar phase, black holes increase their mass in two
different growth regimes: an Eddington-limited phase and a power-law
decline phase of accretion. In the first regime, the black hole
accretes at the Eddington limit until it reaches a critical black hole
mass $M_{\rm BH, crit}$. Following the numerical simulations of
\citet{Hopkins2007a} and the family of light curves defined by
\citet{Hopkins2006a}, the Somerville model assumes that
\begin{equation}
M_{\rm BH, crit} = 1.07 f_{\rm BH,C}  \left({ M_{\rm BH, final} \over 10^9 \msun}\right)^{1.1}~,
\end{equation}
where the coefficient $f_{\rm BH, C}$ determines how much of the black
hole growth occurs in the Eddington-limited versus the power-law
decline phase. It is set to be $f_{\rm BH, C}=0.8$ according to the
merger simulations.  

The model assumes that, when $M_{\rm BH}<M_{\rm BH, crit}$, the black
hole is in the Eddington-limited phase, where the black hole accretes
mass at a fraction of the Eddington limit. 
When the black hole mass exceeds the critical mass $M_{\rm BH, crit}$, the black hole enters the `blow-out' phase, which is modeled by a power-law decline in the accretion rate according to the light curves found in merger simulations from \citet{Hopkins2006a}.
This accretion drives a galactic wind. The outflow mass flux powered
by this galactic wind is
\begin{equation}
\dot{M}_{\rm out, Q} = \epsilon_{\rm wind} \cdot \eta_{\rm rad} \cdot { c \over V_{\rm esc}} \cdot \dot{M}_{\rm acc}~.
\end{equation}
When the black hole reaches the mass $M_{\rm bh, final}$, the
accretion onto the black hole shuts off.
This part of the model is detailed in \citet{Hirschmann2012}.
Readers are referred to this paper for a description.

For the radio mode, the Somerville model assumes a Bondi-Hoyle-type
accretion combined with an isothermal cooling flow solution
\citep{Nulsen2000}. The model calculates the accretion rate in the
radio mode by
\begin{equation}\label{equ:mbh_radio_somerville}
\dot{M}_{\rm radio}=\kappa_{\rm radio} \left[{ kT \over \Lambda(T, Z)}\right] \left({M_{\rm BH} \over 10^8 \msun} \right),
\end{equation}
where $T$ and $Z$ are the temperature and metallicity of the hot halo
gas, and $\Lambda(T, Z)$ is the cooling function. The central black
hole accretes at this rate whenever hot halo gas is present
(``hot-mode'' accretion, $r_{\rm cool} < \rvir$). The energy
associated with this accretion effectively couples to and heats the
hot gas and is given by
\begin{equation}\label{equ:agn_radio_somerville}
L_{\rm heat}= \kappa_{\rm heat}\eta_{\rm rad} \dot{M}_{\rm BH, radio}c^2~.
\end{equation} 
The model assumes a similar heating rate to the Croton model
(Eq. \ref{equ:agn_hr_croton}) but with a slightly different
coefficient,
\begin{equation}\label{equ:agn_hr_somerville}
\dot{m}_{\rm heat} = { L_{\rm heat} \over 3/4 \vvir^2}.
\end{equation}

\subsubsection{The Lu model}
The Lu model does not include an explicit black hole accretion and AGN
feedback model. However, it adopts a halo quenching model to mimic the
effects of AGN feedback that stops radiative cooling in high mass
halos \citep{Cattaneo2006, Bower2006}. In the model, when the halo
mass exceeds a critical quenching mass threshold, $M_{\rm cc}$,
radiative cooling of the hot halo gas ceases. For each merger tree,
the Lu model draws a random number for $\log M_{\rm cc}$ from a normal
distribution with a mean $\log M_{\rm CC,0}$ and a standard deviation
$\sigma_{\rm CC}$. When a halo in the merger tree has a virial mass
higher than $M_{\rm cc}$ the radiative cooling of the halo is switched
off. $\log M_{\rm CC,0}$ and $\sigma_{\rm CC}$ are taken as free
parameters. We find that the rapid declining high-mass end of the
stellar mass function constrains $M_{\rm CC,0}$ to be around
$10^{12}\, \msun$, and $\sigma_{\rm CC}\sim 0.3$ dex.

\subsection{Satellite galaxies: dynamical friction and galaxy mergers}\label{sec:merger}

\subsubsection{The Croton model}
The Croton model makes use of the subhalo information from the
simulation merger tree whenever the subhalo is resolved. The version
of the Croton model we adopt in this paper implements a new
prescription for satellite galaxies (Croton et al. in prep). Here,
subhalos are treated in a similar way to primary halos. Firstly,
subhalos are allowed to host hot gas, which can cool to fuel the
central galaxy of the subhalo. The radiative cooling is treated in the
same way as that described in Section\,\ref{sec:cooling}. Unlike the
primary halos, however, subhalos cannot acquire more baryons from the
hot halo of the host primary halo. When the halo mass of the subhalo
decreases due to tidal stripping, the hot gas of the subhalo is
stripped in proportion to the dark matter mass stripping. Note that,
for simplicity, any gas that is ejected from a subhalo due to SN or
quasar winds is added to the primary halo ejected component, not the
subhalo. This gas can then be reincorporated at a later time into the primary halo. 

The Croton model has also implemented a new prescription to determine
the fate of satellite galaxies that are orbiting within larger
primary halos. First, the model adopts the dynamical friction formula
\citep{Binney1987} to calculate the \emph{average} merger time for a
subhalo upon infall
\begin{equation}\label{equ:df_croton}
t_{\rm fric}={1.17 r_{\rm sub}^2 \vvir \over \ln \Lambda G M_{\rm sub}}~, 
\end{equation}
where $r_{\rm sub} = R_{\rm vir}$ and $M_{\rm sub}$ are the radial distance of the subhalo from
the center of the host and
subhalo mass of the satellite galaxy measured at infall, and $\ln
\Lambda$ is the Coulomb logarithm, approximated as $\ln \Lambda =\ln
(1+M_{\rm vir}/M_{\rm sub})$. Second, the model keeps track of the
dark matter mass of the hosting subhalo to galaxy mass (cold gas mass plus stellar mass) ratio
for every subhalo. 
The model adopts a critical halo mass to galaxy
mass ratio, $f_{\rm ST,crit}$, as a model parameter. When the dark
matter mass hosting the satellite is stripped such that the subhalo
mass to galaxy mass ratio drops below this critical ratio, the model
determines the fate of the subhalo galaxy according to the estimated
dynamical friction timescale. If the subhalo has survived longer than
the expected average dynamical friction times, this implies the
subhalo is more bound than average and hence the model merges the
galaxy with the central galaxy and removes the subhalo from the
tree. If the subhalo reaches the critical mass ratio before the
dynamical friction time, this implies the subhalo was less bound than
average, and the model completely strips the galaxy and adds its
stellar mass into the diffuse stellar mass component of the primary
halo, with any remaining satellite gas going into the primary hot
component. Thus, in the Croton model there are no orphan galaxies
(satellite galaxies without host subhalos), because when a subhalo is
significantly stripped, the hosted galaxy is either merged into the
central galaxy or completely disrupted.

For satellites that do merge, the treatment of the merger remnant
depends on the mass ratio of the two galaxies, $M_{\rm sat}/M_{\rm
  central}$, where $M_{\rm sat}$ and $M_{\rm central}$ are the cold
baryon masses, stellar mass plus cold gas mass, of the satellite and
the central galaxy. Mergers are considered major or minor depending on
whether $M_{\rm sat}/M_{\rm central}$ is larger or smaller than the
value of the free parameter, $f_{\rm MG}$. The Croton model adopts
$f_{\rm MG}=0.3$.

For a minor merger, the satellite's stars are added to the central
bulge, and its gas is added to the central disk. For a major merger,
the model combines all the existing stars from the two merging
galaxies into a central galaxy, which is now assumed to be
spheroidal. All mergers trigger a star-burst, and all stars formed in
the burst are added into the central spheroidal component of the
galaxy. A fraction, $e_{\rm burst}$, of the combined cold gas in the
two merging progenitors becomes stars, and the rest of the gas joins
the gaseous disk. The Croton model assumes that $e_{\rm burst}$
depends on the ratio of the baryon masses of the two galaxies:
\begin{equation}\label{equ:sb}
e_{\rm burst}=\alpha_{\rm burst}(M_{\rm sat}/M_{\rm central})^{\beta_{\rm burst}}~,
\end{equation}
as in the so-called `collisional star-burst model' of \citet{Somerville2001b}.

\subsubsection{The Somerville model}
In the Somerville model, when a lower mass halo merges into a higher
mass halo, the hot gas associated with the lower mass one is
instantaneously stripped and added into the hot gas of the primary
halo.  In the subsequent evolution, the Somerville model does not use
the subhalo information from the $N$-body simulation to follow the
satellite galaxies.  Instead, it adopts a version of the dynamical
friction model proposed by \citet{Boylan-Kolchin2008} right after a
galaxy enters the virial radius of a larger halo.  The dynamical
friction timescale is modeled
\begin{equation}\label{equ:df_somerville}
t_{\rm fric}=A {(M_{\rm vir,1} /M_{\rm vir, 2})^b \over \ln(1+M_{\rm vir,1}/M_{\rm vir,2})} \exp(c\eta) \left[{r_{\rm c}(E) \over \rvir}\right]^d \tau_{\rm dyn}~,
\end{equation}
where $M_{\rm vir, 1}$ and $M_{\rm vir, 2}$ are the virial mass of the
primary and secondary halos before merging, $\eta=j/j_{\rm c}(E)$ is
the specific angular momentum relative to a circular orbit with the
same energy, $r_{\rm c}(E)$ is the circular radius of a circular orbit
with the same orbital energy, and $\tau_{\rm dyn}$ is the dynamical
time scale of the primary halo, $\tau_{\rm dyn}=\rvir / V_{\rm
  c}$. The parameters defined in the formula are fixed to the values
derived by \citet{Boylan-Kolchin2008} based on numerical
simulations, $A=0.216$, $b=1.3$, $c=1.9$, $d=1.0$.

When it first becomes a satellite, each sub-halo is assigned a value
of $\eta$ and $r_{\rm c}(E)$ by choosing $\eta$ in the interval $0\leq
\eta \leq 1$ from a Gaussian distribution with mean $0.5$ and
dispersion $\sigma = 0.214$, and choosing $r_{\rm c}(E)/r_{\rm vir}$
from a uniform distribution on the interval $[0.6-1]$. 
\citet{Zentner2005} showed that this is a good representation of the satellite
orbits in cosmological simulations.

Following \citet{Taylor2001a}, the model accounts for tidal
stripping as the satellite orbits within the host halo.  The rate of
mass loss due to tidal stripping is calculated via:
\begin{equation}
\zeta = -\log [f_{\rm strip} (0.35 \eta^2 -0.2\eta +0.58) ]
\end{equation}
where $f_{\rm strip}$ is an adjustable parameter. The satellite mass
at each time-step is then given by:
\begin{equation}
m_{\rm strip} = m_i \exp(-\zeta t_{\rm acc}/P) 
\end{equation}
where $m_i$ is the mass of the satellite at infall, $t_{\rm acc}$ is
the time since infall, and $P=2\pi \tau_{\rm dyn}$ is the orbital
period. When the mass of satellite falls below a fraction, $f_{\rm
  dis}$, of the mass enclosed by $r_s$, the Navarro-Frenk-White scale
radius, the satellite is considered `disrupted'.  The parameter,
$f_{\rm dis}$, is here set equal to unity.  The cold gas within a disrupted
satellite is added to the hot gas reservoir of the host halo, the
stars are added to a `diffuse' stellar halo, and the halo is removed
from all further calculations. The treatment of subhalo merging and
disruption has been tested by comparing with the results from high
resolution dissipationless cosmological N-body simulations. The model
reproduces the subhalo conditional mass function and radial
distribution of surviving subhalos.

In the Somerville model, a more sophisticated starburst model is
adopted. At the beginning of a galaxy merger, the model allocates a
reservoir of `burst fuel' $m_{\rm burst}= e_{\rm burst} M_{\rm cold}$,
where $M_{\rm cold}$ is the combined cold gas from both of the
progenitor galaxies. The burst continuously converts a fraction of
this fuel into stars until it is exhausted. The efficiency parameter
$e_{\rm burst}$ and the timescale $\tau_{\rm burst}$ for the
burst-mode star formation depend on the baryonic mass ratio of the two
progenitors, the combined cold gas content of the merging galaxies,
the bulge-to-total stellar mass ratio, virial velocity of the host
halo and redshift of the merger. The recipe for the starburst
efficiency and timescale functions of these variables is based on
hydrodynamic simulations of binary galaxy mergers
\citep{Robertson2006,Cox2008}. We refer to \citet[][]{Somerville2008}
(Section 2.5.2 and reference therein) for details of the model. 
In the models used here, we use the updated model for the burst
efficiency presented in \citet{Hopkins2009}.

\subsubsection{The Lu model}
Like the Croton model, the Lu model uses the subhalo information in
the merger trees to follow the satellite population. 
Unlike the Croton model, however, the model estimates the dynamical
fraction timescale from the time
 when the subhalo is no longer resolved in the
simulation. When the identification of a subhalo disappears in the
merger tree, the model keeps the satellite galaxy hosted by the
subhalo orbiting in the main halo for a dynamical friction time,
$t_{\rm fric}$, defined above by Equation~\ref{equ:df_croton}
\citep{Binney1987}. To calculate this time the properties of the
satellite and the host halo at the time when the subhalo was last
identified are used (this differs from the Croton model where $t_{\rm
  fric}$ is calculated upon initial infall of the subhalo). If the
main halo merges into another halo before this merger time is reached,
a new value for the merger time is calculated and the merger clock is
restarted.

The Lu model assumes that satellite galaxies will merge into the
central galaxy of the host halo after a time 
$t_{\rm orb}$ since the
satellite's host subhalo is no longer resolved in the simulation. To
allow uncertainty for the estimation of the dynamical friction time
scale, the model adopts a factor $f_{\rm DF}$ of $t_{\rm fric}$ to be
the orbiting time scale, i.e.
\begin{equation}\label{equ:df_lu}
t_{\rm orb}=f_{\rm DF}t_{\rm fric}
\end{equation}
where $f_{\rm DF}$ is a free parameter. When two galaxies merge, the
Lu model also assumes that a certain amount of cold gas from the two
merging galaxies is converted into stars in a starburst. To do this,
the Lu model follows the same model assumed by Eq. \ref{equ:sb}, with
$\alpha_{\rm burst}$ and $\beta_{\rm burst}$ as free parameters.

\subsection{Chemical enrichment}\label{sec:chemical}
All three models assume that when a parcel of new stars form, d$M_*$,
a certain mass of metals, ${\rm d}M_Z = y{\rm d}M_*$, is created 
and instantaneously mixed back into the cold gas in the
disk. A \citet{Chabrier2003} initial mass function (IMF) and a
constant yield of $y=1.5Z_{\odot}$ is taken throughout.

New stars have the average metallicity of the cold gas at the time they
formed. When cold gas is blown out of the disk due to SN or quasar
winds, the associated metals are mixed with either the hot gas or
ejected from the halo altogether in the same proportion as they were
re-heated. Note that all the three models adopt the instantaneous
recycling assumption, meaning that some fraction of the 
mass that is turned into stars
at each time-step is instantaneously returned to the cold
gas disk due to very short lived stars and mass lost from stellar
winds. For a Chabrier IMF
the recycling fraction is $f_{\rm rec}=0.43$.
We also deposit the new metals into the cold gas.

\subsection{Semi-empirical model}
In this paper we also compare some of the SAM predictions with the
semi-empirical model of \citet{Behroozi2012}, which is built on the
same halo merger trees. The semi-empirical results are derived in a
very different way than the traditional semi-analytic models discussed
above. Rather than parameterizing the physics of star formation,
\citet{Behroozi2012} use a flexible parametrization for the stellar
mass--halo mass relationship, $SM(M_h, z)$. This comprises six
parameters to control the relationship at fixed redshift, including a
characteristic stellar mass, halo mass, faint-end slope, massive-end
cutoff, transition region shape, and scatter. For each parameter set,
three secondary parameters control the redshift scaling at low
($z=0$), mid ($z\sim 1$) and high ($z>3$) redshift.  Additional
nuisance parameters account for systematic uncertainties in converting
between galaxy luminosities and stellar masses/star formation rates.

A specific choice of $SM(M_h, z)$ applied to the halo merger trees
will result in predictions for observed stellar mass functions, cosmic
star formation rates, and specific star formation rates.  Comparing
these predictions to observed data from $z=0$ to $z\sim8$ results in a
likelihood for a given choice of $SM(M_h, z)$.  The posterior
distribution for $SM(M_h,z)$, along with derived values for the
average star formation rate as a function of halo mass and redshift,
are then inferred from observed data using an MCMC approach.  The full
details of this method are presented in \citet{Behroozi2012}.

\section{Methodology}\label{sec:method}
\subsection{Merger trees}
The merger trees employed for the SAMs in this paper are extracted
from the Bolshoi $N$-body cosmological simulation \citep{Klypin2011}.
It was run in a
volume 250 $h^{-1}$\,Mpc on a side using $\sim8$ billion particles
with mass and force resolution adequate to follow subhalos down to the
completeness limit of $V_{\rm max} = 50 {\rm km \,s}^{-1}$ (halo
maximum circular velocity).  There are 180 stored time steps for
merger tree construction, which should yield stable SAM predictions
for low-z galaxy properties \citep{Benson2012}.  Dark matter halo
finding was done with the Rockstar code \citep{Behroozi2013a}.  The
approach is based on adaptive hierarchical refinement of
friends-of-friends groups in six phase-space dimensions and one time
dimension, which allows for robust (grid-independent,
shape-independent, and noise-resilient) tracking of substructure.  The
merger trees were generated using the Consistent Trees algorithm,
which simulates the gravitational motion of halos, to improve the
completeness and purity of both merger trees and halo catalogs
\citep{Behroozi2013}.

\subsection{Model Tuning}
All models are tuned independently to the same calibration data set,
either by ``hand'' or by ``machinery''. We choose the stellar mass
function of local galaxies estimated by \citet{Moustakas2013} as our
primary calibration data set. The mass function is derived from a
sample of 504,437 galaxies selected from the SDSS Date Release 7
\citep[DR7;][]{Adelman-McCarthy2008}. The stellar mass measurements
are determined from SED fitting of multiple photometric bands, including
SDSS optical bands, GALEX UV bands, and 2MASS $JHK_{\rm s}$ bands and
the photometry at 3.4$\mu$m and 4.6$\mu$m from the WISE All-Sky Data
Release. The authors adopted the 
\citet[][]{Chabrier2003} IMF from
0.1-100 $\msun$, as adopted in 
the SAMs presented here.
They assumed exponentially
declining star formation histories with stochastic bursts and allowed
a wide range of galaxy ages and histories, adopting reasonable priors
on stellar metallicity and dust attenuation. The error bars in the
data include both the uncertainty due to sample size (i.e., Poisson
error) and sample variance. The authors also empirically determined
the stellar mass completeness limits of their sample. Galaxies with
stellar mass larger than $10^9\, \msun$ are above the surface brightness
and stellar mass-to-light ratio completeness limits
\citep{Blanton2005, Baldry2008}. The main constraint of the mass
function for the models is from bins with $M_*\geq10^9\msun$. For
further details, readers should refer to \citet{Moustakas2013}.

The Lu model has been calibrated with a MCMC engine, which supports
efficient model parameter space exploration. To do this, the method
first defines a likelihood function, which quantifies the closeness
between the model and calibration data. As the low-mass end of the
stellar mass function potentially involves considerable
incompleteness, they also treat this incompleteness in the likelihood
model. The Lu model defines the completeness fraction as the ratio
between the observed number density and the underlying number density
of galaxies. For a given stellar mass bin $i$, the completeness
fraction is $p_i={\phi_{{\rm obs},i}\over \phi_i}$. If a model
predicts $\phi_i$ for a mass bin, the expected number density of
galaxies to be observed is $p_i \phi_i$. The completeness fraction
$p_i$ is assumed to be a power-law function of stellar mass as implied
by the estimate of \citet{Baldry2008} who showed that the slope of the
galaxy stellar mass function for mass lower than $10^{8.6}\msun$ can
be as steep as $-1.8$ (solid gray line in Figure 1). Following this
estimate, the Lu model assumes that $p_i$ remains unity for
$M_{*}>10^9\msun$ but decreases toward the low-mass end following a
power law of stellar mass as
\begin{equation}
p_i=\left({M_{{*},i} \over 10^9\msun}\right)^{\alpha_{\rm IN}}~,
\end{equation}
where $\alpha_{\rm IN}$ is the difference of the faint-end slope
between the incomplete and the complete sample. The model adopts
$\alpha_{\rm IN}$ as a free parameter to allow it to change with a
uniform prior between 0 and $1.80-1.05=0.75$. With this approach, if
the SAM predicts a faint-end slope of $-1.8$, we would expect the
observed faint-end slope to be between $-1.8$ and $-1.05$.

Taking into account the incompleteness, the $\chi^2$-like logarithmic
likelihood for the stellar mass function is
\begin{equation}
\ln L= \sum_{i=0}^k {(\phi_{i,obs} - p_i\phi_{i,mod}(\theta))^2 \over \sigma_i^2},
\end{equation}
where $\phi_{i, obs}$ denotes the data points from
\citet{Moustakas2013}, and $\phi_{i,mod}$ denotes the model prediction
from the SAM.  Thus, the likelihood is not only a function of the model
parameter vector $\theta$ but also a function of $\alpha_{\rm
  IN}$. $\alpha_{\rm IN}$ is treated as a nuisance parameter and its
distribution is sampled along with other parameters using
MCMC. Finally, the distribution of $\alpha_{\rm IN}$ is marginalized
over when deriving the posterior of other model parameters
\citep{Lu2012}.

For the Lu results, the MCMC is run for 6,000 iterations with 256
parallel chains using the differential evolution algorithm
\citep{Braak2006} until converged. The convergence test is done with
the Gelman-Rubin test \citep{Gelman1992}. We obtain 750,000
posterior samples from the MCMC to make predictions for the Lu
model. For this particular run, the ``best fit'' model (maximum
likelihood) is close to the median model of the full posterior.  For
the Croton model and the Somerville model, the choices of parameters
are kept close to those in the published papers,
but some hand-tuning has been done to get as close a fit as possible to the 
\citet{Moustakas2013} mass function.  The exact parameter
values are listed in Table \ref{tab:model}. For the Lu model, except
for a few fixed parameters, the 95\% posterior range for each of free
parameters is given in the parentheses in the table.

\section{Model results}\label{sec:results}

\subsection{Stellar mass functions at $z=0-6$}

\begin{figure*}[htb]
\begin{center}
\begin{tabular}{cc}
\includegraphics[width=0.35\textwidth]{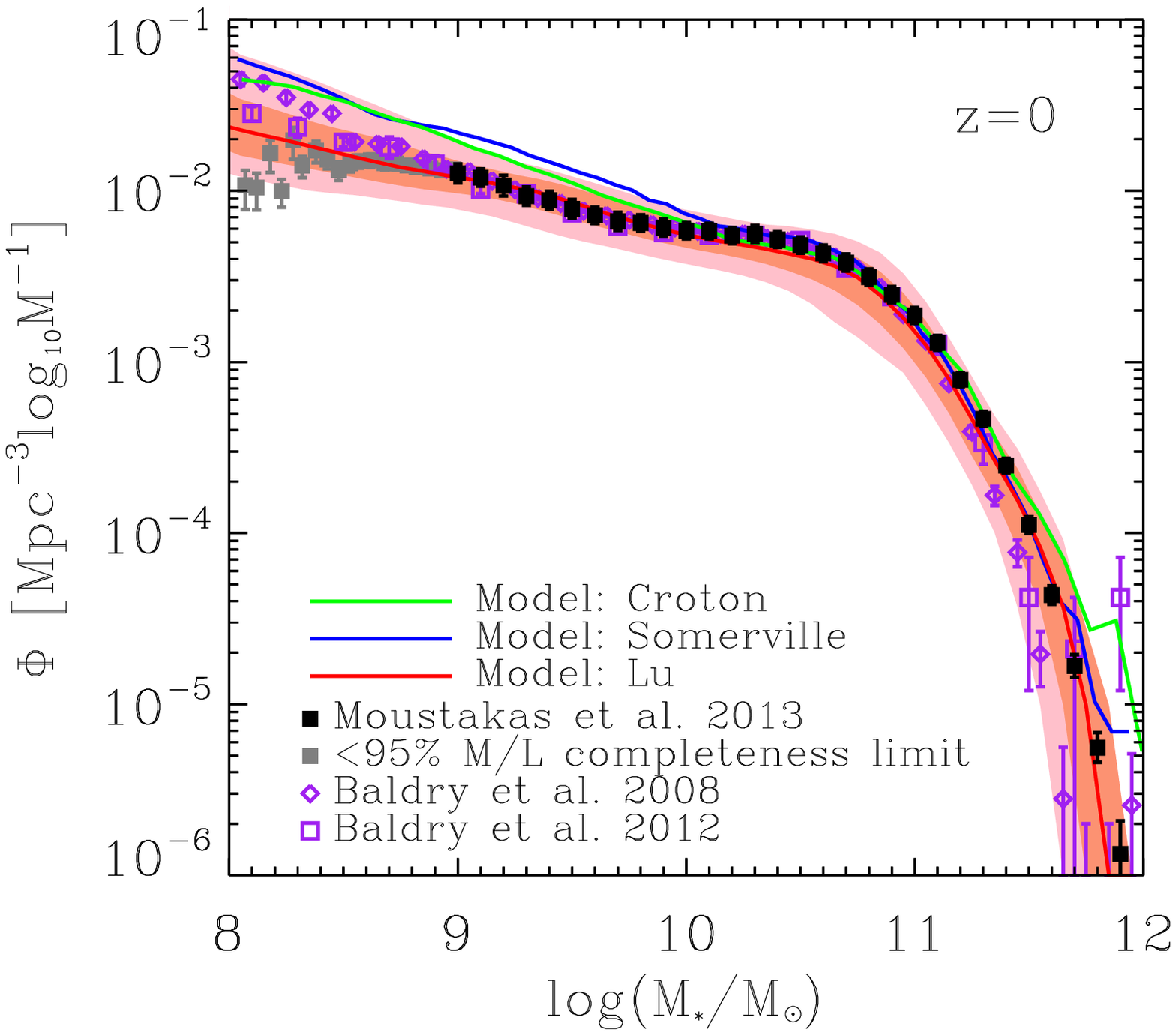} &
\includegraphics[width=0.35\textwidth]{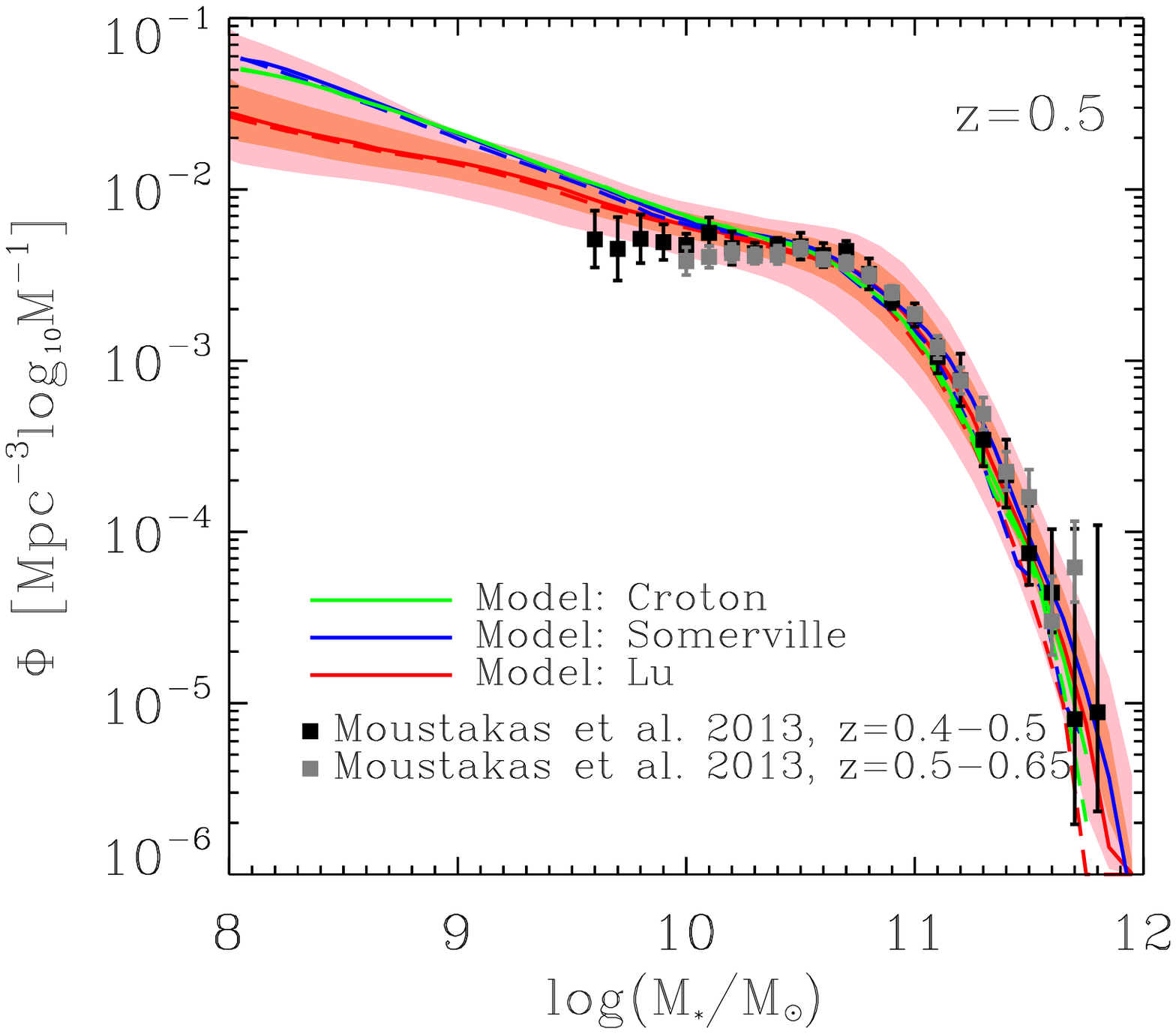} \\
\includegraphics[width=0.35\textwidth]{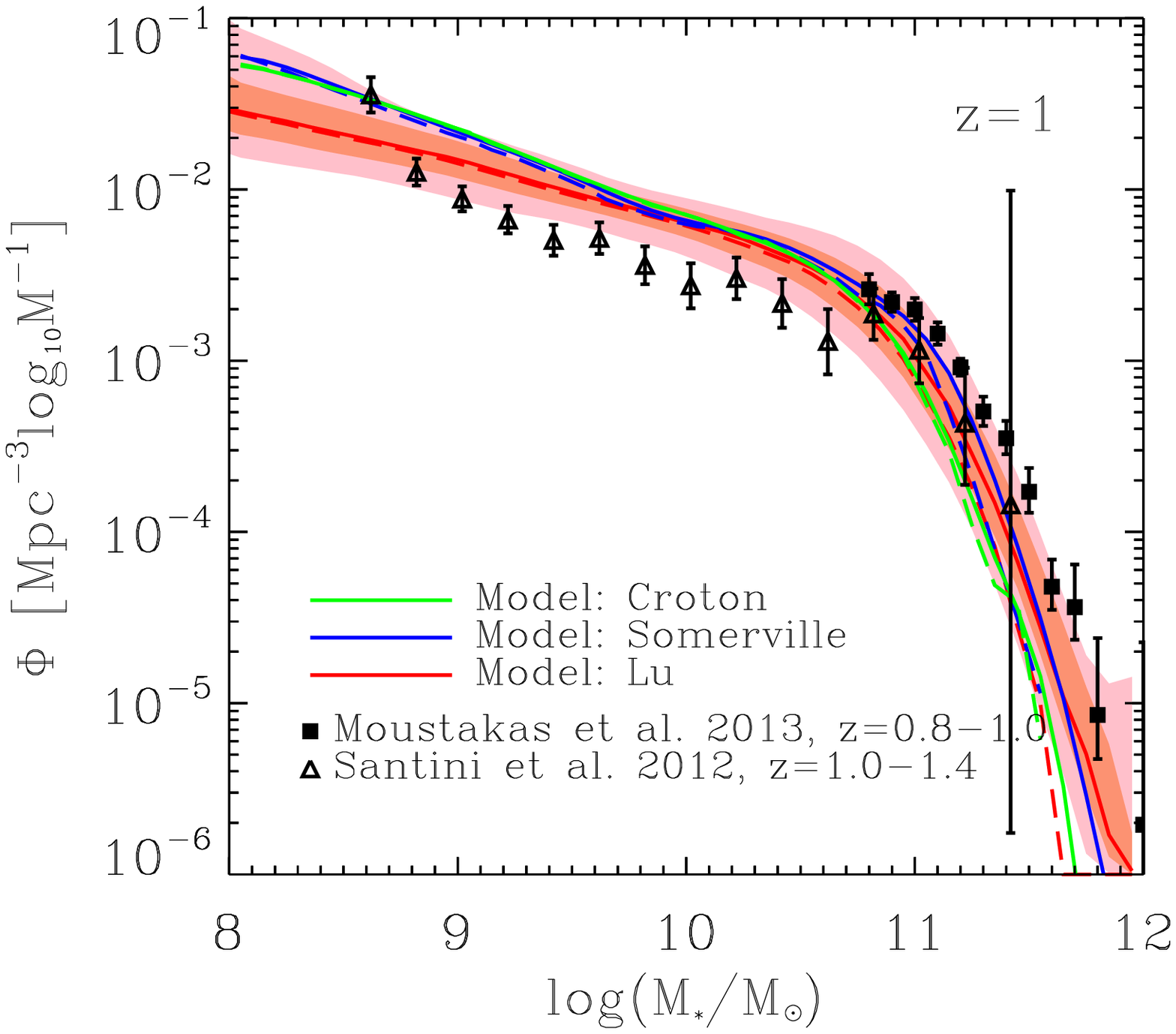} &
\includegraphics[width=0.35\textwidth]{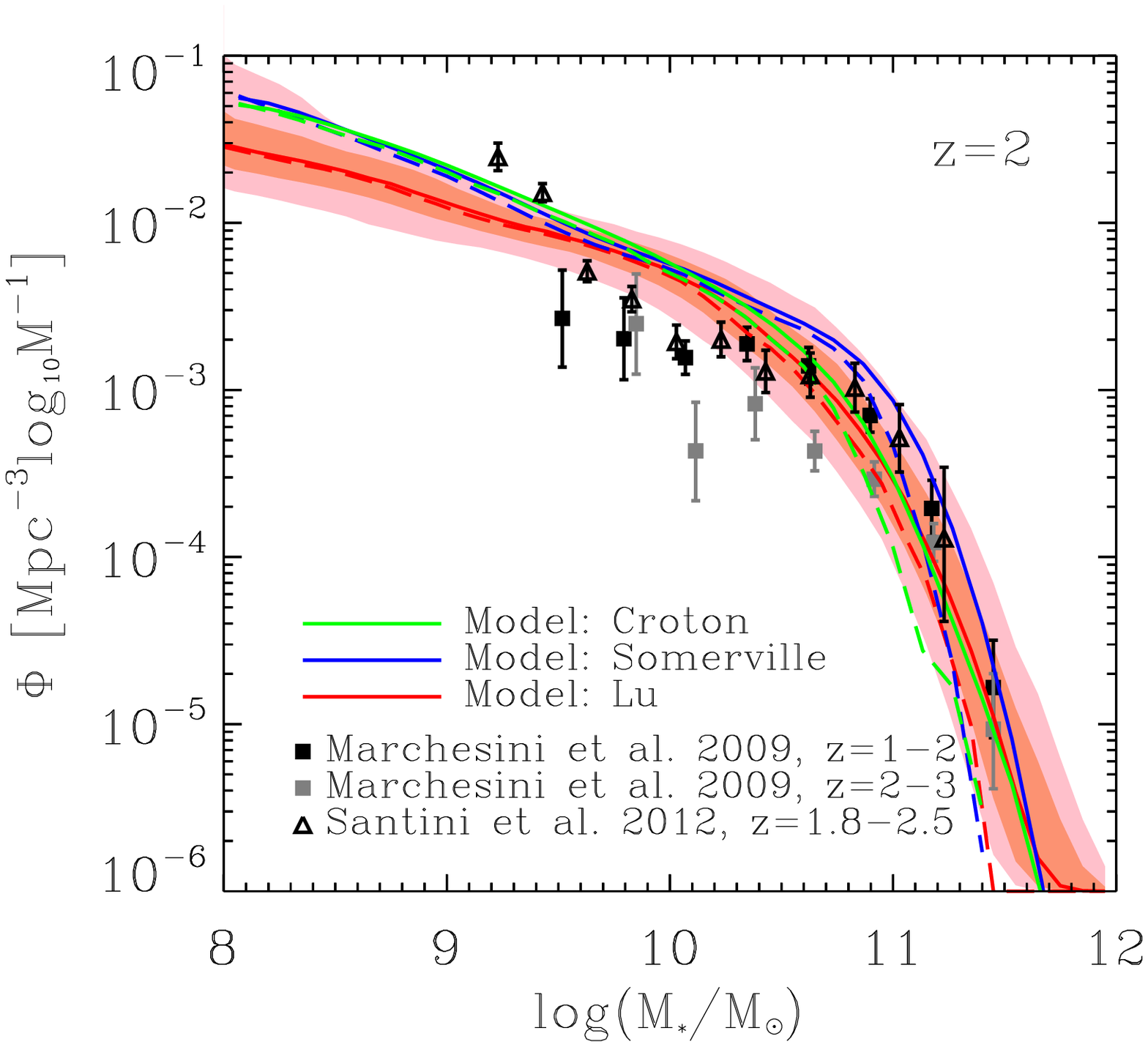} \\
\includegraphics[width=0.35\textwidth]{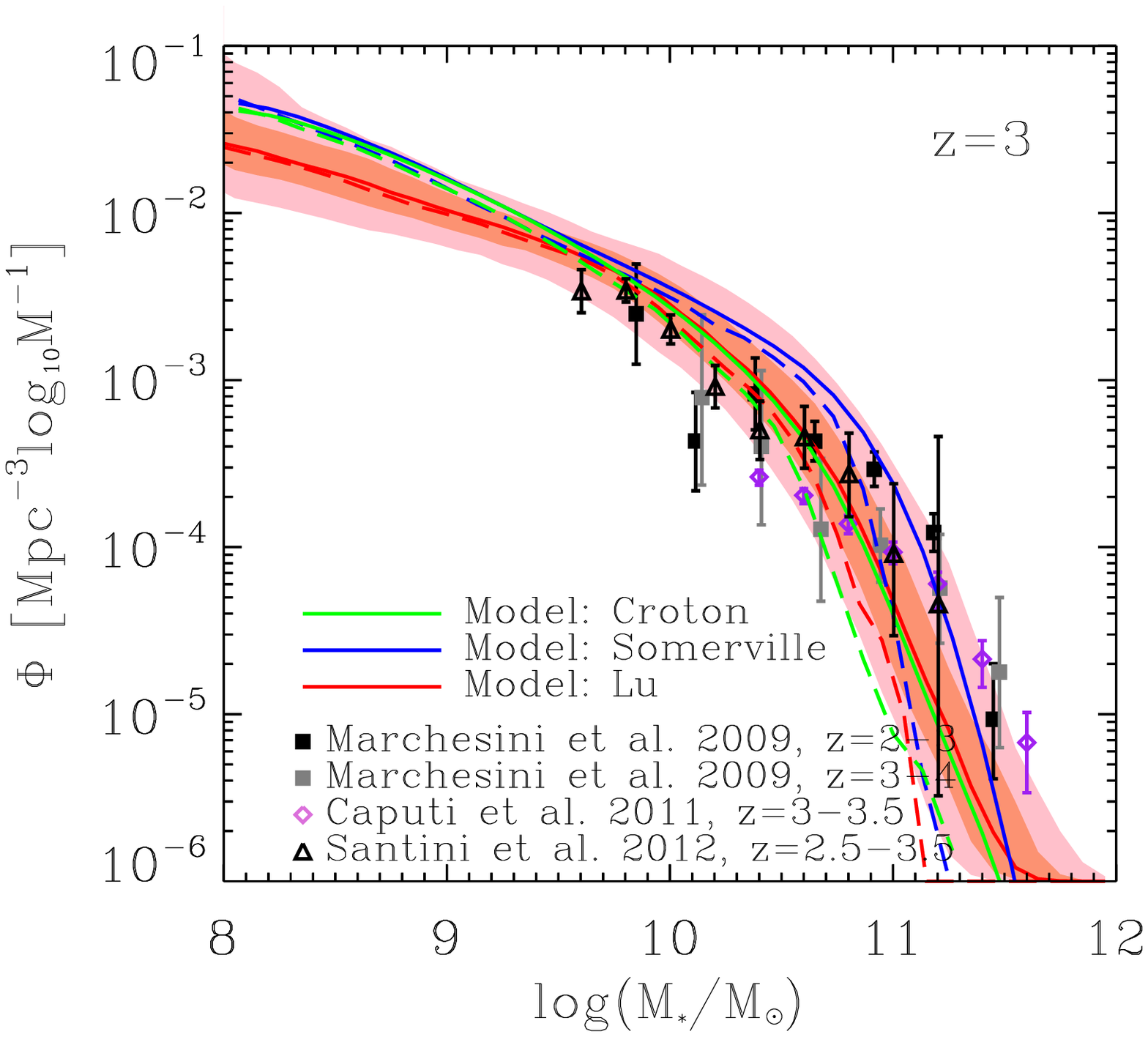} &
\includegraphics[width=0.35\textwidth]{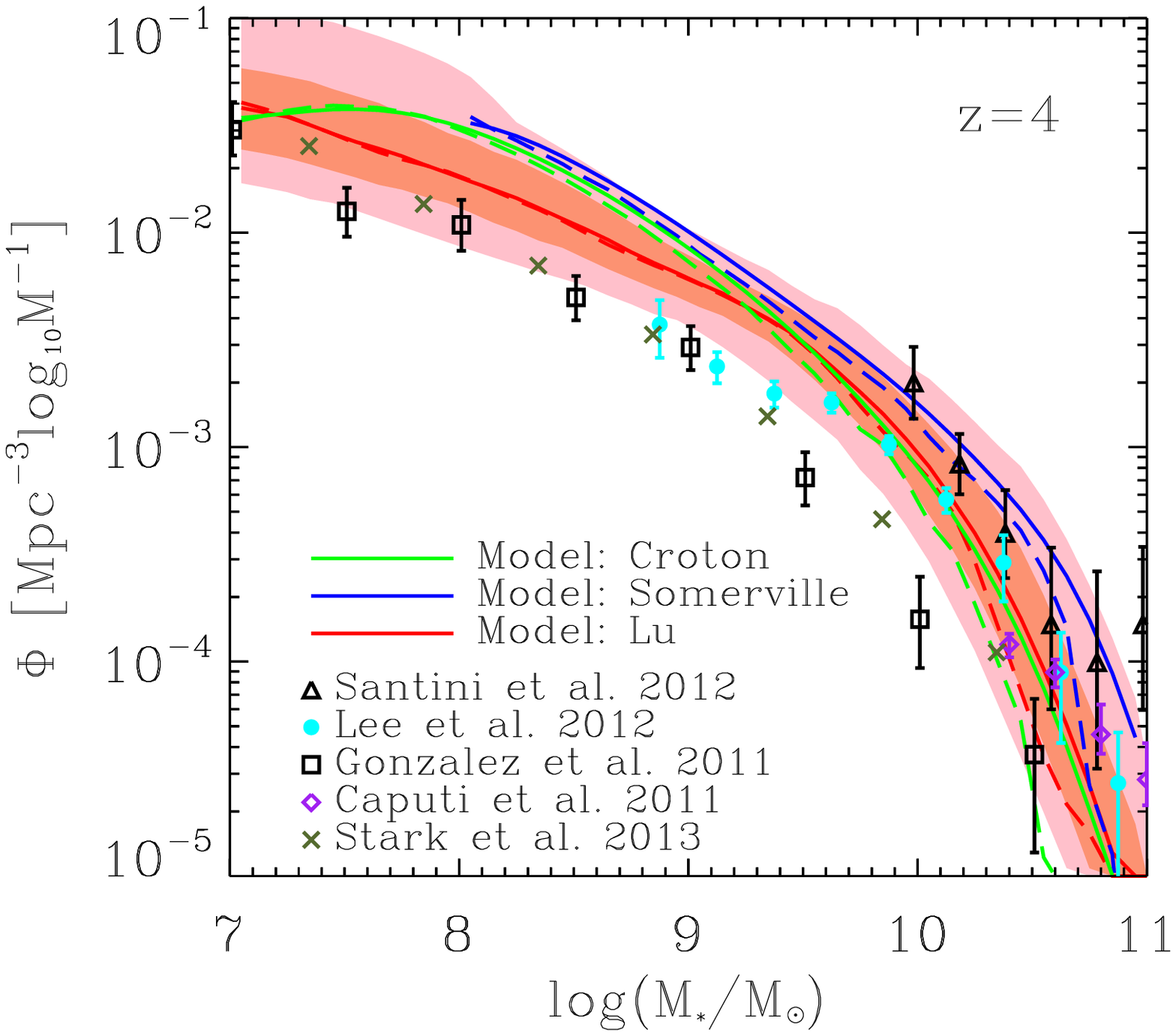} \\
\includegraphics[width=0.35\textwidth]{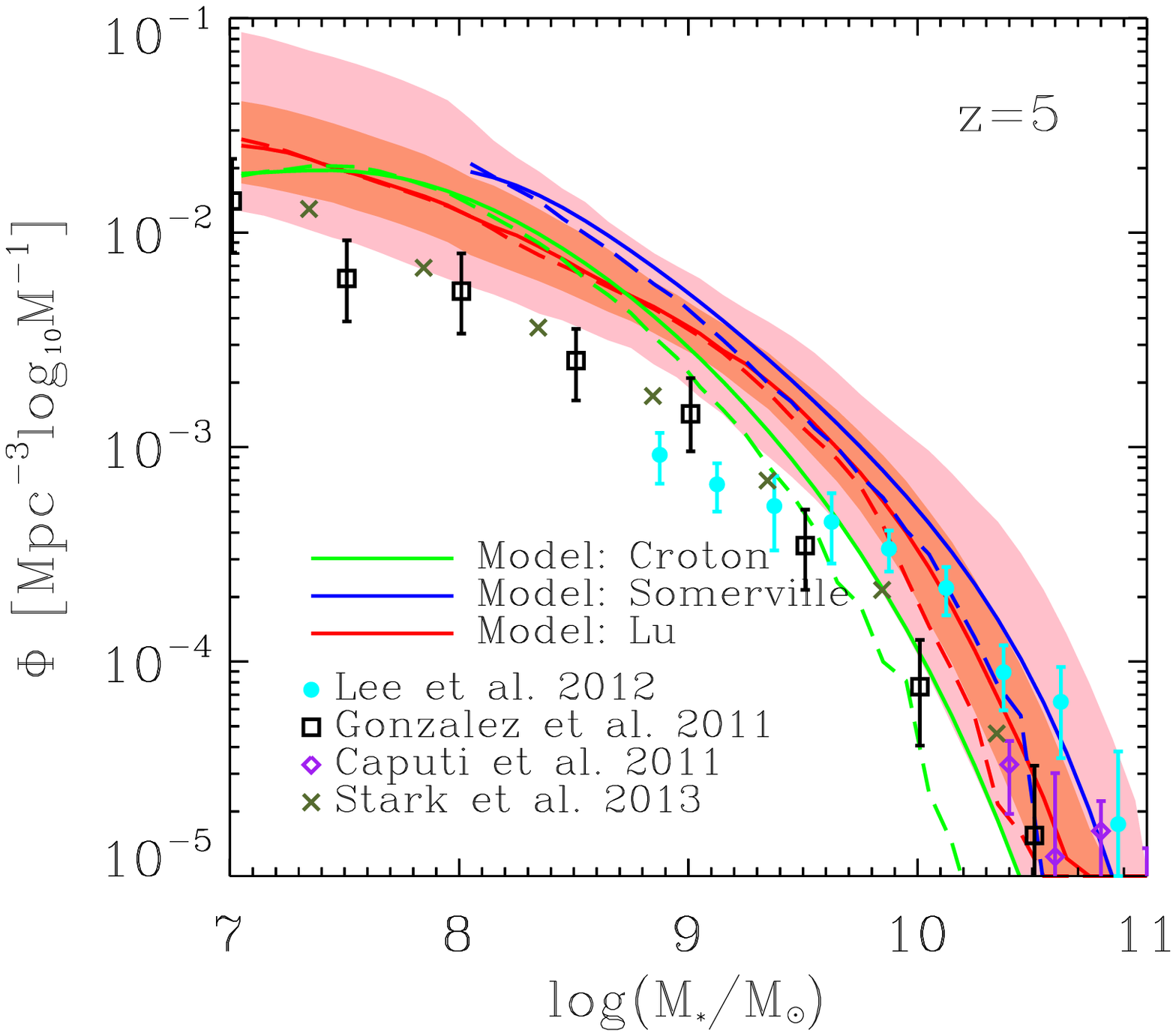} &
\includegraphics[width=0.35\textwidth]{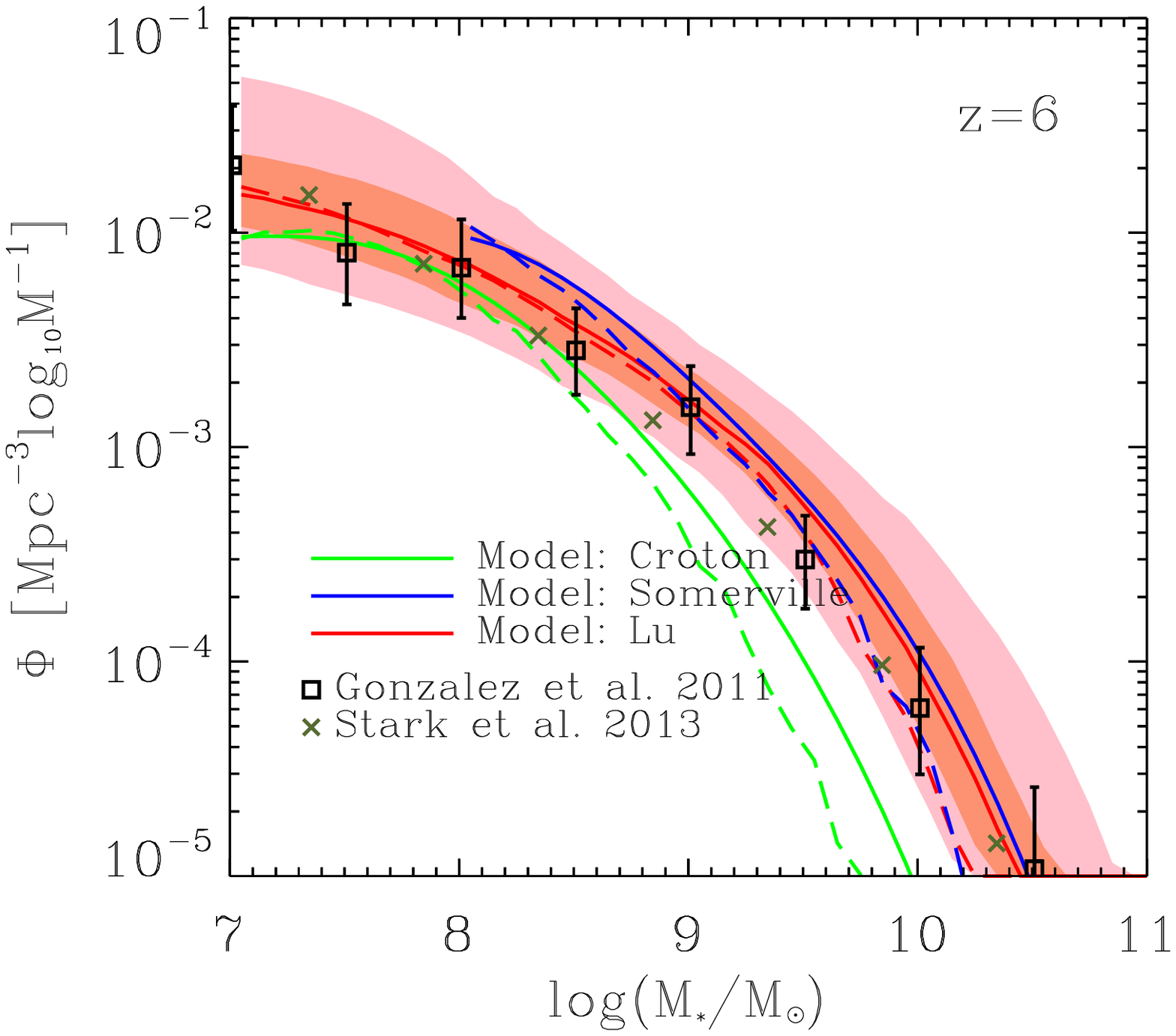} 
\end{tabular}
\caption{Galaxy stellar mass functions at $z=0, 0.5, 1, 2, 3, 4, 5$,
  and 6 predicted by the Croton model (green), the Somerville model
  (blue) and the Lu model, for which the dark and light red bands
  enclose 67\% and 95\% posterior probabilities, and the red solid
  line denotes the median prediction. 
  The solid lines show the model
  predictions with stellar mass uncertainties convolved, and the
  dashed lines show the raw model predictions with no uncertainties
  convolved.
The observational estimates for $z=0$ are from \citet{Moustakas2013},
\citet{Baldry2008}, and \citet{Baldry2012}.  
For $z=0.5-6$, the data are from \citet{Moustakas2013,
  Marchesini2009, Santini2012, Caputi2011, Lee2012, Gonzalez2011} and
\citet{Stark2013} for each relevant redshift as noted in each panel.
Note that only the $z=0$ stellar mass function \citep{Moustakas2013}
is used to calibrate the models.  }\label{fig:smf_allz}
\end{center}
\end{figure*}

The stellar mass function of galaxies describes one of the most
basic statistical properties of the galaxy population. We use the
local galaxy stellar mass function \citep{Moustakas2013} to calibrate
each of the models, and then make predictions at higher
redshifts. There are always uncertainties when estimating stellar masses
from observational data. These uncertainties can significantly bias the
stellar mass function at the high-mass end by pushing low-mass
galaxies into high mass bins, thus elevating the number density estimates 
for the
high-mass bins. In our predictions we mimic this bias by applying the
simple prescription given by \citet{Behroozi2012}, who assume that the
error on each model galaxy stellar mass is log normal, with a redshift
dependent standard deviation, $\sigma(z)=\sigma_0+\sigma_z z$, 
for the logarithmic stellar mass, where $\sigma_0=0.07$ and $\sigma_z=0.04$. 
We convolve this error with the predicted stellar mass when computing the mass functions
to compare with the observed functions
at each redshift \citep[see][for further discussion]{Behroozi2012}.
In Figure~\ref{fig:smf_allz}, we show the stellar mass functions
produced by each model at several redshifts between $z=0$ and 6. Each
panel shows both the convolved and unconvolved versions of the
predicted stellar mass functions for the three models (Croton,
Somerville, and Lu). 

For the Lu model, which uses MCMC to
explore the model parameter space and to sample the posterior
distribution of model parameters, we marginalize over the posterior
probability distribution and plot the predictive distributions that
enclose 67\% and 95\% 
posterior probabilities of the predicted mass function, as well as the
median model posterior predictive distribution.

The first panel (upper left) shows the stellar mass function at $z=0$,
which is used to tune the models. In the panel, the black square
points with error bars are the observational data from
\citet{Moustakas2013}. The gray squares show the stellar mass range that is below
the 95\% completeness limit 
(J. Moustakas, private communication). 
We also overplot the mass functions of \citet{Baldry2008} and \citet{Baldry2012}, 
which show steeper slopes of the mass function at the low-mass end. 
The posterior predictive distribution from the Lu model shows that the MCMC can find
models that match the constraining data remarkably well across the
entire mass range. This is likewise true for the hand-tuned Croton
model and the Somerville models, at least for $\log M_*/\msun
\geq10$. At lower stellar masses these two models overpredict the mass
function by up to a factor of two. Even after doing many iterations of
hand tuning, we find that it is difficult to perfectly match the
low-mass end in the parameter ranges. On the other hand, the Lu model,
aided by the MCMC calibration, is able to produce a low-mass-end
slope as shallow as the data.
 We note, however, that the best models
found in the Croton model and the Somerville model are generally
encompassed by the 95\% posterior predictive distribution of the Lu
model.

Before continuing to higher redshifts, it is interesting to understand
some of the key differences between the different models. The most
obvious difference between the Lu model and the other two is the
outflow mass-loading factor. For the Croton model, the mass-loading is
assumed to be a constant at 3.5 times the SFR for all galaxies. For
the Somerville model, the mass-loading factor scales with the halo
maximum velocity as $\sim V_{\rm max}^{-2.25}$. The mass-loading
factor changes from 1.5 for halos with $V_{\rm max}=200{\rm
  km\,s^{-1}}$ to 7.1 for halos with $V_{\rm max}=100{\rm
  km\,s^{-1}}$. In the Lu model, the MCMC prefers a (perhaps extreme)
model where the mass-loading factor increases with decreasing halo
circular velocity by $\sim \vvir^{-6}$, and star formation is more
strongly suppressed. For a halo with a circular velocity $V_{\rm
  c}=100{\rm km\,s^{-1}}$, a typical mass-loading factor in the
posterior of the Lu model is about 17.

The remaining panels in Figure \ref{fig:smf_allz} show the predicted
stellar mass functions at seven other redshifts, along with
accompanying observed data from each redshift range (as
labeled). These observations are not used to tune any model; we simply
take each $z=0$ calibrated model and plot their higher redshift mass
functions. It is encouraging that all model predictions are fairly
consistent with data in all panels, out to $z \sim 6$, especially when
the uncertainty in the stellar mass estimation is taken into
account. As the redshift increases, the three models gradually diverge
from each other, and the posterior predictive distributions of the Lu
model also become broader. However, because of this broadening, the
diverging predictions of the two hand-tuned models remain within the
95\% posterior range, with the exception of the Croton model, which
predicts a relatively lower stellar mass function for
$M_*>10^{9}\msun$ at $z\geq5$.

In general, the divergence of the three models illustrates how
different choices for the parameterization of the physical processes,
and their accompanying parameter values, can lead to clear differences
in higher redshift predictions. These choices are degenerate for the
low-redshift stellar mass functions solely because the models are
tuned against such data. This highlights the importance of obtaining
accurate high-z galaxy statistics, especially for galaxy properties
that can separate out the different assumptions built into SAMs
through their earlier epoch consequences. In particular, the current
best data sets are inconsistent with each other at the highest
redshifts, making it hard to know what the ``correct'' answer is. For
example, the model predictions for $z=4$ are between the observational
data of \citet{Gonzalez2011} and \citet{Santini2012}, and are fairly
consistent with the data of \citet{Lee2012} and
\citet{Caputi2011}. The overprediction of the mass function at
$z=4\sim5$ compared to \citet{Gonzalez2011} is possibly because the
stellar mass functions of \citet{Gonzalez2011} at these redshifts are
derived from UV selected galaxy samples, which may potentially miss
non-star-forming or dusty galaxies.  Moreover, there is evidence that
the nebular emission is likely to contaminate the rest-optical
broadband light for galaxies at $z\sim$6-7 \citep{Stark2013}. It has
been shown that the treatment of nebular lines and assumed star
formation histories can affect the stellar masses estimated from SED
fitting \citep[e.g.][]{Schaerer2013}.  Data from the full CANDELS
survey, analyzed in a self-consistent way, will tighten the
constraints of the stellar mass functions at these redshift ranges.

\begin{figure}[htb]
\begin{center}
\includegraphics[width=0.45\textwidth]{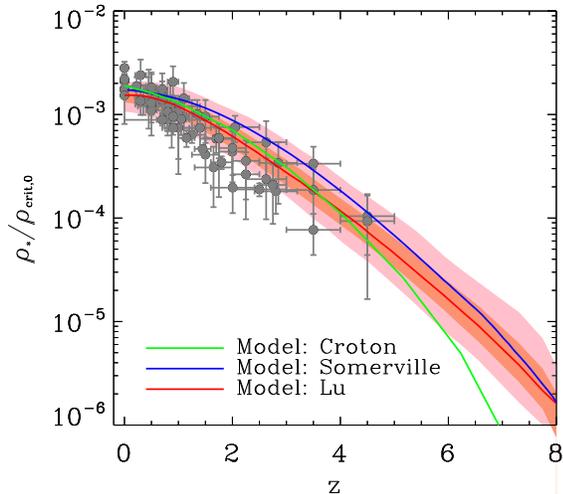}
\caption{The volume averaged stellar mass density of the universe,
  normalized by the critical density of the universe at the present
  time, as a function of redshift.  The filled bands denote the direct
  SAM predictions of 67\% (dark red) and 95\% (light red) posterior
  regions from the Lu model.  The blue and green lines are predicted
  by the Somerville model and the Croton model.  The points with error
  bars are various observational estimates, taken from the compilation
  of \citet{Wilkins2008}.  }
\label{fig:smh}
\end{center}
\end{figure}

In Figure \ref{fig:smh} we integrate the stellar masses of galaxies
larger than $10^8 \,\msun$ at each redshift to show the cosmic stellar
mass density evolution. The three models produce similar trends and
the two individual models of the Croton model and the Somerville model
stay within the 95\% posterior range of the Lu model. Because all the
models are carefully tuned to match the stellar mass function at
$z=0$, the predicted cosmic stellar mass densities at $z=0$ agree with
each other within 10\% and are consistent with the ensemble of
observational data (as marked). Since the Croton model displays a more
rapid evolution of the high-mass end of the stellar mass function than
the Somerville model and the median of the Lu model, it predicts lower
cosmic stellar mass densities at high redshifts for $z>4$. Similarly,
the Somerville model has a larger number of high-mass galaxies at
higher redshifts than the Lu model, and so predicts higher cosmic
stellar mass densities at those redshifts.

It is worth pointing out that although the three models make diverging
predictions for the evolution of the stellar mass function at high
redshift, their predicted cosmic stellar mass density as a function of
redshift is in broad agreement with the data, given the large
scatter between different observational estimates and their large
error bars. It is obvious that the galaxy stellar mass function and
its evolution is more constraining than the evolution of the global
stellar mass density.

\subsection{Star formation rates}


\begin{figure*}[htb]
\begin{center}
\includegraphics[width=0.95\textwidth]{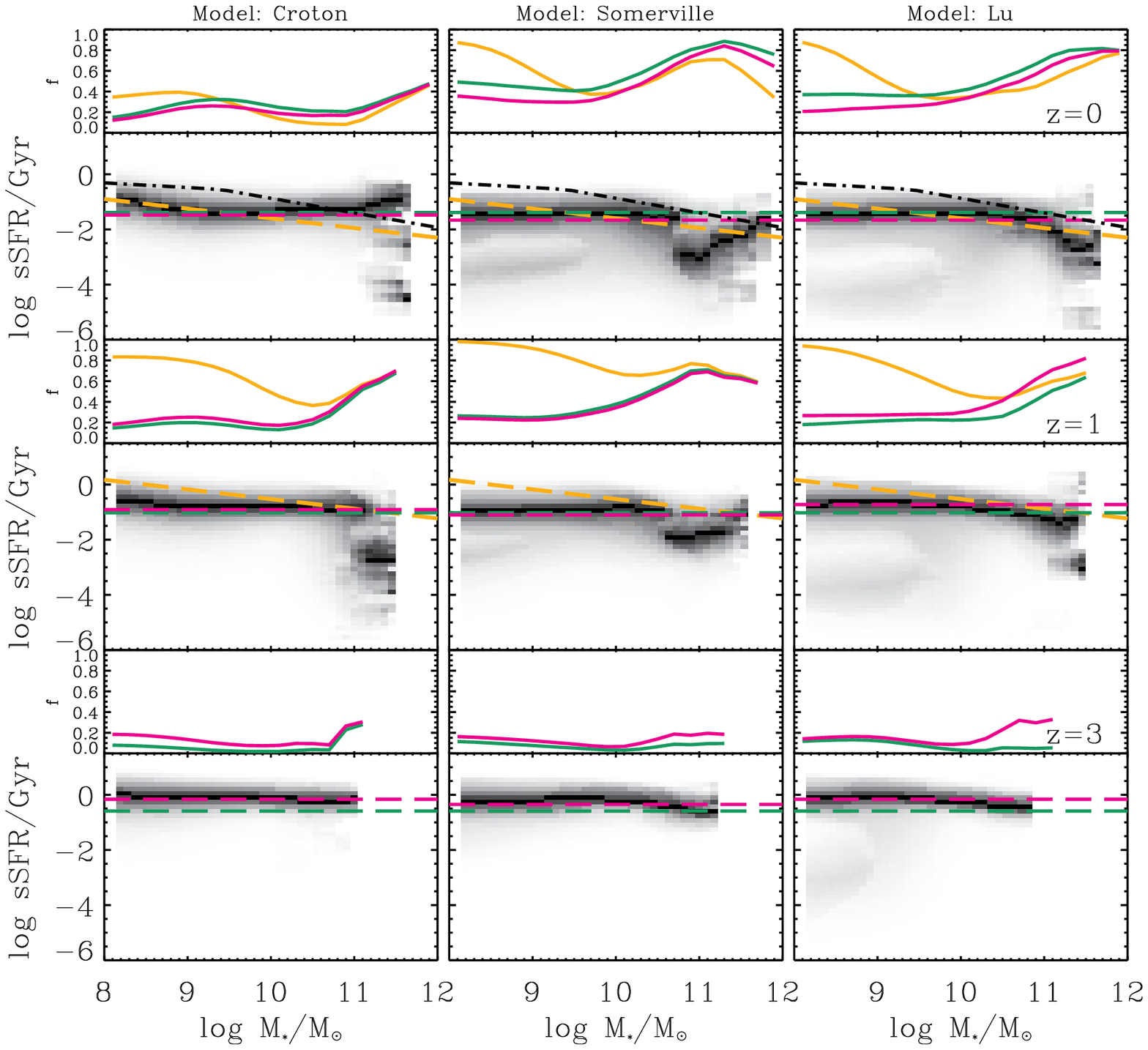}
\caption{The joint distribution of specific star formation rate and stellar mass of model galaxies at $z=0, 1$, and 3.
The first column shows the predictions of the Croton model, the second
column the Somerville model, and the third column the Lu model.  The
black dash-dot line in the $z=0$ panels denotes the star forming sequence from the
observational study of \citet{Salim2007} for local galaxies.  The orange dashed lines in 
the $z=0$ and $z=1$ panels denote the division line separating ``star forming'' galaxies and
``quiescent" galaxies adopted in \citet{Moustakas2013}. 
The magenta dashed lines denote the division lines that mark the sSFR 0.35 dex lower than 
star forming sequence produced by each model at the corresponding redshift. 
The green dashed lines denote the characteristic sSFR defined by Eq.~\ref{equ:sf_qu_hubble} corresponding 
to the age of the universe at each redshift.   
On the top of each panel, we show the fraction of galaxies that have a sSFR lower than 
each division line as a function of galaxy stellar mass. The color coding of the lines 
is the same as that of the division lines in the sSSFR-$M_*$ diagram. 
 }
\label{fig:ssfr_map}
\end{center}
\end{figure*}

\begin{figure*}[htb]
\begin{center}
\includegraphics[width=0.95\textwidth]{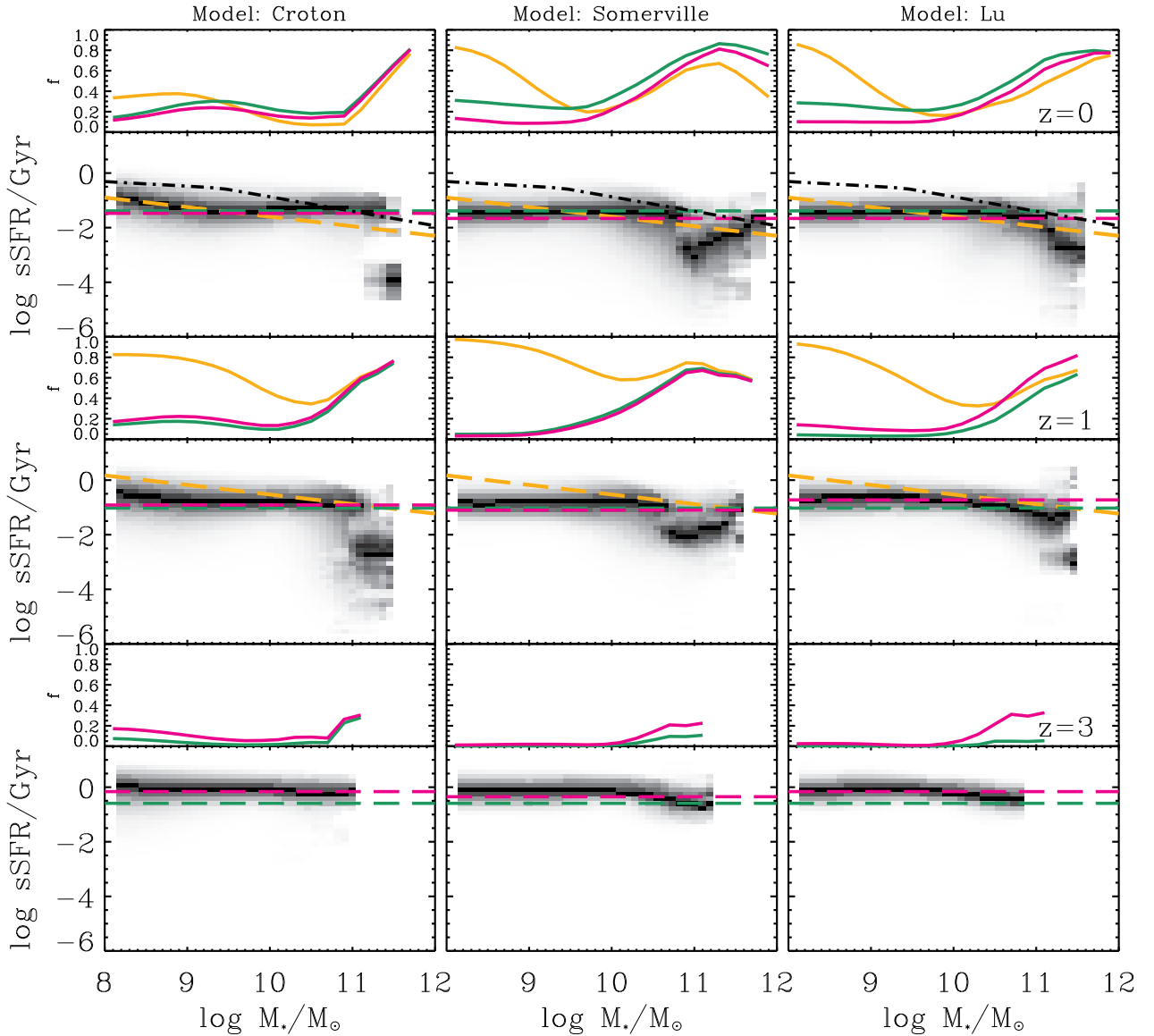}
\caption{Same as Figure \ref{fig:ssfr_map}, but for central galaxies only.}
\label{fig:ssfr_cen_map}
\end{center}
\end{figure*}

\begin{figure*}[htb]
\begin{center}
\includegraphics[width=0.95\textwidth]{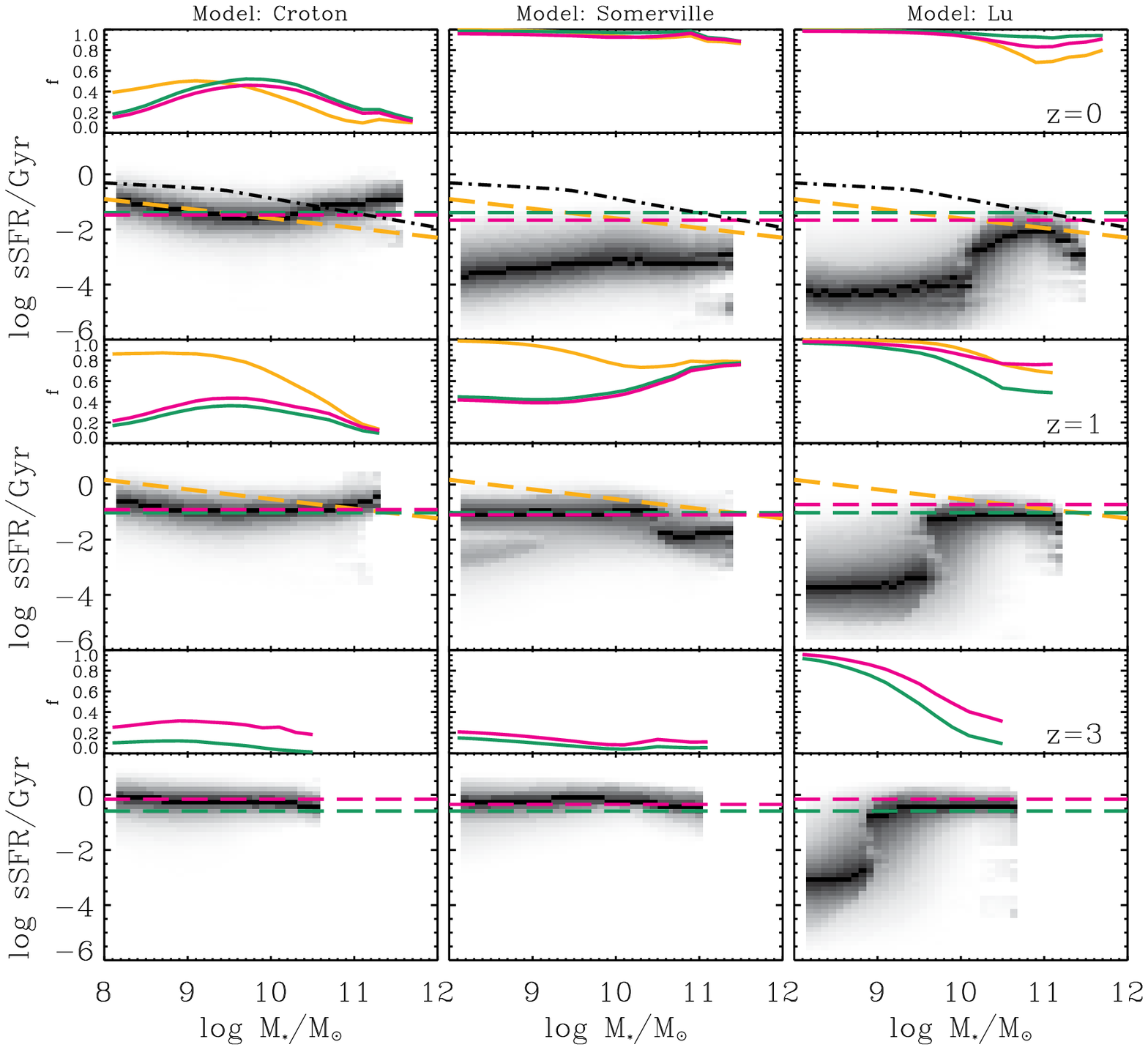}
\caption{Same as Figure \ref{fig:ssfr_map}, but for satellite galaxies only.}
\label{fig:ssfr_sat_map}
\end{center}
\end{figure*}

In Figure \ref{fig:ssfr_map} we show the specific star formation rate
(sSFR) vs. stellar mass for our three galaxy models (as marked),
focusing on redshifts of 0, 1 and 3.  The sSFR is defined as the star
formation rate averaged over a time interval $\sim100$ Myr normalized
by the stellar mass of the galaxy.
It is clear that all the models predict a clear star forming sequence, which appears to
be a nearly horizontal ridge. 
We overplot the sSFR as a function of stellar
mass of the star forming sequence of local galaxies in
\citet{Salim2007} 
in each $z=0$ panel of
Figure \ref{fig:ssfr_map}.  
In observations, the star forming sequence 
has a width of about 0.35 dex in sSFR at $z<1.1$ \citep{Noeske2007}, which is similar to the model predictions. 
However, we find that both the normalization
and the slope of the star forming sequence predicted by the models
differ from the observational results in detail.  The overall sSFRs of the star
forming galaxies in the model are systematically lower, especially for
low mass galaxies.  Moreover, the slope of the star forming sequence
predicted by the models is shallower than the observations, especially for
the Somerville model and the Lu model.

To characterize the star forming sequence predicted in the models, we
find the peak sSFR on the star forming sequence for galaxies with
stellar mass between $10^8\msun$ and $10^{10}\msun$ in each model. We
then draw a line at a sSFR 0.35 dex lower than the peak sSFR.  We show
it by a dashed magenta line in each panel of Figure
\ref{fig:ssfr_map}.  All models predict a star forming sequence that
bends downward to lower sSFRs at high stellar masses, and an
increasing number of galaxies below the magenta line when
$M_*>10^{11}\msun$.  The trend holds at higher redshifts up to $z=3$
as we show in the figure.

We also define another characteristic sSFR as a function of redshift defined as 
\begin{equation}\label{equ:sf_qu_hubble}
{sSFR} = {1 \over \tau_{\rm H}(z) (1-f_{\rm rec})},
\end{equation}
where $t_{\rm H}(z)$ is the age of the universe at the redshift $z$,
and $f_{\rm rec}=0.43$ is the assumed recycling fraction of star
formation for the adopted Chabrier IMF.  If a galaxy has a sSFR lower
than this characteristic sSFR, it means that the time for the galaxy
to accumulate its stellar mass by keeping its current star formation
rate is longer than the age of the universe, suggesting that the galaxy
must have had a higher star formation rate in the past to form the
bulk of its stellar mass, and its SFR must have decreased. On the other hand, if a galaxy has a sSFR higher than the
sSFR defined by Eq.~\ref{equ:sf_qu_hubble}, it means that the galaxy
can form its stellar mass in a timescale shorter than the age of the
universe with its current star formation rate, suggesting its SFR has
recently increased.  We also overplot this characteristic sSFR in
Figure \ref{fig:ssfr_map}. 
We find that this
characteristic sSFR closely follows 
the star forming sequence in each model.  There is a tendency that, at
high redshifts ($z>1$), the SAM predicted star forming sequence
evolves faster and becomes higher than the characteristic sSFR defined
by Eq.~\ref{equ:sf_qu_hubble} at higher redshifts.  This indicates that
galaxies in the models typically have a rising star formation
history at high-z.  We also note that the evolution of the
normalization of the star forming sequence is consistent with the
observational data of \citet{Noeske2007}
and similar to the redshift dependence
of the specific halo mass accretion rate predicted in
cosmological simulations \citep{Dekel2009}. 
This suggests that the star formation rate of star forming galaxies at a
given stellar mass follows the halo accretion rate. This is similar to
what is found in empirical models \citep{Behroozi2012, Behroozi2013b,
  Yang2013, Mutch2013}, which also match the evolution of the stellar
mass function across a large redshift range.

In observations, galaxies are split into a star forming population and
a quiescent population according to the sSFR for a given stellar mass
\citep[e.g.][]{Noeske2007, Salim2007, Moustakas2013}.
\citet{Moustakas2013} found that their galaxy sample at $z\sim0$ to 1
can be divided into two separate populations according to a division
line defined as
\begin{equation}\label{equ:sf_qu_moustakas}
\begin{split}
\log \left( {SFR \over \msun {\rm yr}^{-1}}\right)=-0.49 + 0.65 \log \left({M_* \over 10^{10}\,\msun}\right) \\
 + 1.07 \left(z-0.1\right).
\end{split}
\end{equation}
We overplot these lines 
in Figure \ref{fig:ssfr_map} for $z=0$ and $z=1$ 
and find that because the model predicted star forming sequence has a
shallower slope than the dividing line, the dividing line cuts through
the star forming sequence predicted by all the models at
$\sim10^9\,\msun$. In addition, this dividing line evolves more rapidly
to higher sSFRs with redshift than the models predict. For example, at
$z\geq1$ the dividing line is above the star forming sequence over a
large range of stellar masses.

On the top of each sSFR-$M_*$ diagram, we plot the fraction of galaxies 
that have a sSFR lower than each of the characteristic sSFR lines
as a function of stellar mass. 
In general, all models predict a trend that the fraction of galaxies
below the sSFR dividing line increases with increasing stellar masses.
The fractions defined by the two dividing lines are similar for all
the models.  In all the models, the fraction is lower than 50\% for
low-mass galaxies, and the fraction goes up to nearly 90\% for
high-mass galaxies.  The Croton model produces a relatively weaker
trend at $z=0$, but similar trend at higher redshifts.  At the very
highest stellar masses, the prediction of the Somerville model starts
to to turn over, which reflects the increasing sSFR at the very
high-mass end predicted by the model.  At higher redshifts, the models
predict fewer galaxies below the division lines, making the predicted
fractions increasingly lower. At $z=3$, all the models predict that
nearly all low-mass galaxies ($M_*<10^{10}\,\msun$) have a sSFR higher
than the characteristic sSFRs.  The fraction keeps the increasing
trend at high redshifts, indicating that high-mass galaxies are
getting quenched at an early cosmic epoch.

We also show the fraction based on the dividing line defined by
Eq.~\ref{equ:sf_qu_moustakas}, adopted in observations
\citep{Salim2007, Moustakas2013}. 
\citet{Moustakas2013} found that in the observations the number density of
galaxies is increasingly dominated by galaxies with relatively lower
sSFR as stellar mass increases. In their data, for galaxies with mass
$>10^{11}\,\msun$, more than 70\% are classified as ``quiescent''
galaxies, while for low mass galaxies, $M_*<10^{9}\msun$, fewer than
30\% are ``quiescent''.  We find, however, if we adopt this dividing
line, the fraction of galaxies that are below the dividing sSFR
follows a different trend.  Although the models reproduce the trend of
the data at intermediate and higher masses, they exhibit an increasing
fraction as a function of decreasing stellar mass at lower stellar
masses.  In all models, the number of galaxies below the dividing line
increases for decreasing stellar mass at the low mass end
($<10^{9.5}\,\msun$) at both $z=0$ and $z=1$.  As we discussed, the
increasing trend at the low-mass end is because the star forming
sequence predicted by the models is much flatter than that in
observations.  The dividing line adopted in observations cuts through
the model star forming sequence at the low-mass end. Therefore,
although the models predict a clear star forming sequence, the
low-mass star forming galaxies in the models are still classified as
``quiescent'' galaxies by the definition of
Eq. \ref{equ:sf_qu_moustakas}.  
These results provide a cautionary note that the ``quiescent''
fraction sensitively depends on the definition of ``quiescent''
galaxy.  The comparison also demonstrates that the evolution and the
mass dependence of the quiescent fraction \citep[see,
  e.g.][]{Brammer2011, Mutch2013a} provides strong constraints on
models, and we plan to investigate this in future work.

To understand how central galaxies and satellite galaxies evolve
differently in the models, we show the sSFR-$M_*$ diagram for central
galaxies only in Figure~\ref{fig:ssfr_cen_map}, and satellite galaxies
only in Figure~\ref{fig:ssfr_sat_map}.  In fact, the star forming
sequence in all models is mainly populated by central galaxies.  For
the highest stellar mass bins, $M_*>10^{11}\, \msun$, a fraction of
central galaxies appear below the dividing line for $z\leq1$,
indicating that star formation has been truncated in these
galaxies. While the Somerville model predicts that the sSFR
distribution of high mass galaxies is shifted about one order of
magnitude lower, the peak sSFR starts to increase again for higher
stellar masses.  Among the three models, the Croton model predicts the
largest separation in sSFR between massive quiescent galaxies and star
forming galaxies. At higher redshifts, no model predicts a significant
population of high-mass quiescent galaxies; almost all central
galaxies at such early times are star forming.

Interestingly, the distributions of satellite galaxies in the
sSFR-$M_*$ diagram predicted by the three models are very
different. The Croton model predicts a broader and only slightly lower
sSFR for satellites than centrals at a given stellar mass. In this
model, the distribution does not evolve with redshift
significantly. The reason for this similarity is that this model
adopts a similar treatment for radiative cooling in satellite galaxies
as for central galaxies. As we described earlier, the Croton model
does not instantaneously strip all the hot gas from subhalos, but
allows them to continue to accrete gas from their own hot halo to fuel
star formation. For this reason, the majority of the satellite
galaxies are star forming and the distribution of satellite galaxies
in the sSFR-stellar mass diagram is similar to centrals with slightly
reduced star formation rates.

In the Somerville model, the sSFR distribution of satellite galaxies
strongly depends on redshift. At $z=0$, most satellite galaxies are
distributed below the dividing line and are quiescent galaxies. At
higher redshifts, the sSFR, regardless of stellar mass, is as high as
central galaxies at the same epoch. The distribution of satellite
galaxies in the diagram is similar to that of central galaxies.

The Lu model predicts a bimodal distribution of sSFRs. At $z=0$, the
sSFR is low for satellites with mass lower than $\sim 10^{10}\,\msun$,
and it jumps to higher values when the stellar mass is higher than
$10^{10}\,\msun$. According to the dividing line, low-mass satellites
are quiescent and high-mass satellites are star forming. At higher
redshifts, the satellite galaxies still have a similar bimodal
distribution in their sSFRs. The transition stellar mass, however,
decreases with increasing redshift. At $z=3$ the transition stellar
mass is $\sim 10^9\,\msun$.

\begin{figure}[htb]
\begin{center}
\includegraphics[width=0.45\textwidth]{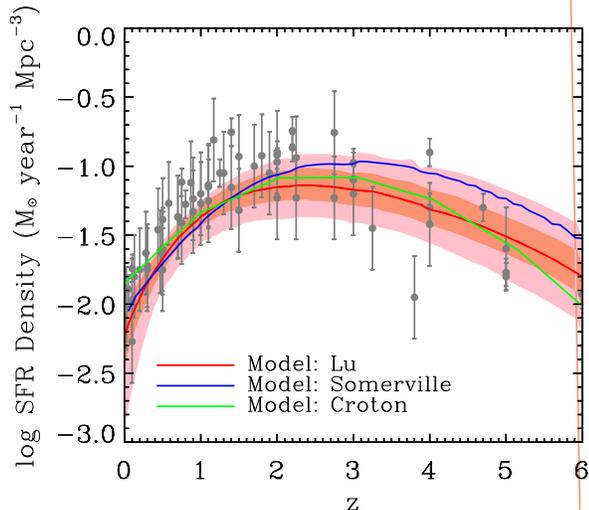}
\caption{Star formation rate density of the universe as a function of
  redshift. The green line denotes the prediction of the Croton model,
  the blue line denotes the prediction of the Somerville model, and
  the dark and light red bands encompass 67\% and 95\% predictive
  posterior regions of the Lu model.  The data points with error bars
  are from the compilation of various data sources given in
  \cite{Behroozi2012}. }
\label{fig:sfrh}
\end{center}
\end{figure}

In Figure \ref{fig:sfrh} we integrate the star formation rate from
galaxies with stellar masses larger than $10^8\msun$ at each redshift
to plot the cosmic star formation rate density as a function of
redshift. Again, although the three models have different sSFR-stellar
mass distributions in detail, they produce similar trends for the
volume averaged star formation rate density. The Lu model predicts a
large variation at all redshifts for the kinds of acceptable
histories, while the two other models fall into the 95\% posterior
region of the Lu model prediction. This indicates that all three
models are consistent with each other, even though they adopt
different parameterizations and parameter values. The comparison shows
that while more accurate data is needed at high redshift, 
better measurements of the
volume averaged cosmic star formation rate density do not provide
enough constraining power to discriminate between the models and to
break model degeneracies. However, the comparisons show that detailed
observational data, such as the evolution of the star forming sequence
and the quiescent fractions at different redshifts, can strongly
constrain galaxy formation models.  We expect that these data from
CANDELS will be helpful to discriminate between the existing models.

\subsection{Build-up of stellar mass in central galaxies}

\begin{figure}[htb]
\begin{center}
\includegraphics[width=0.45\textwidth]{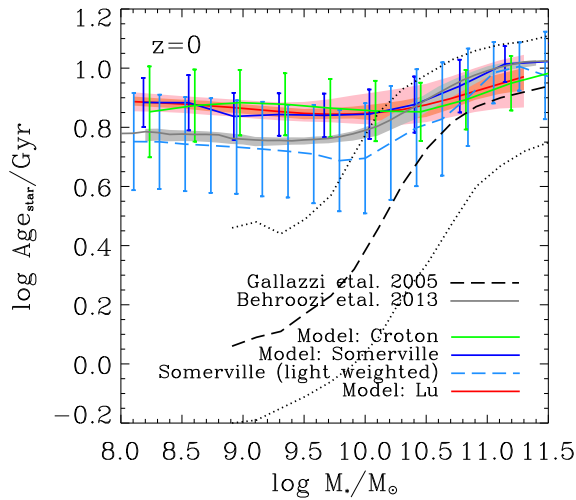}
\caption{Stellar age as a function of stellar mass at $z=0$ predicted
  by the three models. The green line denotes the prediction of the
  Croton model, the blue line denotes the prediction of the Somerville
  model, and the error bars on them show 1-$\sigma$ scatter of the
  model galaxy samples. The dark and light red bands encompass 67\%
  and 95\% predictive posterior regions of the Lu model. All three of those
  predictions are stellar mass-weighted stellar ages. The dashed blue
  line shows the V-band light-weighted stellar ages predicted by the
  Somerville model. The grey solid line shows the stellar
  mass-weighted ages of the Behroozi model. The dashed black line is
  the observational estimates of \citet{Gallazzi2005} and dotted
  lines are the standard deviations of the observational estimates.  }
\label{fig:sage}
\end{center}
\end{figure}

To examine the build-up of stellar mass in model galaxies, we keep
track of the star formation history of every galaxy along each merger
tree. We select central galaxies at $z=0$ and then calculate their
stellar mass-weighted stellar age. In Figure \ref{fig:sage} we show
the mean relation of the stellar mass-weighted age as a function of
stellar mass, for $z=0$ galaxies predicted by the three models.

We find that all the models predict similar trends for the
age--stellar mass relation. The models predict that the stellar
mass-weighted age is nearly constant for all stellar masses. The
Somerville model predicts that the age of high mass galaxies is
slightly higher (0.1 dex) than the two other models, but for galaxies
with mass lower than $10^{10}\,\msun$, the relation becomes flat, with
an age of approximately $8\times10^9$ yr.

The observational estimates of \citet{Gallazzi2005} indicate that the
light-weighted stellar age decreases rapidly with decreasing galaxy
stellar mass, from $\sim9$ Gyr to $\sim1$ Gyr, in contrast to the
behavior of all three models. This is, in general, a much younger
stellar population than the models predict, except for the highest
mass galaxies. To understand the discrepancy we have done further
tests. First, the gray line in Figure \ref{fig:sage} shows the
averaged stellar mass-weighted age from the empirical model of
\citet{Behroozi2012}, which also fits the stellar mass functions of
galaxies up to $z = 8$, but is free of the more complex details and
parameterizations of galaxy formation modeling. Their results are
consistent with that of the SAMs (but not the
 ``fossil'' data). Second, including the BC03 stellar population
synthesis model, we use the Somerville model to predict the V-band
light-weighted age as a function of stellar mass, plotted in light
blue. The light-weighted stellar ages are about 0.15 dex lower than
the stellar mass-weighted ages. However, this difference cannot
explain the large offset between the model predictions and the
data. These tests suggest that there may be unaccounted for degeneracies
between the stellar age and mass measurements, or that the stellar age
estimates may be biased. It may even highlight a deeper problem in
reproducing the star formation histories of low-mass galaxies in the
SAMs and empirical models.

The light-weighted stellar ages
of \citet{Gallazzi2005} were derived from Balmer-line indices, which
are sensitive to stellar age \citep{Worthey1994,
  Worthey1997}. However, such indices are very sensitive to recent
star formation, which can strongly bias the inferred
values. \citet{Trager2009} carried out tests, which coupled a SAM to
stellar population models to produce synthetic spectra, and computed
line strengths from these spectra. They then used the line strengths
to determine ages in the same way as for observed spectra and found
that the simple stellar population equivalent ages determined by the
Balmer-line indices were always younger than mass-weighted ages by
more than 40 per cent on average, and younger than light-weighted ages
by roughly 25 per cent on average. As shown schematically by
\citet{Trager2000} and quantitatively by \citet{Serra2007}, the
addition of a small fraction of young stars to an old population
strongly biases the apparent age of a galaxy. This is because hot,
young stars contribute much more strongly per unit mass to the Balmer
lines than do old stars \citep{Trager2000a, Serra2007}. Our results
are consistent with such trends, and highlight the importance of
considering more complex and realistic star formation histories
in computing line-strength-derived ages.

\begin{figure*}[htb]
\begin{center}
\begin{tabular}{cc}
\includegraphics[width=0.4\textwidth]{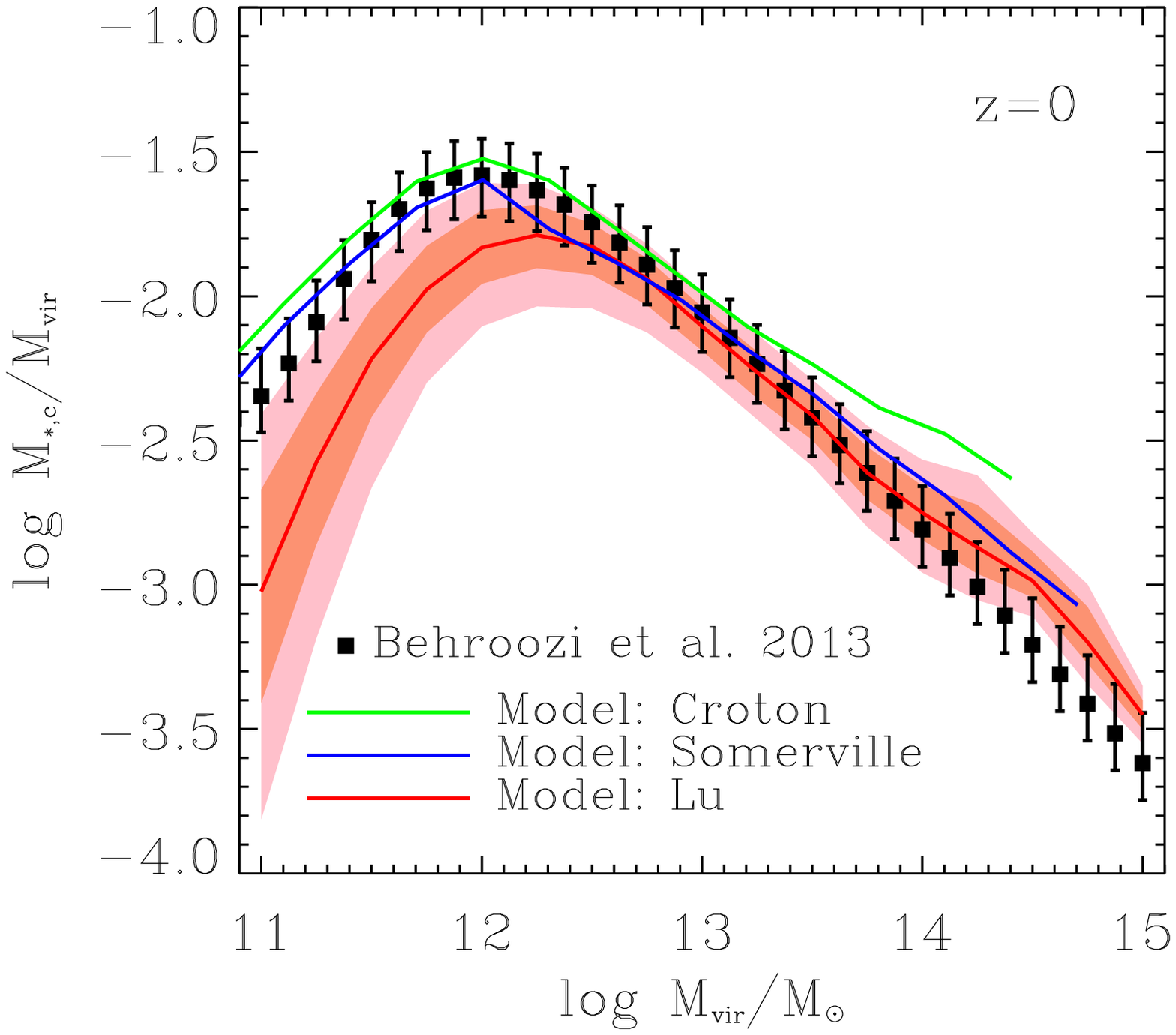} &
\includegraphics[width=0.4\textwidth]{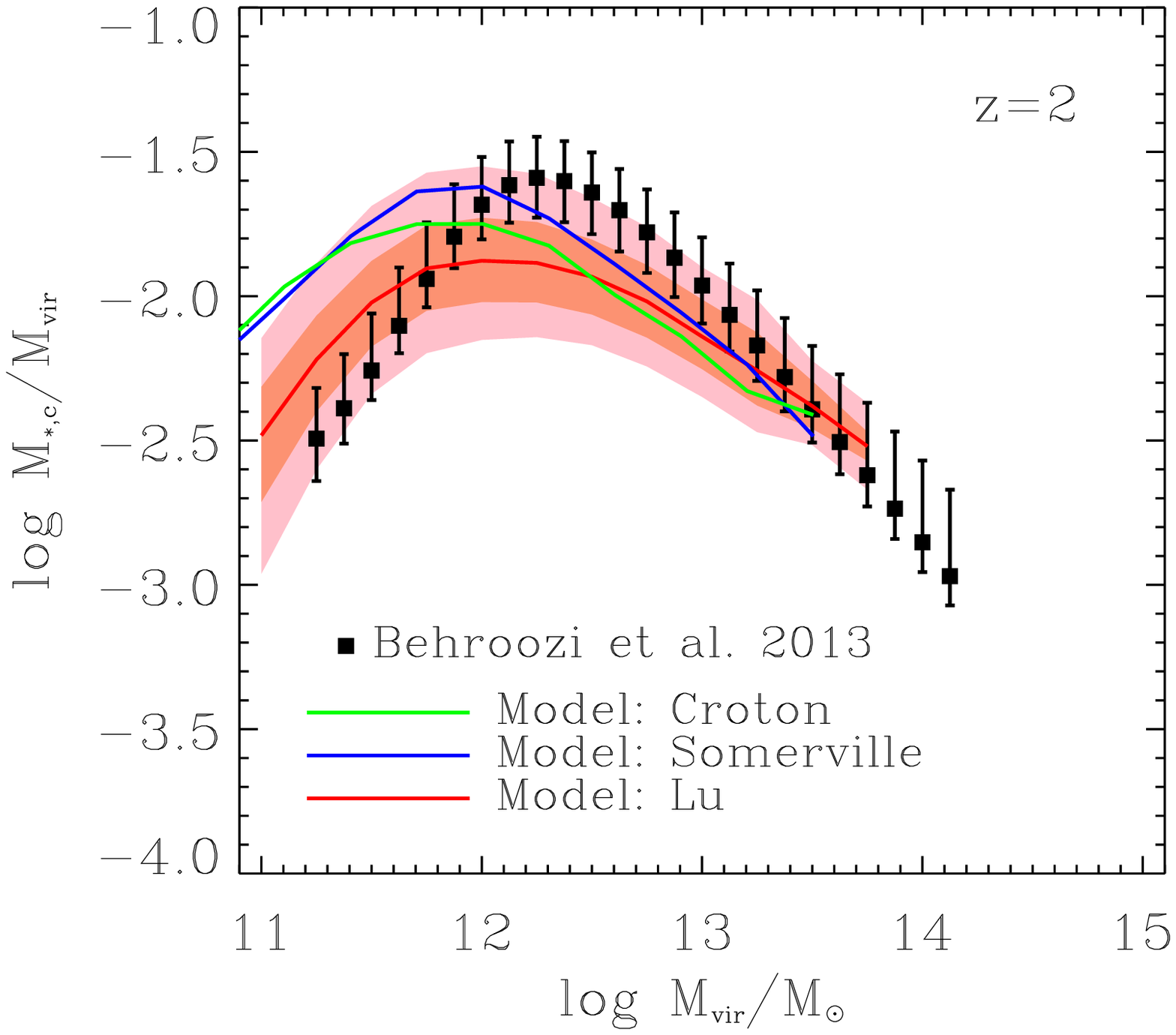} 
\end{tabular}
\caption{The ratio between the central galaxy stellar mass and the
  host halo virial mass as a function of halo virial mass at z=0 and
  2. The green line denotes the prediction of the Croton model, the
  blue line denotes the prediction of the Somerville model, and the
  dark and light red bands encompass 67\% and 95\% predictive
  posterior regions of the Lu model.  The black squares with error
  bars are the empirical constraints derived by
\citet{Behroozi2012}. }
\label{fig:smhm}
\end{center}
\end{figure*}

To further explore how central galaxies build up their mass in the
models, we select halos of a given virial mass at various redshifts
and study how stellar mass and star formation rate in such halos
change. First, we consider central galaxy mass as a function of halo
mass at redshift $z=0$ and 2 (Figure \ref{fig:smhm}). We compare the
three SAM predictions (solid lines) with the empirically derived
results of \citet{Behroozi2012}.
At $z=0$ (left panel) the models agree with each other
for halo mass larger than $10^{12}\msun$, 
but the Croton model predicts a higher stellar mass-halo mass ratio for
high-mass halos. At $M_{\rm vir}=10^{14}\,\msun$, the mass ratio predicted 
by the Croton model is about a factor of two higher than other models. 
At halo masses lower than $10^{12}\,\msun$, the models diverge from each
other, which is directly related to the difference in the low-mass end of
the predicted stellar mass functions. The Lu model produces the lowest
stellar mass function at the low-mass end and, therefore, it produces
the lowest stellar mass-halo mass ratio. On the other hand, the Croton
model, which shows the steepest low-mass end for the stellar mass
function, produces the highest stellar mass halo mass ratio.

More interestingly, the models show similar evolution with redshift
between $z=0$ and 2 but systematically differ from the results of the
empirical model, as seen by comparing with the right panel in Figure
\ref{fig:smhm}. In the empirical model, the evolution of the stellar
mass ratio for high-mass halos is very mild. In contrast, all SAMs,
especially the Croton model, predict a relatively stronger evolution,
in the sense that high-mass halos build up their stellar mass fairly
late. At the low-mass end, the empirical model results suggest that
the stellar mass ratio has evolved a lot since $z=2$, while the SAMs
predict a relatively slower evolution. This again shows that, in SAMs,
low-mass halos grow their stellar mass at early times more rapidly
than they do in such empirical models, and high-mass halos grow their
stellar mass more slowly at high redshift.

\begin{figure*}[htb]
\begin{center}
\begin{tabular}{cc}
\includegraphics[width=0.4\textwidth]{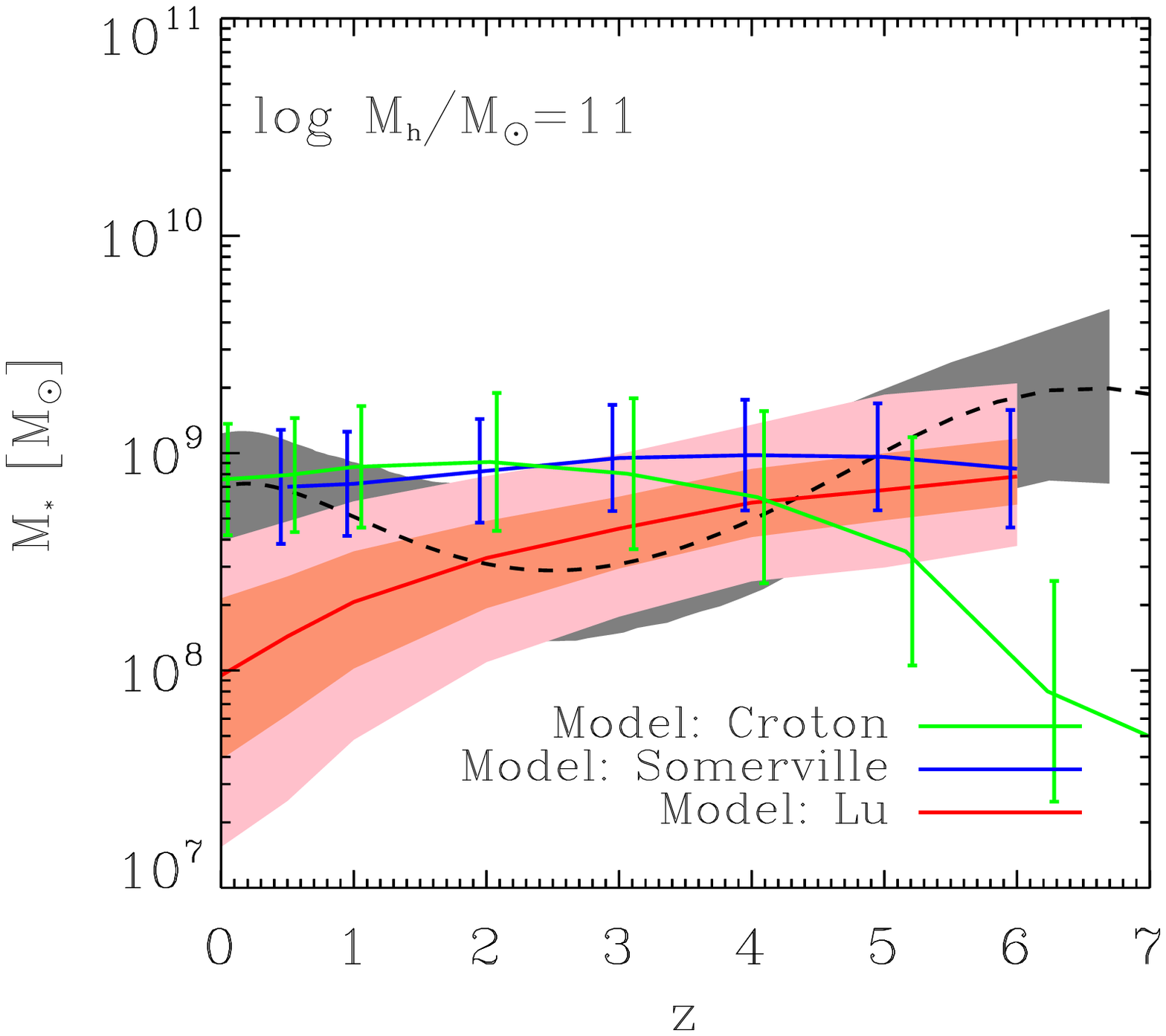} &
\includegraphics[width=0.4\textwidth]{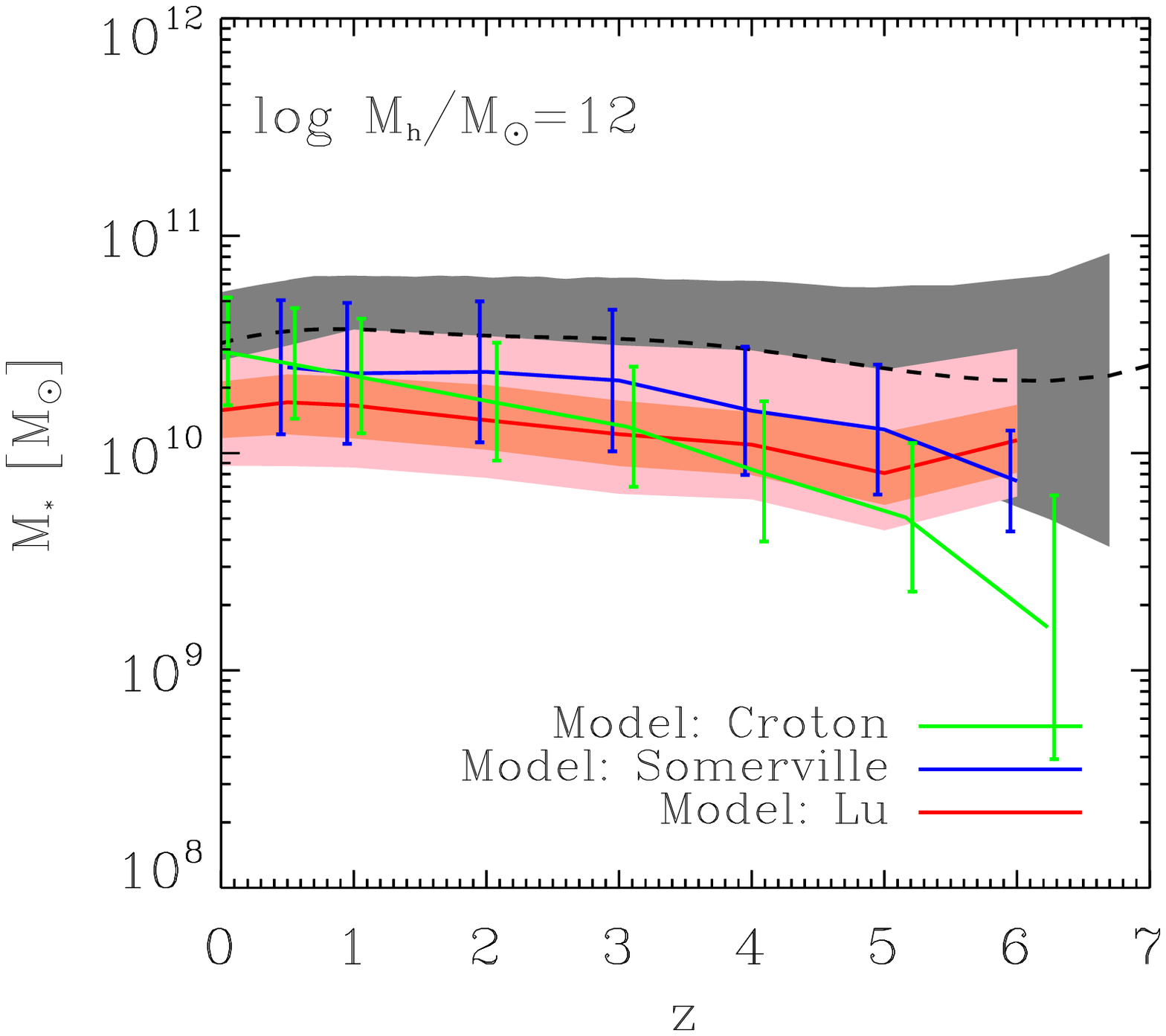} \\
\includegraphics[width=0.4\textwidth]{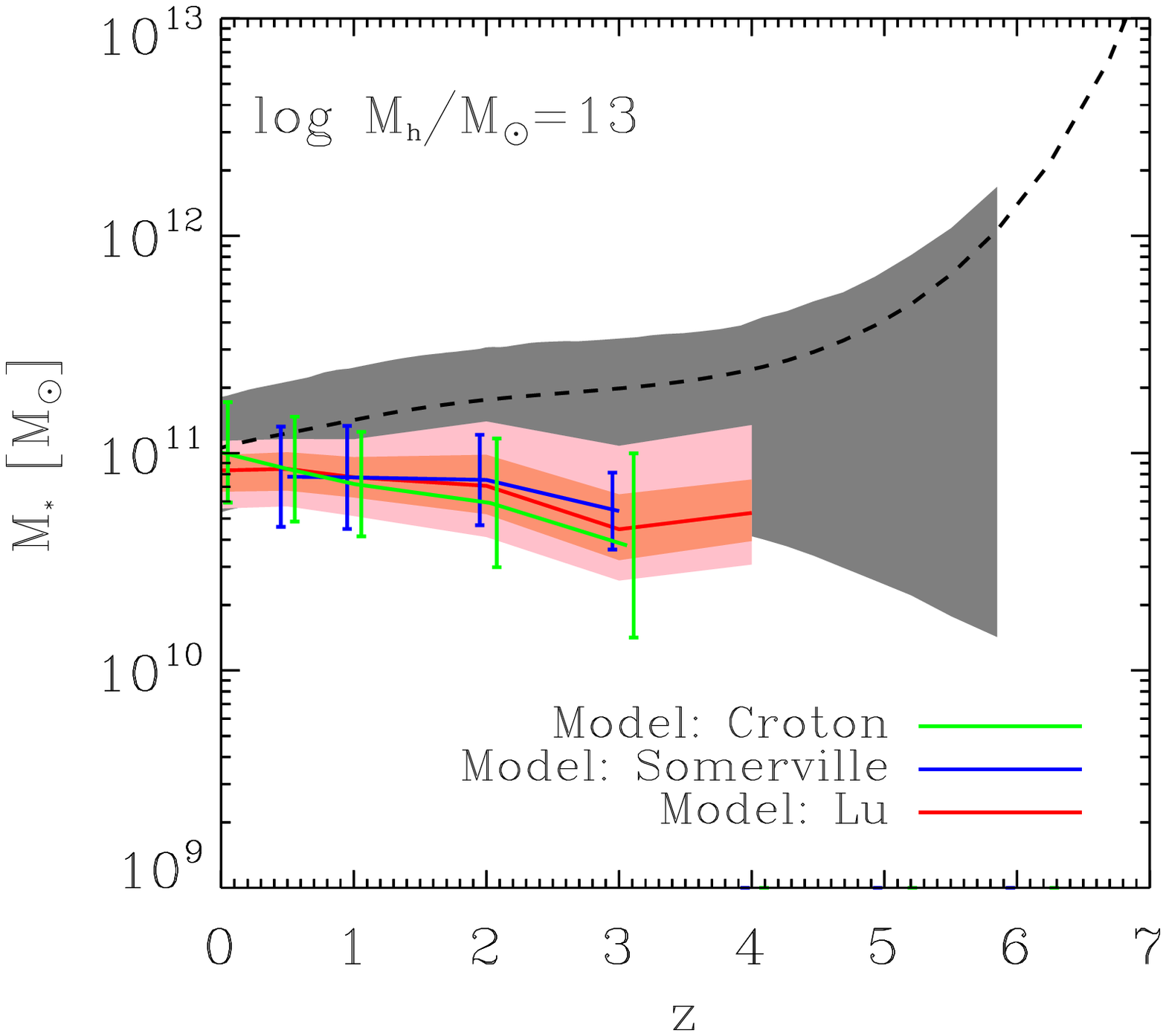} &
\includegraphics[width=0.4\textwidth]{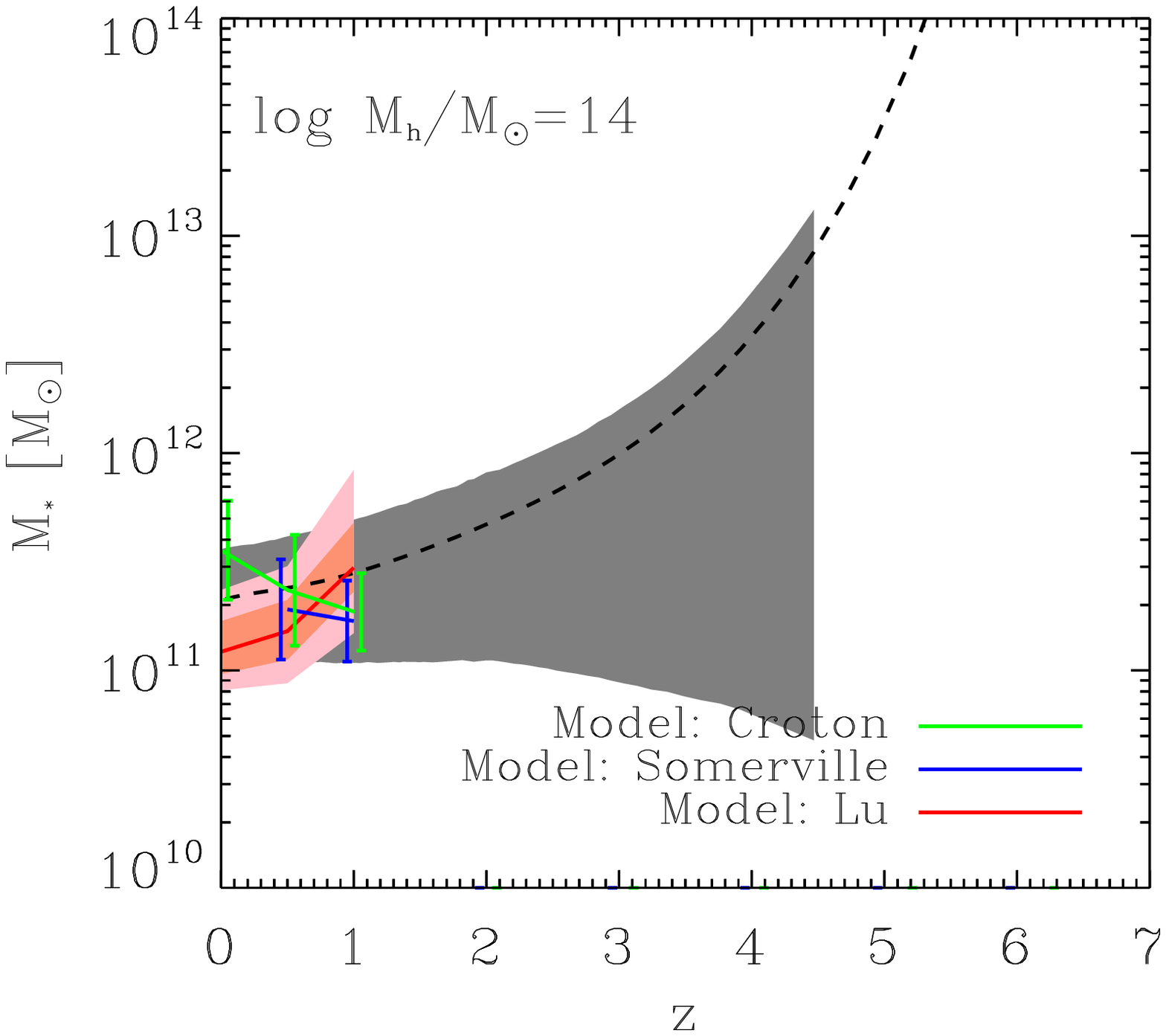} 
\end{tabular}
\caption{Central galaxy stellar mass as a function of redshift for
  different halo masses. The green line denotes the prediction of the
  Croton model, the blue line denotes the prediction of the Somerville
  model, and the error bars on them show the 1-$\sigma$ scatter of the
  model galaxy samples. The dark and light red bands encompass 67\%
  and 95\% predictive posterior regions of the Lu model. The black
  dashed line and grey band are the results of the empirical model of
  \citet{Behroozi2012}.  }
\label{fig:mstar_z}
\end{center}
\end{figure*}

We now focus our investigation on four halo mass bins, $M_{\rm
  vir}=10^{11}, 10^{12}, 10^{13}$, and $10^{14}\,\msun$, with a
$\pm0.1$ dex width, and select central galaxies hosted by such halos at
each epoch. In Figure~\ref{fig:mstar_z} we plot the median central
galaxy mass vs. redshift for each SAM (colored solid lines with error
bars), separated into these halo mass bins. Also included are the
empirical results of \citet{Behroozi2012} (black line with gray shaded
region).

We find that, while the SAM results agree with each other in a broad
sense, they disagree with each other in detail, and also show
discrepancies with the \citet{Behroozi2012} empirical model. For
low-mass halos ($10^{11}\,\msun$) in the top left panel of
Figure~\ref{fig:mstar_z}, the Lu model predicts a mildly decreasing
stellar mass with decreasing redshift, the Somerville model predicts a
roughly constant stellar mass over all redshifts, and the Croton model
predicts a strongly increasing trend with decreasing redshift for
$z>2$ and flat for $z<2$. In contrast, the Behroozi result suggests a
minimum in the characteristic stellar mass hosted in $10^{11}\,\msun$
halos at $z\sim 2-3$, and the SFR increases mildly at earlier and
later times. In higher mass bins the three SAMs tend to display better
agreement with each other across all redshifts plotted, but they
uniformly have lower characteristic stellar masses when compared to
the Behroozi empirical model.

\begin{figure*}[htb]
\begin{center}
\begin{tabular}{cc}
\includegraphics[width=0.4\textwidth]{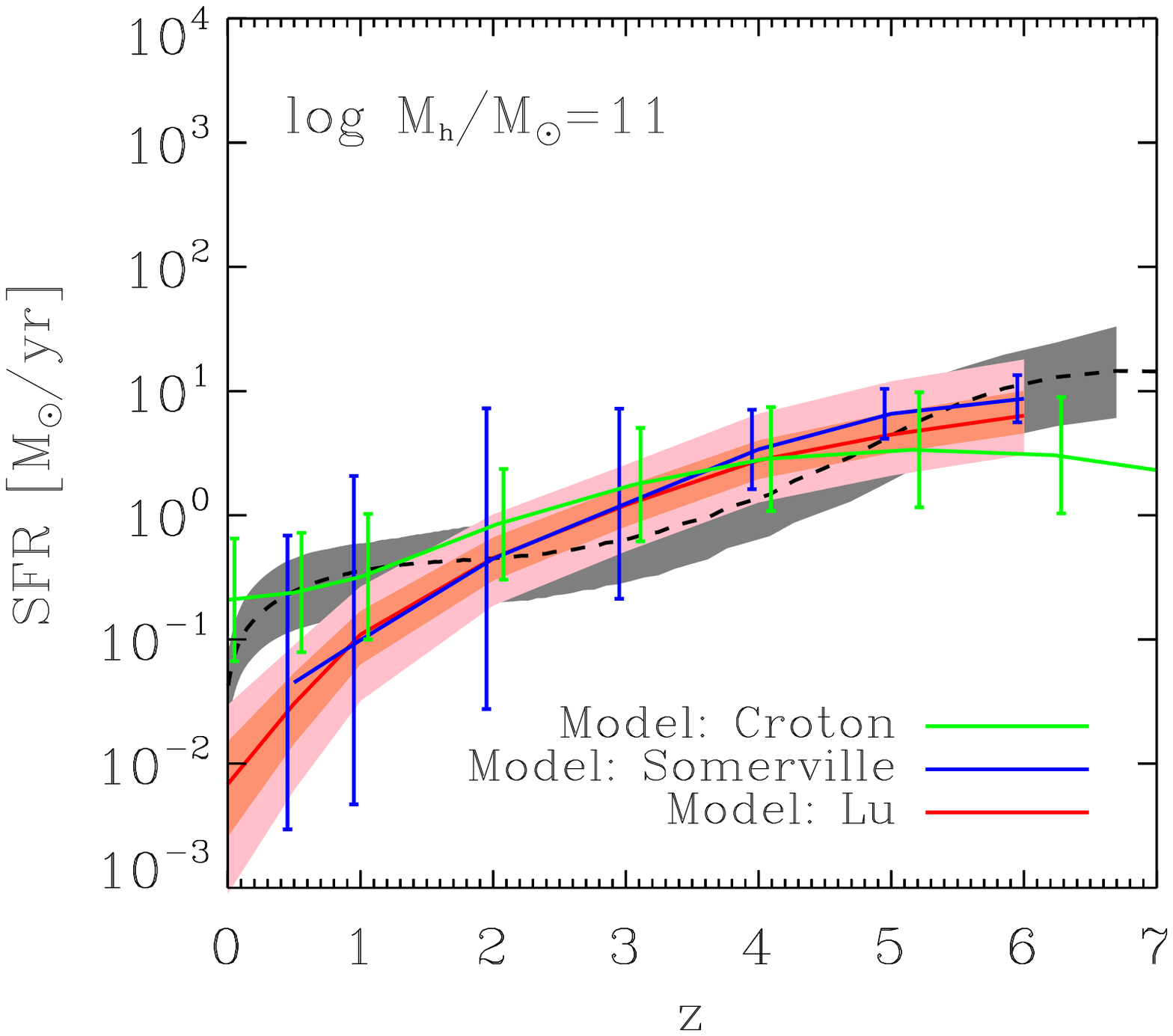} &
\includegraphics[width=0.4\textwidth]{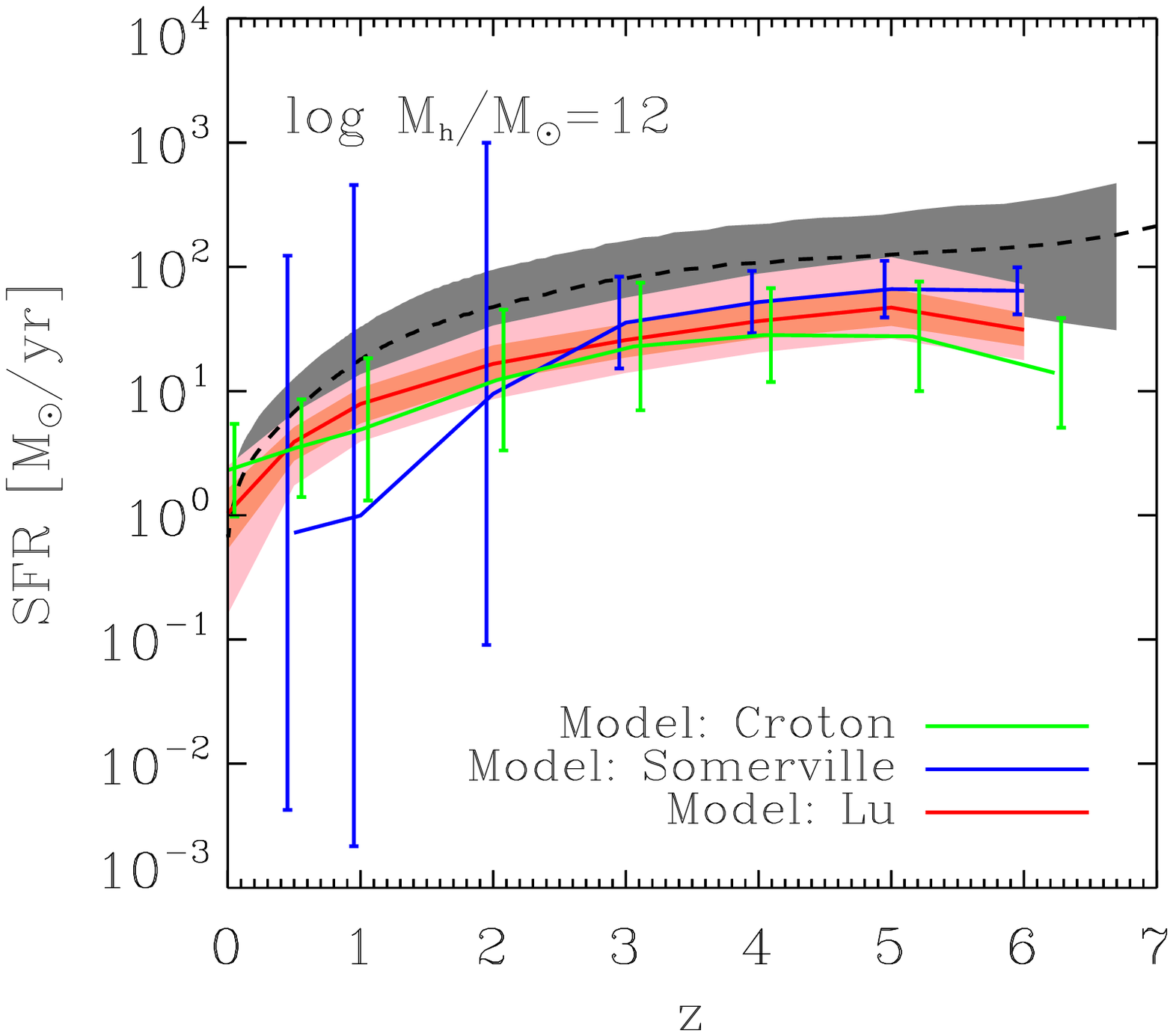} \\
\includegraphics[width=0.4\textwidth]{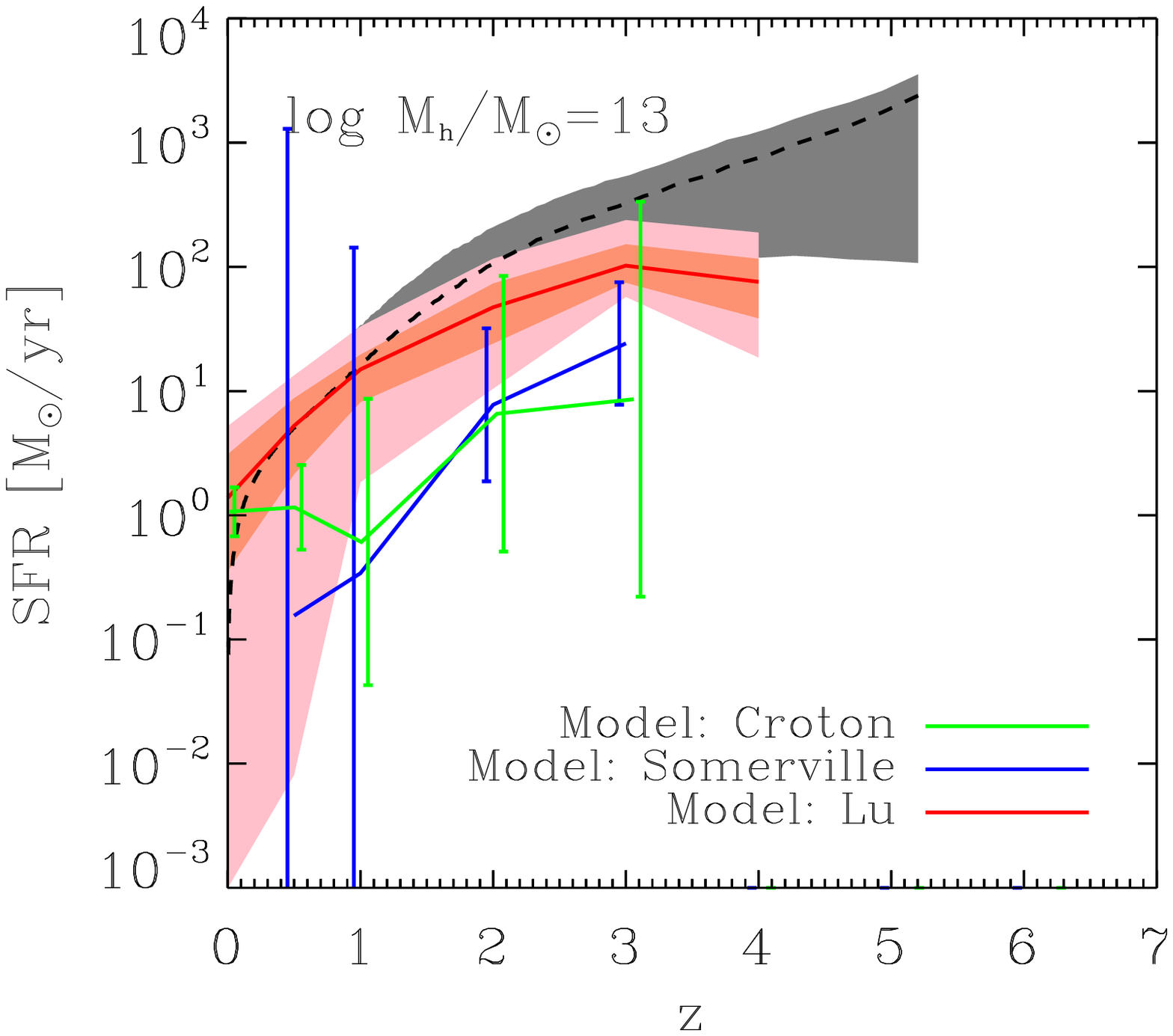} &
\includegraphics[width=0.4\textwidth]{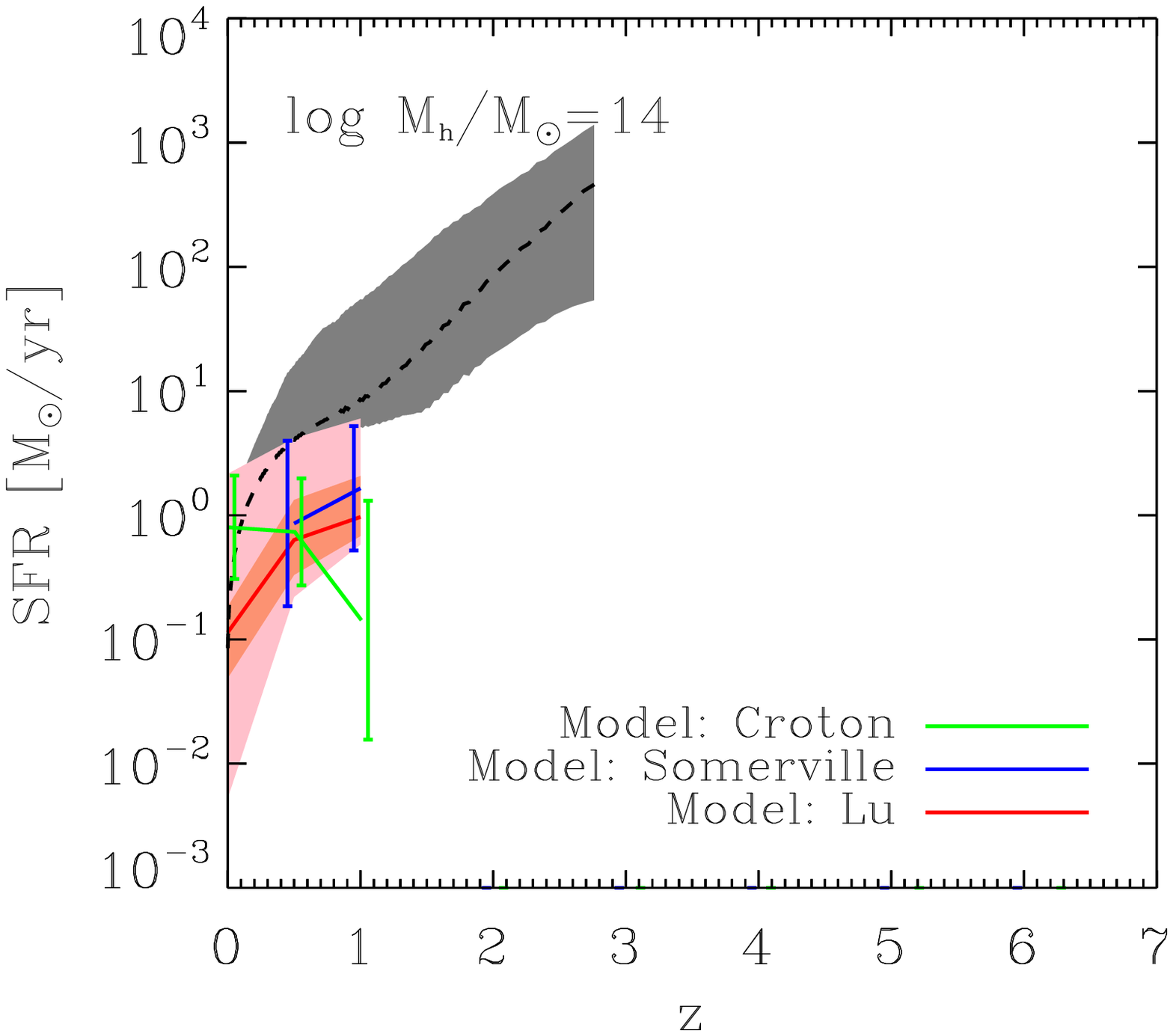} 
\end{tabular}
\caption{Central galaxy star formation rate as a function of redshift
  for different halo masses. The green line denotes the prediction of
  the Croton model, the blue line denotes the prediction of the
  Somerville model, and the error bars on them show the 1-$\sigma$ scatter
  of the model galaxy samples. The dark and light red bands encompass
  67\% and 95\% predictive posterior regions of the Lu model. The
  black dashed line and grey band are the results of the empirical
  model of \citet{Behroozi2012}.  }
\label{fig:sfr_z}
\end{center}
\end{figure*}

We also compare the star formation rates in central galaxies for
the same halo mass binning and redshift ranges, as shown in four
panels in Figure~\ref{fig:sfr_z}. The same plotting schema as used in
Figure~\ref{fig:mstar_z} has been adopted. Here, all models are
consistent, although the Somerville SAM displays a larger SFR
dispersion at later times (typically $z<2$) than the other models.
This is because the quenching of star formation due to AGN feedback is
not explicitly tied to the halo mass in the Somerville model, while in
the Croton model and Lu model, halo mass enters explicitly in the
recipe for quenching. As a result, in halos near the `transition' mass
($10^{12}$--$10^{13} \msun$), some galaxies are quenched, and some are
still on the star forming main sequence, leading to a very broad range
of SFR at a given halo mass, in spite of the tighter connection
between halo mass and stellar mass.

\subsection{Build-up of cold gas in central galaxies}\label{sec:coldgas}

Another important aspect of galaxy evolution, especially at high
redshift, is the cold gas content. SAMs make explicit predictions for
gas in galaxies, as has been recently studied with other models
\citep{Power2010, Fu2010, Lagos2011, Popping2013, Lu2013a}. 
Cold gas is a key component of
the modeling process because it is the fuel for star formation and is
sensitively affected by various key feedback processes.

For the local galaxy population, there are existing observations that
directly probe the cold gas in galaxies \citep{Zwaan2003, Zwaan2005,
  Giovanelli2005, Keres2003}. At high redshift, indirect estimates can
be obtained based on knowledge of the relationship between star
formation and cold gas obtained from local galaxies
\citep[e.g.][]{Kennicutt1998}. In this way, one can infer the cold gas
content from the observed star formation surface density in high
redshift galaxies \citep{Erb2006,Popping2012}. Therefore, it is
important to understand how the cold gas content is predicted in SAMs.
The cold gas fraction in galaxies is dichotomous: quiescent galaxies
are generally gas poor and elliptical, while star forming galaxies are
relatively gas rich and disk dominated \citep[e.g.][]{Martig2009}. We
therefore select model galaxies based on their bulge mass to total
stellar mass ratio, $B/T\equiv {M_{\rm *, B} / (M_{\rm *, B} + M_{\rm
    *, D})}$, where $M_{\rm *, B}$ is the bulge stellar mass, and
$M_{\rm *, D}$ is the disk stellar mass. We only select disk dominated
galaxies with $B/T<0.3$ to be analyzed in this section.

\begin{figure*}[htb]
\begin{center}
\begin{tabular}{cc}
\includegraphics[width=0.45\textwidth]{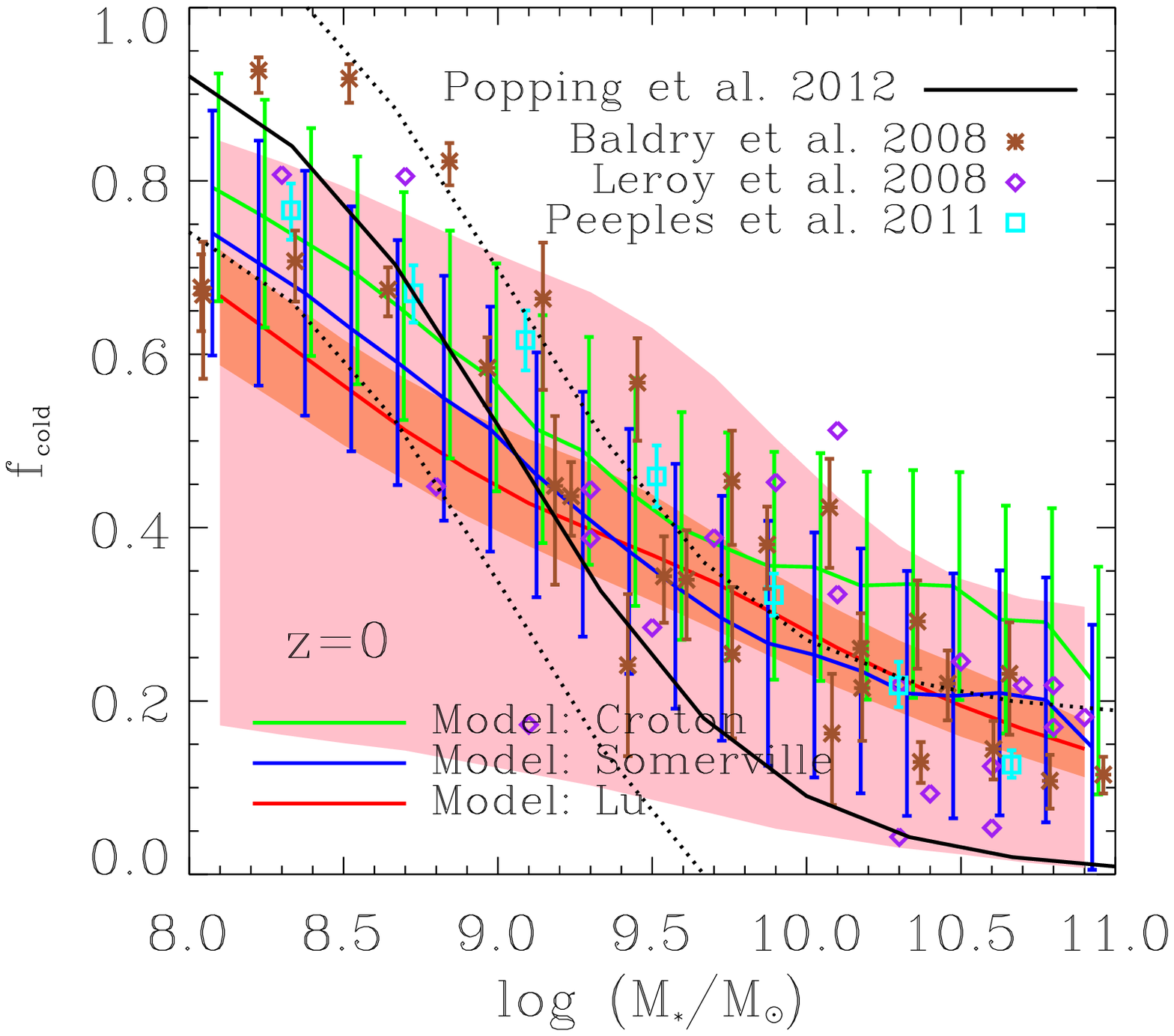} &
\includegraphics[width=0.45\textwidth]{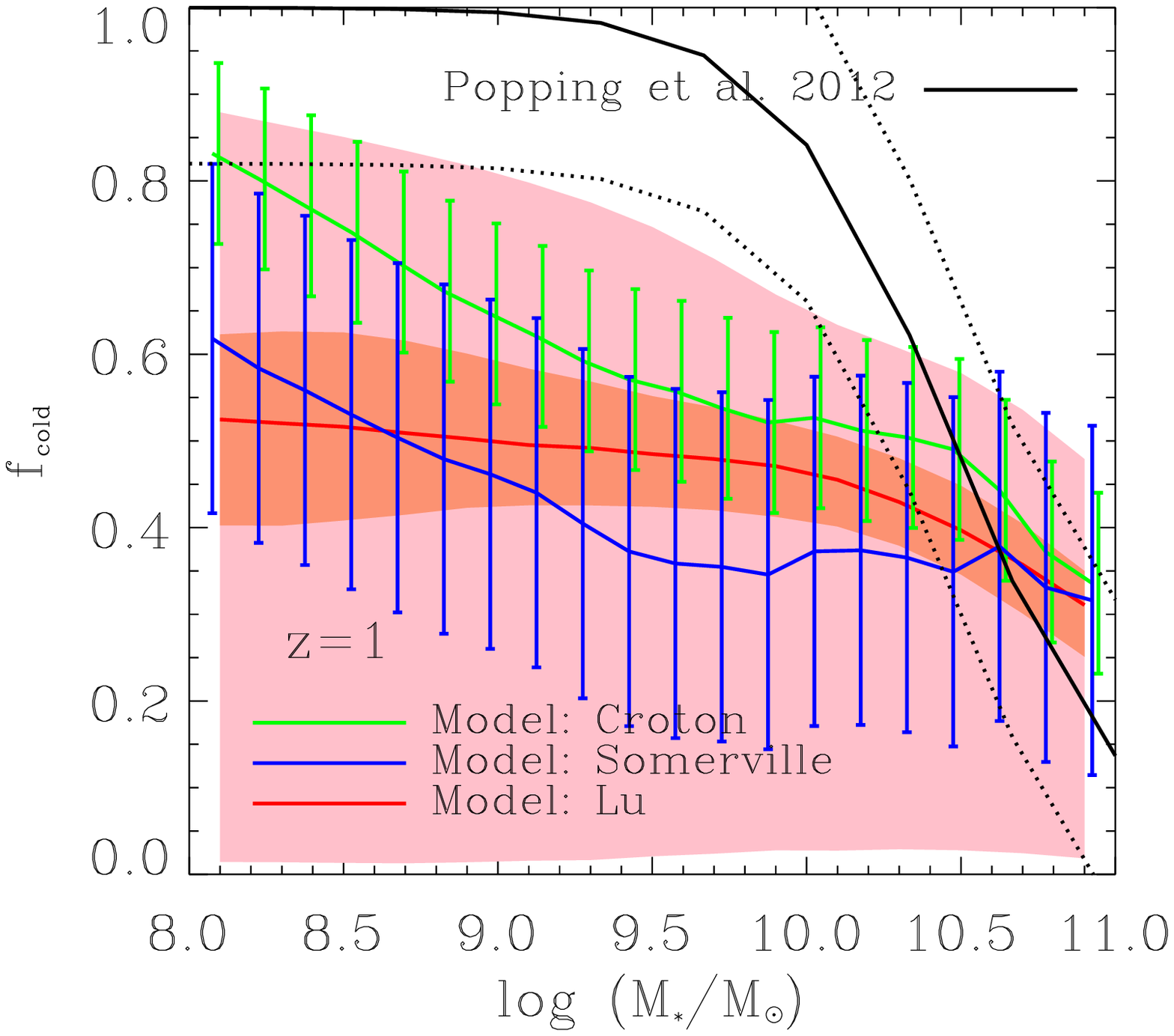} \\
\includegraphics[width=0.45\textwidth]{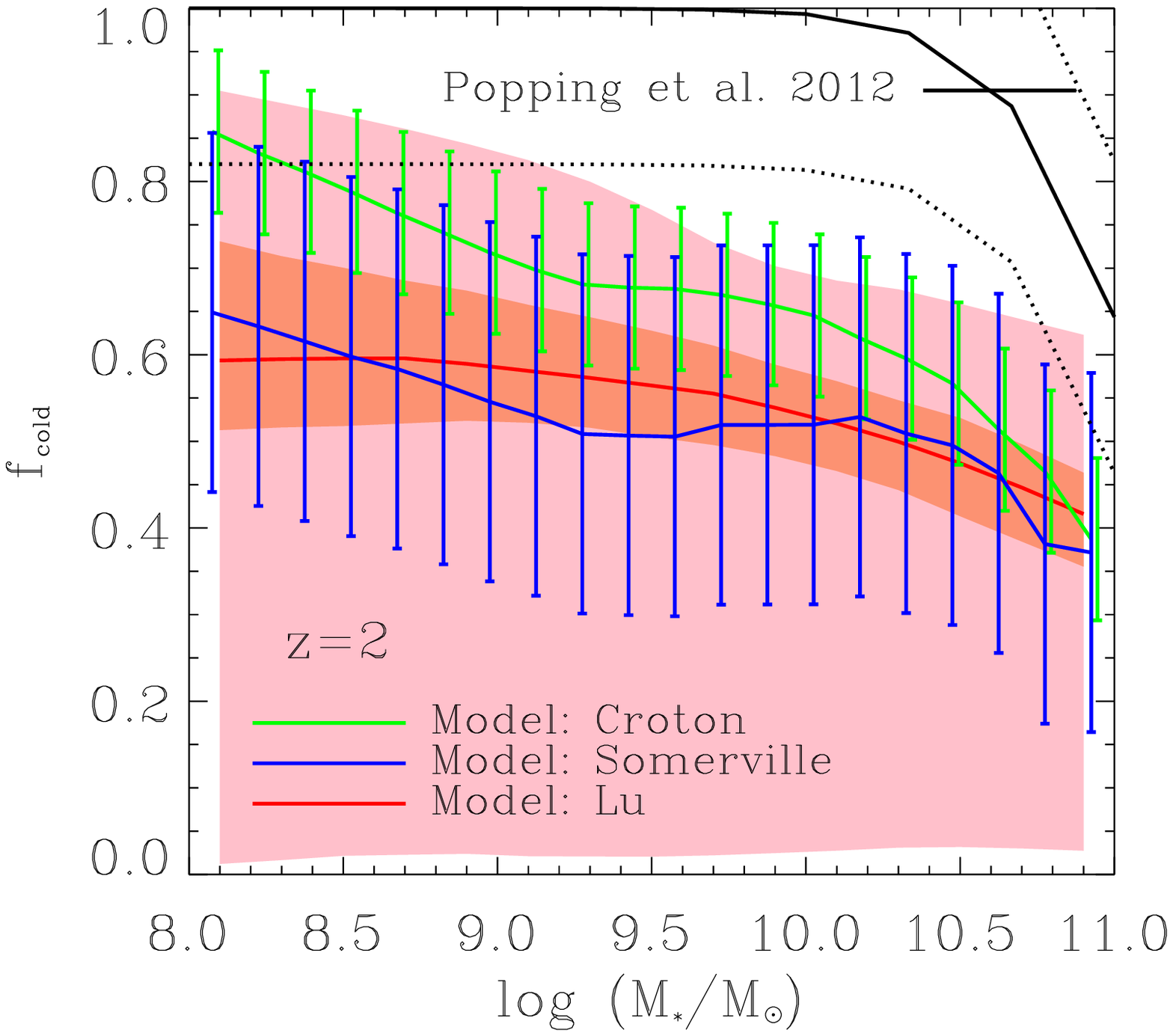} &
\includegraphics[width=0.45\textwidth]{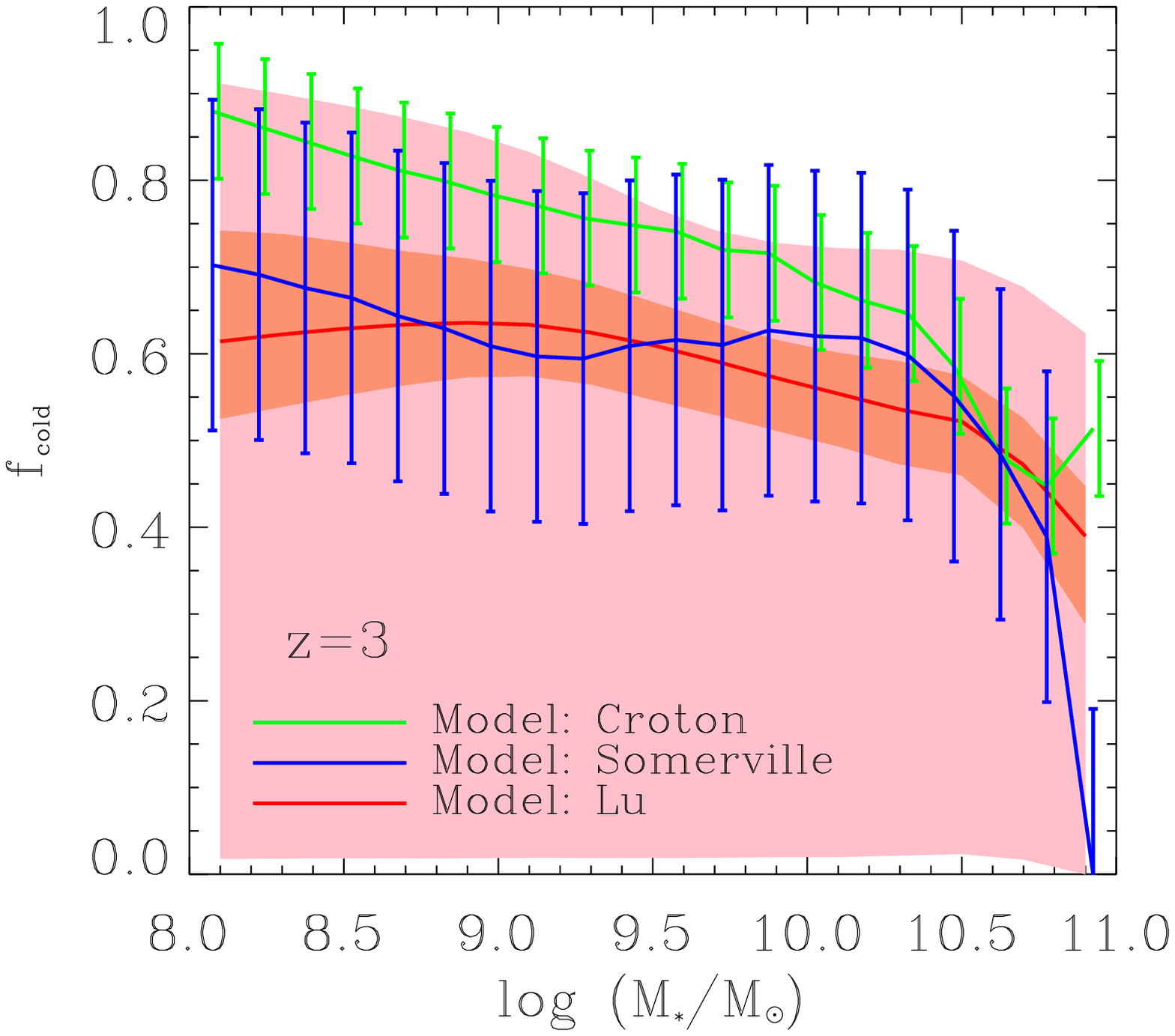}
\end{tabular}
\caption{Cold gas mass fraction, $f_{\rm cold}=M_{\rm cold} / (M_{\rm
    cold}+M_*)$, as a function of stellar mass for central disk
  galaxies at $z=0, 1, 2$, and 3. The green line denotes the
  prediction of the Croton model, the blue line denotes the prediction
  of the Somerville model, and the error bars on them show the 1-$\sigma$
  scatter of the model galaxy samples. The dark and light red bands show
  67\% and 95\% predictive posterior regions of the Lu model. The
  black solid line and dotted lines show the indirect empirical estimates of
  \citet{Popping2012} and their $1-\sigma$ deviations for $z\leq2$. 
  The color symbols in the $z=0$ panel show compilations of cold gas fraction measurements for local galaxies
  from \citet{Baldry2008}, \citet{Leroy2008}, and \citet{Peeples2011}. }
\label{fig:fcold}
\end{center}
\end{figure*}

In Figure~\ref{fig:fcold}, we show the cold gas fraction of disk
dominated galaxies, defined as $f_{\rm cold}=M_{\rm cold} / (M_{\rm
  cold}+M_*)$, as a function of stellar mass at $z=0, 1, 2,$ and
3. While there are differences in the $f_{\rm cold}-M_*$ predictions
between each of the models, in general, the cold gas fraction decreases
with increasing stellar mass. At $z=0$, all models predict that low
mass disk galaxies are dominated by cold gas, with gas fractions
commonly reaching above 60\%. At higher masses, above $10^{9.5}\,\msun$,
the majority of the model galaxies have much lower gas fractions, less
than 30\%, even when classified as disk galaxies. At higher redshifts,
the cold gas fraction tends to have larger values at a given stellar
mass for all models, especially the high-mass end.

There are also some differences between the models.  At all four
redshifts, the Croton model predicts a higher cold gas mass fraction
than the two other models, and the median predictions of the Lu model are
similar to the Somerville model with a slightly lower fraction
($<10$\%) at very low mass end ($<10^9\,\msun$). At higher redshift,
$z\geq1$, the Lu model predicts a flatter cold gas mass fraction as a
function of stellar mass. The Croton model and the Somerville model
have a steeper cold gas mass fraction-stellar mass relation below
$10^9\,\msun$. The posterior prediction of the Lu model covers a large
range of cold gas fractions. The two other models are always
encompassed by the 95\% confidence range.

In Figure~\ref{fig:fcold}, we also compare the model predictions with
observational measurements of the cold gas fraction for local galaxies
and an indirect determination of the gas fraction by
\citet{Popping2012} for $z\leq2$.  For $z=0$, we have included
compilations of the cold gas fraction from \citet{Baldry2008},
\citet{Leroy2008}, and \citet{Peeples2011}.  It is worth noting the
data points are from different data sources and measure different
components of the cold gas.  The data from \citet{Leroy2008} are
measurements of both HI and H$_2$ including a correction for helium.
The \citet{Baldry2008} data set include the atomic gas masses derived
from the Westerbork HI Survey \citep{Swaters2002, Noordermeer2005} and
the HIPASS catalogue \citep{Meyer2004, Wong2006}, and the literature
compilation of \citet{Garnett2002}.  These data also include a
correction of 1.33 for helium, but do not include the molecular
hydrogen gas.  The compilation of \citet{Peeples2011} includes the
total HI gas masses measured from 21 cm line fluxes
\citep{McGaugh2005}, the HI gas masses from \citet{West2009,
  West2010}, and the total cold gas measurements from
\citet{Leroy2008}.  
At $z=0$ all models are in general agreement with
the data, but with a tendency to overpredict the gas fraction at the
intermediate to high-mass end. At higher redshifts, the
\citet{Popping2012} results reveal how the cold gas becomes
increasingly dominant. Typical galaxies with stellar mass below
$10^{9.5}\,\msun$ at $z=1$, and $10^{10.5}\,\msun$ at $z=2$ have more than
95\% of their cold baryon mass in cold gas. In the models, however,
although the cold gas fraction does increase with redshift, the
amplitude is still significantly lower than the data.  The large
differences between the models and the broad posterior predictive
distribution from the constrained Lu model indicate that the
prediction of the cold baryon fraction is sensitive to
the parameterizations of star formation and feedback implemented in a
model.  The comparison demonstrates that the gas fraction in the models
are not strongly constrained, and observational data for the cold
baryon fraction will be very useful to tighten the constraint.

\subsection{Build-up of metals in central galaxies}

\begin{figure*}[htb]
\begin{center}
\begin{tabular}{cc}
\includegraphics[width=0.45\textwidth]{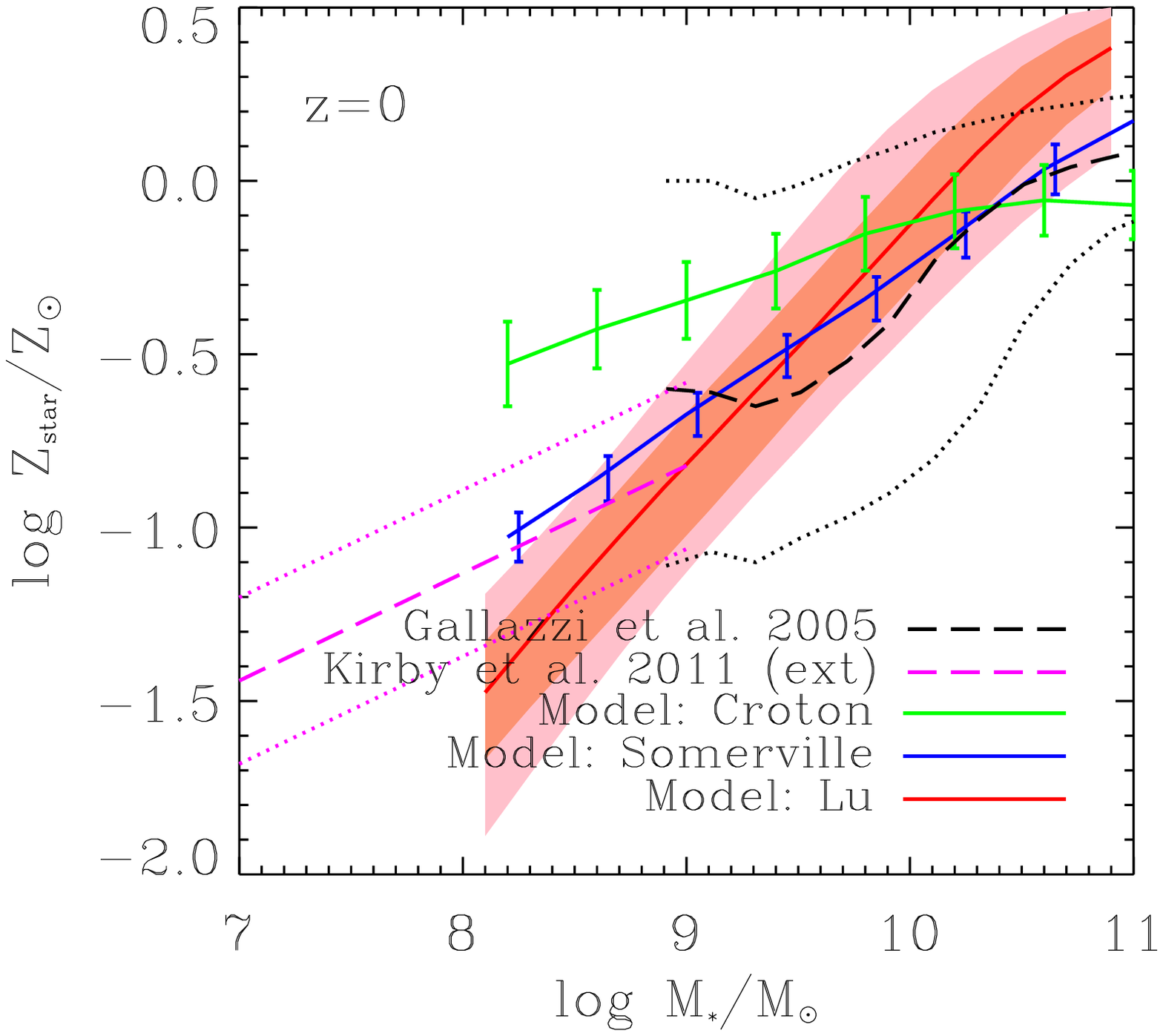} &
\includegraphics[width=0.45\textwidth]{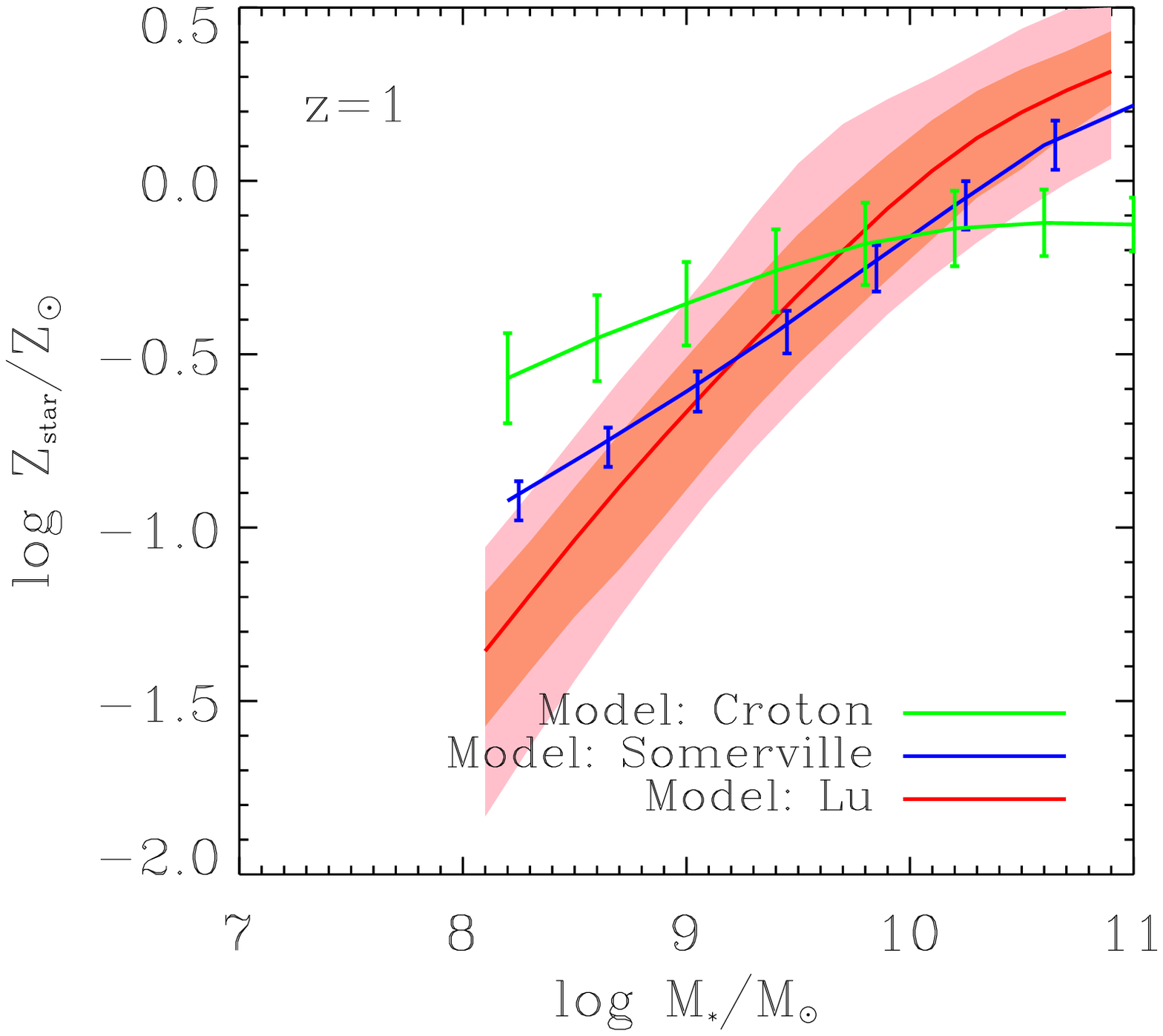} \\
\includegraphics[width=0.45\textwidth]{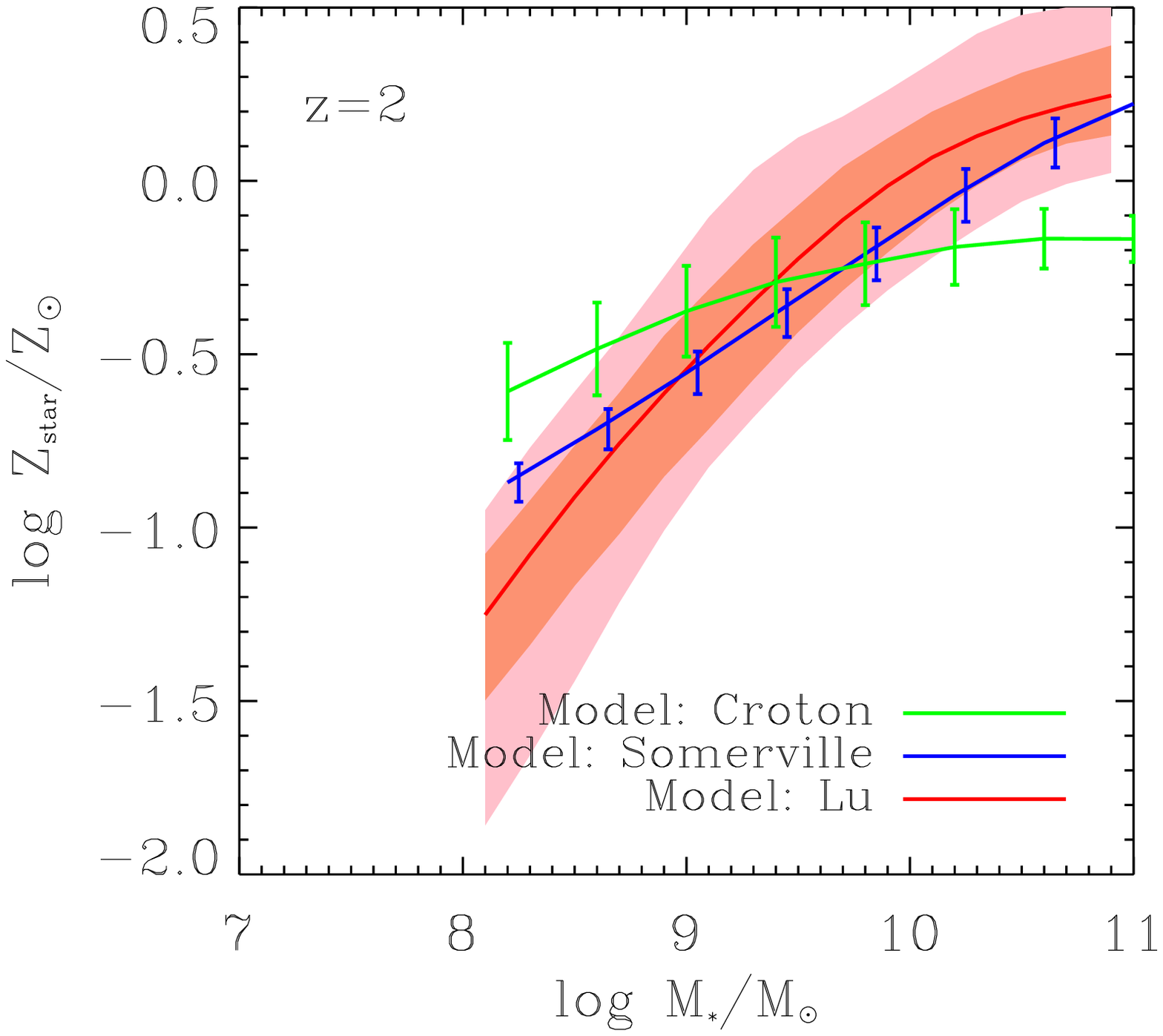} &
\includegraphics[width=0.45\textwidth]{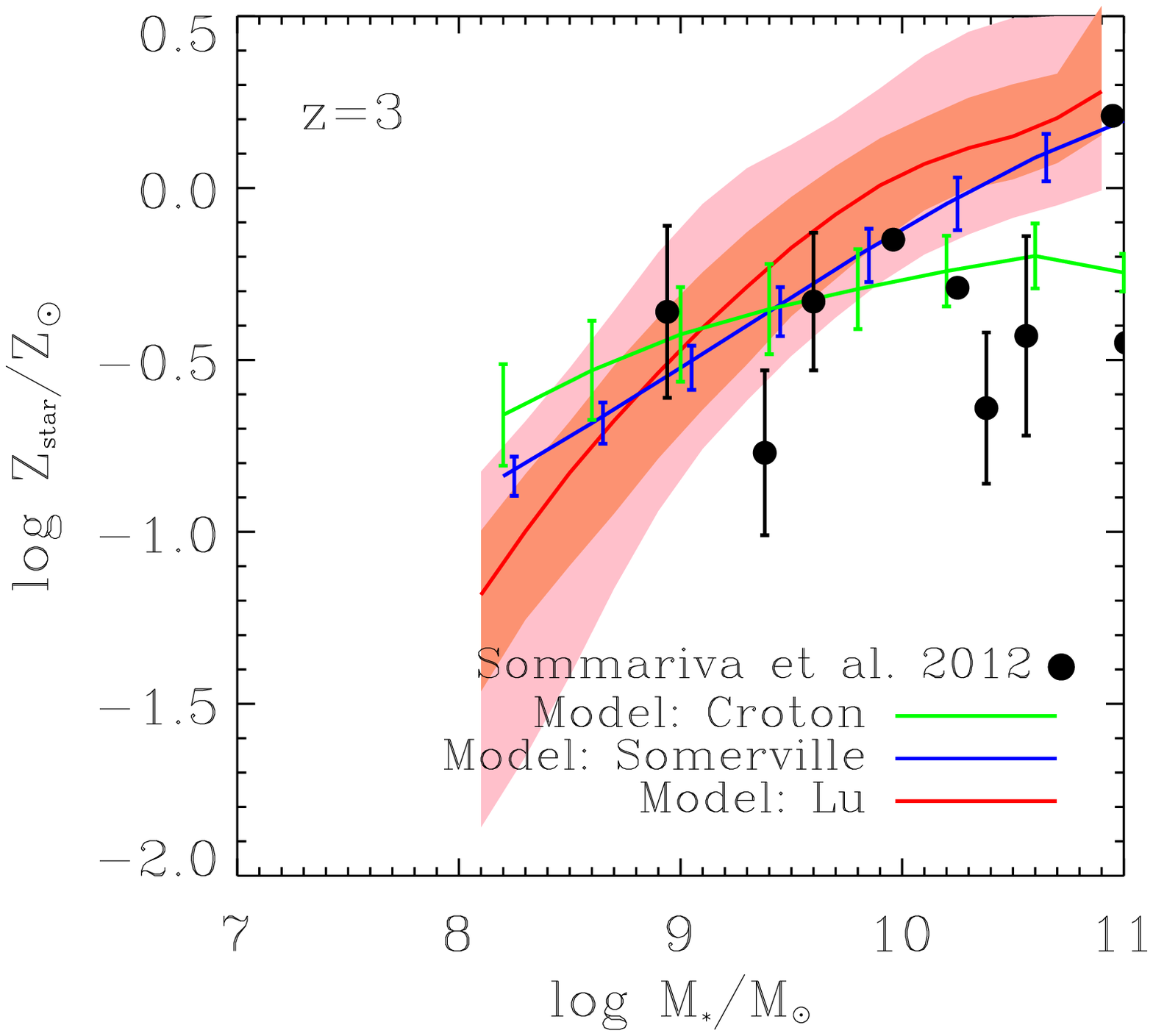}
\end{tabular}
\caption{Stellar metallicity as a function of stellar mass at $z=0, 1,
  2$, and 3 predicted by the three SAMs.  The green line denotes the prediction
  of the Croton model, the blue line denotes the prediction of the
  Somerville model, and the error bars on them show the 1-$\sigma$ scatter
  of the model galaxy samples. The dark and light red bands encompass
  67\% and 95\% predictive posterior regions of the Lu model.  The
  black dashed line shows the observational estimates of
  \citet{Gallazzi2005} for local galaxies.  The solid magenta line
  shows the extrapolation of the results of \citet{Kirby2011} to
  higher stellar masses.  
The dotted lines are 1-$\sigma$ dispersions
  of the observational estimates. The black dots in the $z=3$ panel
  denote the observational measurements of the stellar metallicity of
  a sample of galaxies at $z\sim3$ presented in \citet{Sommariva2012}.
}
\label{fig:smetal}
\end{center}
\end{figure*}

For each model galaxy we predict the amount of metals that are locked
into stars. In Figure \ref{fig:smetal}, the stellar metallicity of
model central galaxies as a function of stellar mass at $z=0, 1, 2,$
and 3 is plotted, where we adopt a recent calibration of the solar
metallicity, $Z_{\odot}=0.0142$ \citep{Asplund2009}. All models
predict a trend that the stellar metallicity increases with stellar
mass, which is broadly consistent with the observational estimate of
\citet{Gallazzi2005}. \citet{Trager2009} have shown that, unlike the
Balmer-line indices 
which are strongly biased by the presence of young stars,
stellar metal-line indices correlate almost perfectly with mass- and
light-weighted metallicity \citep[as shown in][]{Serra2007}, with
very small scatter. The equivalent metallicity derived from fitting
the metal-line indices with simple stellar population models is
therefore a very good tracer of the light- or even mass-weighted
metallicity of a galaxy.  This explains why the models can agree with
the metallicity-stellar mass relation of \citet{Gallazzi2005} but seem
to be inconsistent with their stellar age-stellar mass relation.

It is clear, however, that the models predict somewhat different
slopes for the metallicity-stellar mass relation.  The Croton model
shows a relatively shallow slope, while the Lu model shows the
steepest slope among the three models. The Somerville model matches
the data reasonably well. For the high-mass end, $M_*>10^{10}\,\msun$,
the metallicity predicted by the median Lu model is about 0.2-0.3 dex
higher than the observation of \citet{Gallazzi2005}. At lower masses,
the Lu model predicts a rapidly decreasing trend for decreasing stellar
masses.

We extrapolate the results of \citet{Kirby2011} from dwarf galaxies to
compare with our model predictions. It is clear that the Lu model
underpredicts the metallicity for low-mass galaxies. In contrast, the
Croton model overpredicts the metallicity in the stellar mass range
$10^8-10^{9}\,\msun$. For higher redshifts, the metallicity-stellar mass
relation predicted by the three models barely evolves with time. This
is consistent with observational measurements of stellar metallicity
for a sample of galaxies at $z\sim3$ by \citet{Sommariva2012}. As a
result of the weak redshift evolution, the model predictions at high
redshifts keep the same trend, in that the Croton model consistently
predicts the shallowest stellar metallicity-stellar mass relation and
the Lu model has the steepest relation.

The discrepancy between the models sheds light on the
metal enrichment processes in galaxy formation. As we described earlier,
each model assumes a different mass-loading factor for low-mass
galaxies in their calibration. The Croton model assumes a constant
mass-loading factor, the Somerville model assumes a modest velocity
dependence for the loading factor, and the Lu model adopts extremely
strong velocity dependence for the mass-loading factor. As a result,
the stronger galactic wind in low-mass galaxies in the Lu model tends
to blow out a larger amount of metal-enriched cold gas. This explains
the trend that, at all redshifts, the Lu model predicts the steepest
metallicity-stellar mass relation, and the Croton model predicts the
shallowest.

\begin{figure*}[htb]
\begin{center}
\begin{tabular}{cc}
\includegraphics[width=0.45\textwidth]{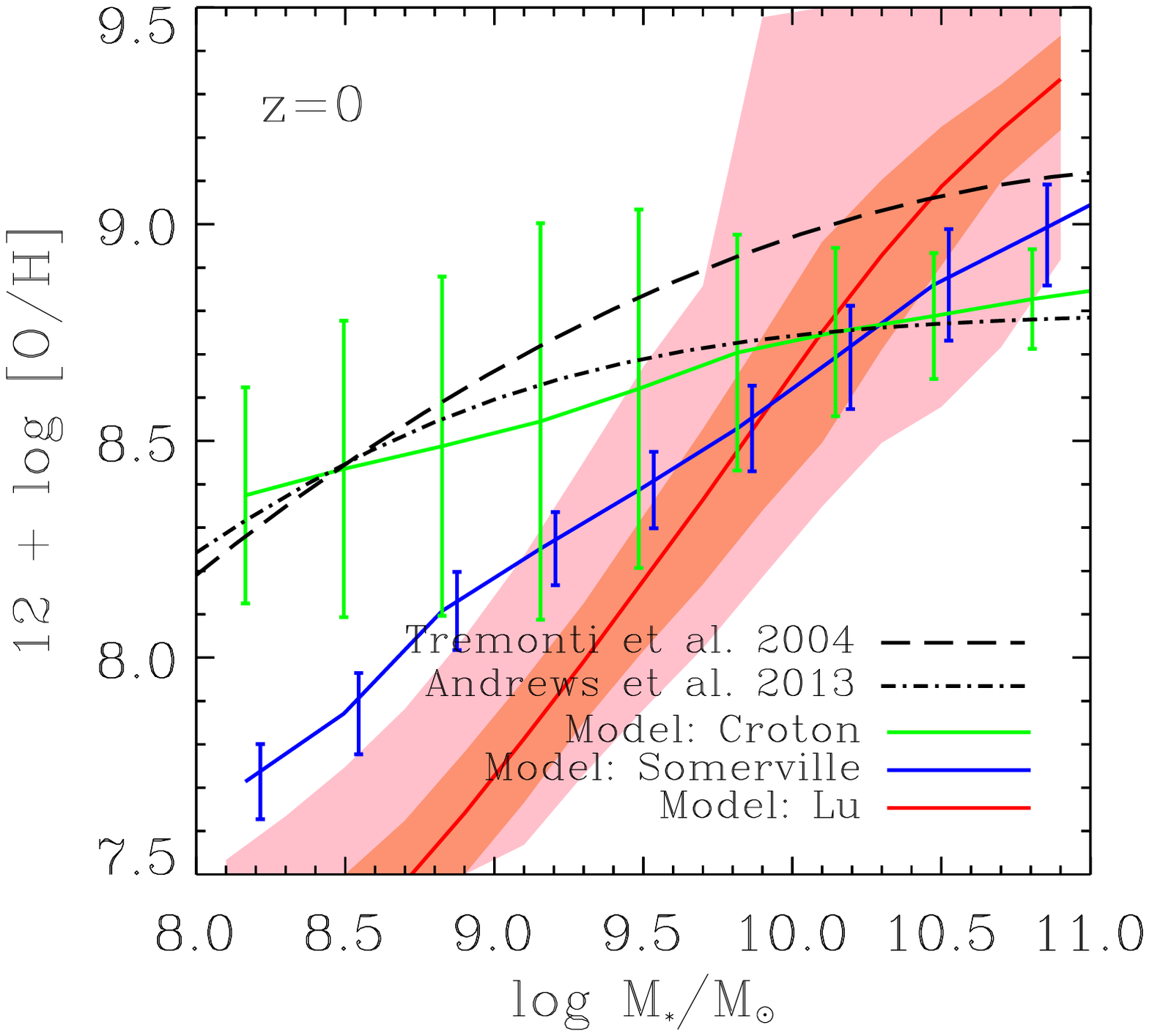} &
\includegraphics[width=0.45\textwidth]{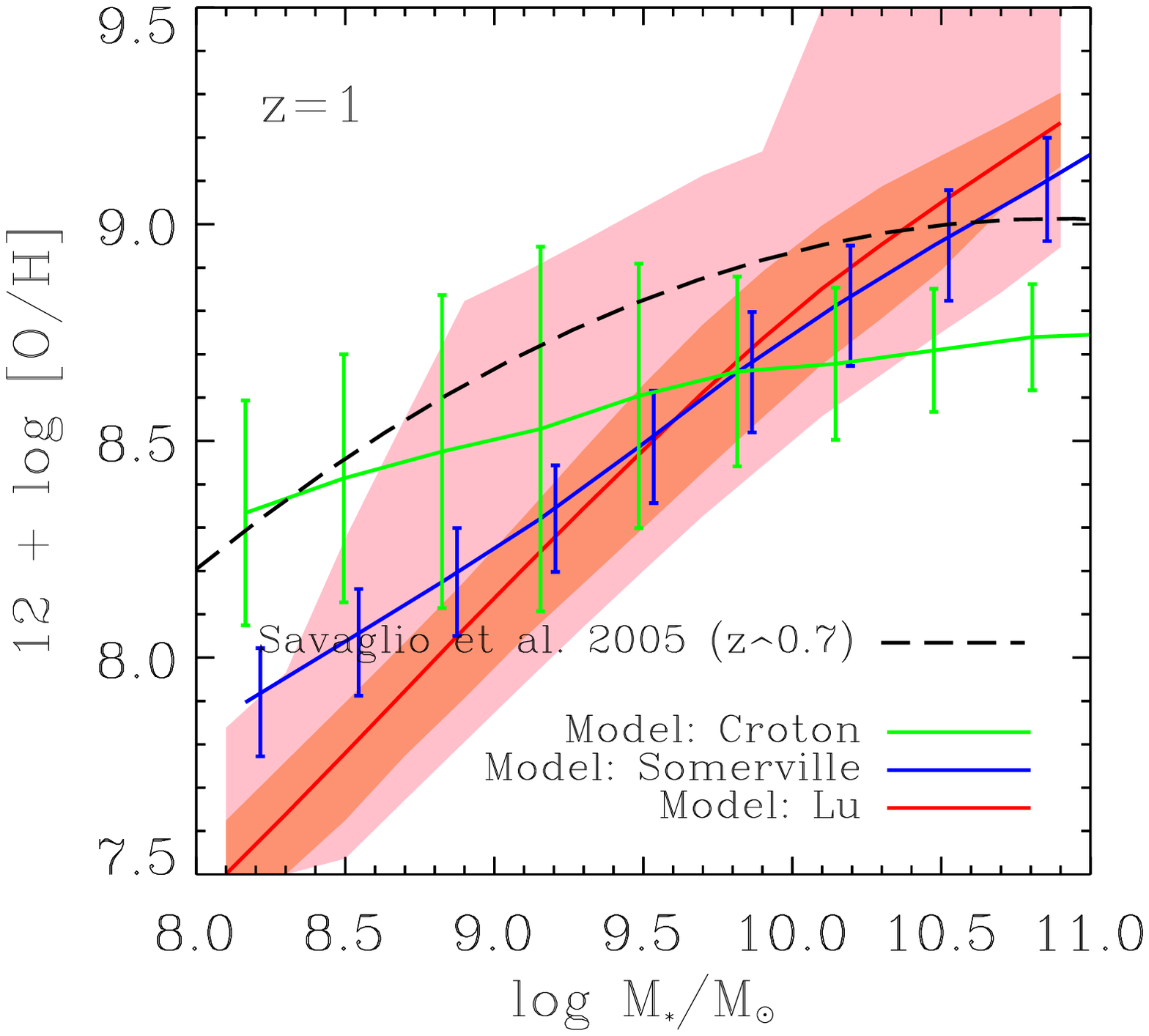} \\
\includegraphics[width=0.45\textwidth]{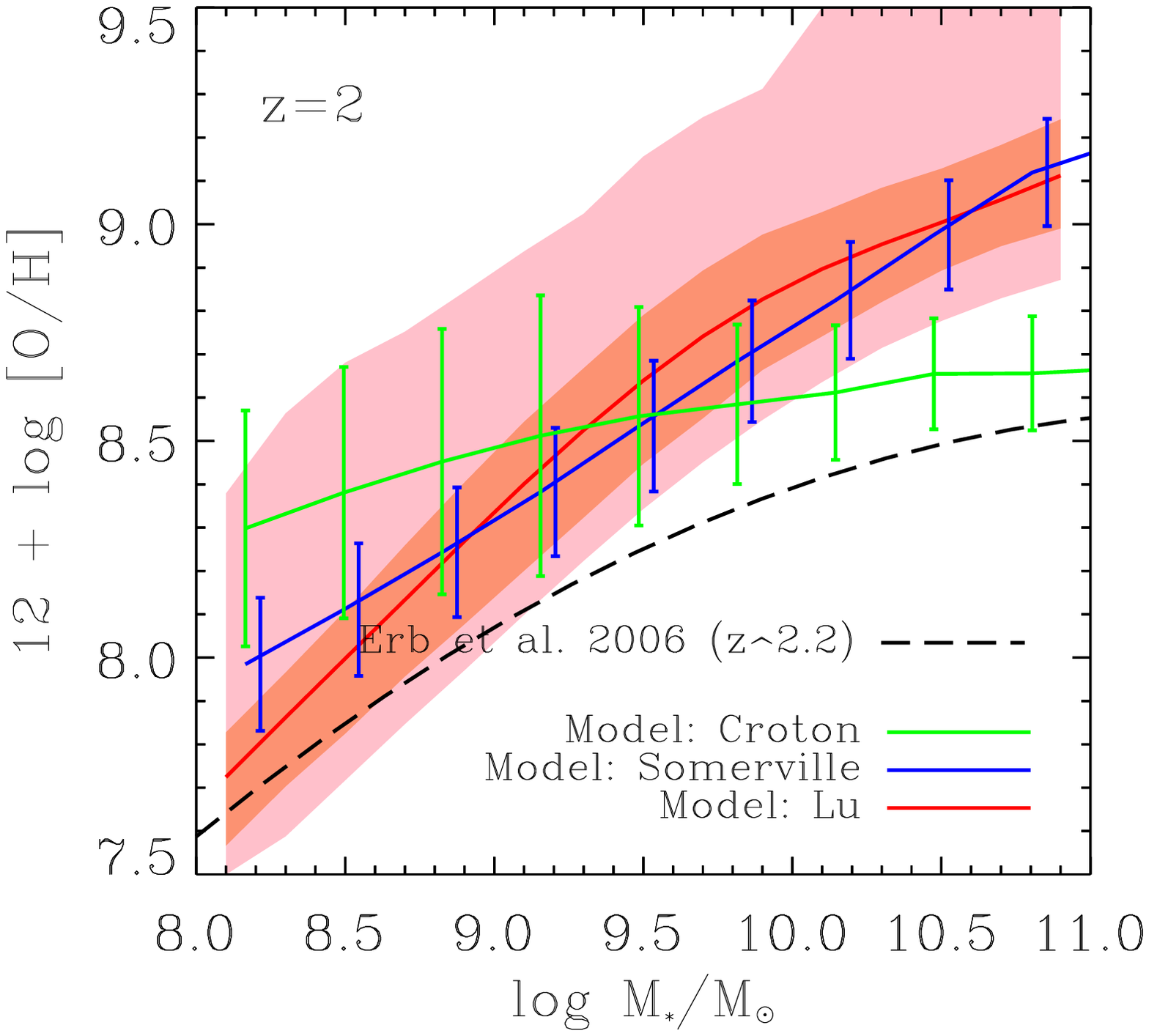} &
\includegraphics[width=0.45\textwidth]{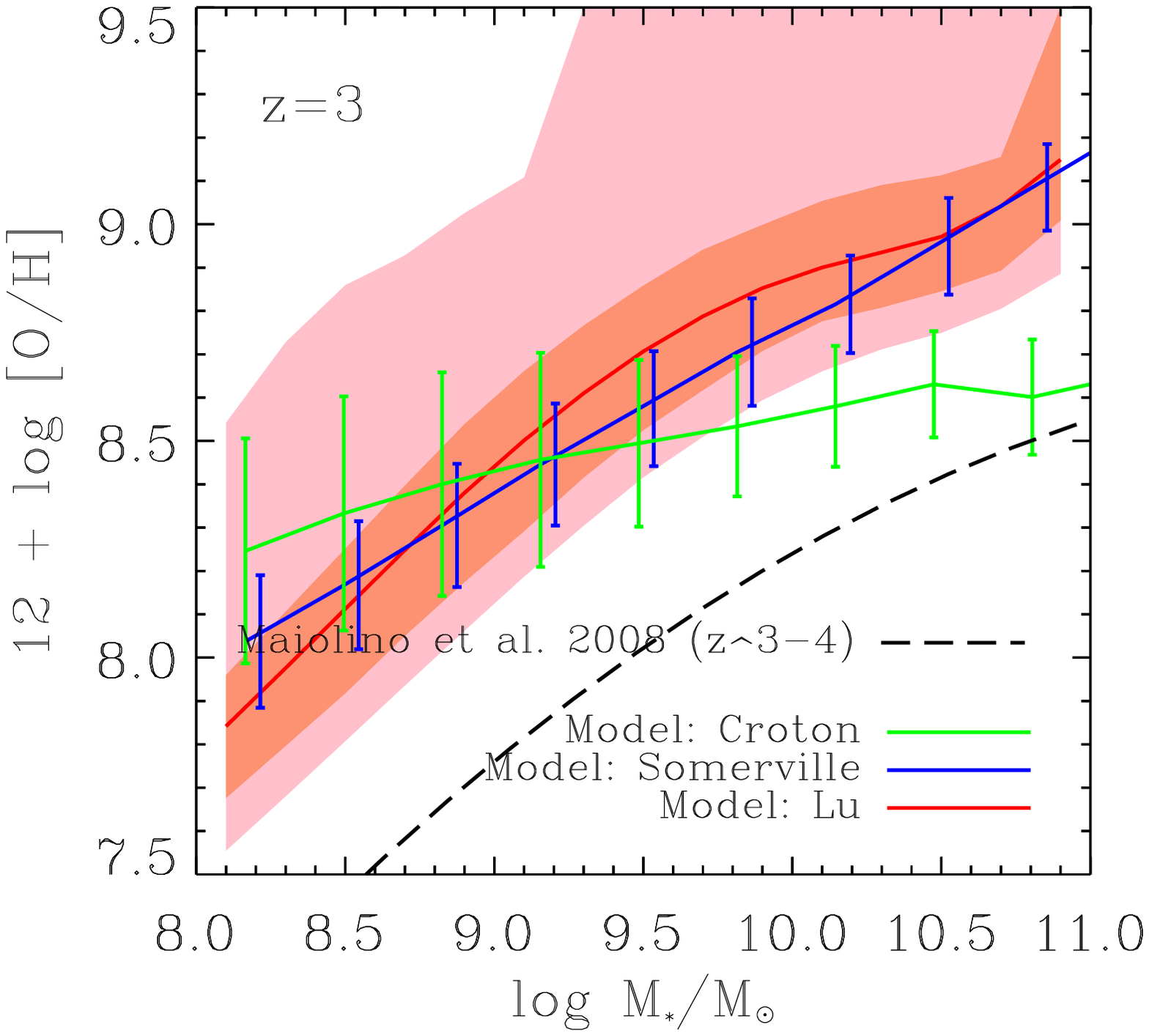}
\end{tabular}
\caption{ISM metallicity as a function of stellar mass at $z=0, 1, 2,
  $ and 3 predicted by the SAMs.  The green line denotes the
  prediction of the Croton model, the blue line denotes the prediction
  of the Somerville model, and the error bars on them show the 1-$\sigma$
  scatter of the model galaxy samples. The dark and light red bands
  encompass 67\% and 95\% predictive posterior regions of the Lu
  model.  The black dashed lines in each panel are the observational
  results of \citet{Tremonti2004} for $z=0.07$, \citet{Savaglio2005}
  for $z=0.7$, \citet{Erb2006} for $z=2.2$, and \citet{Maiolino2008}
  for $z\sim 3-4$.}
\label{fig:cgmetal}
\end{center}
\end{figure*}

We also make predictions for the gas phase metallicity. Figure
\ref{fig:cgmetal} shows the metallicity of cold gas for model galaxies
as a function of stellar mass at four redshifts, $z=0, 1, 2,$ and 3,
using the same schema employed previously. To compare our predicted
metallicities with observational data, we normalize the predicted
metallicities by the solar metallicity, $\left[12+\log({\rm
    O/H})\right]_{\rm \odot}=8.69$ \citep{AllendePrieto2001} and
$Z_{\odot}=0.0142$ \citep{Asplund2009}
\footnote{Note that our models
  only track the total metallicity, and do not actually track
  the Oxygen abundance. We are effectively assuming that all galaxies
  have solar abundance ratios.}. The cold gas in SAMs represents the
ISM for disc galaxies. 
Thus, we take only central galaxies with bulge
masses less than 0.3 of the total stellar mass.
A number of observations are overplotted to be compared with the models:
\citet{Tremonti2004} and \citet{Andrews2013} locally, \citet{Savaglio2005} at $z\sim0.7$,
\citet{Erb2006} at $z\sim2.2$, and \citet{Maiolino2008} at $z\sim3-4$
\citep[also see][]{Mannucci2009}.

At all redshifts, Figure \ref{fig:cgmetal} shows an increasing trend
of metallicity with stellar mass for all three models. However, as was
the case for the stellar metallicity, we find different slopes and
different evolution from the present to $z=3$. In particular, the
Croton model predicts a shallower slope for the gas phase metallicity
- stellar mass relation at all epochs, and the Lu model predicts the
steepest. At $z>0$, the Lu model and the Somerville model display very
similar results.

Comparing models to data, we find that the Lu and Somerville models
lie significantly under the observed relations for all but the highest
mass galaxies at $z\le 1$, while the shallower slope of the relation
in the Croton model is in better agreement with the low redshift observations, 
especially the recent observational results of \citet{Andrews2013}. 
However, at higher redshifts all models produce ISM metallicities that
are higher than the observational estimates, and significantly so at
$z=3$. It is worth noting that observations of the metallicities of
high-redshift galaxies have significant disagreement. For example,
\citet{Richard2011} found a weaker redshift evolution in the
stellar mass-metallicity relation in a gravitationally lensed galaxy
sample at $z\sim 2-3$. The authors found that their samples are
$\sim0.25$dex more metal-rich than those studied in
\citet{Maiolino2008} and
\citet{Mannucci2009}. Moreover,\citet{Mannucci2010} showed that the
cold gas metallicity also depends on SFR using local and $z\sim 2.5$
galaxies. The observed evolution of the metallicity-stellar mass
relation could be due to the fact that the SFR for galaxies with a given
stellar mass is increasing with redshift. A selection effect, that
high SFR galaxies tend to be selected in the high redshift
samples, could also result in 
the metallicities being biased low. 
  
As we discussed earlier, the different mass-loading factor scaling
relations adopted in the models strongly affect the resulting metal
content of model galaxies.  On the other hand, the cold gas mass also
affects the gas phase metallicity. As we showed in Section
\ref{sec:coldgas}, all models underpredict the cold gas mass at
redshift $z>0$ with respect to the observations. This means the model
galaxies at higher redshift have less cold gas to dilute the
metallicity at a fixed stellar mass than 
the observed galaxy population. At high redshifts, the lack of an
evolution in the normalization of the gas phase metallicity-stellar
mass relations may simply reflect a lack of evolution in the cold gas
fraction in the models. If high-z galaxies indeed have higher gas
fraction, for the same metal yield and loss rates, the models could
produce lower metallicities if the cold gas mass fraction increases.

\subsection{History of galaxy outflow}

When star formation occurs, SAMs generally assume that strong outflows
are generated by star formation feedback. Star formation feedback
plays an important role in shaping the low-mass end of the stellar
mass function by blowing out cold gas from the disk that would
otherwise be available to form the next generation of
stars. Furthermore, supernovae that drive outflows additionally
pollute the ISM, ICM and IGM with metals that alter the subsequent
cooling rate of gas back into the galactic disk.

\begin{figure*}[htb]
\begin{center}
\begin{tabular}{cc}
\includegraphics[width=0.45\textwidth]{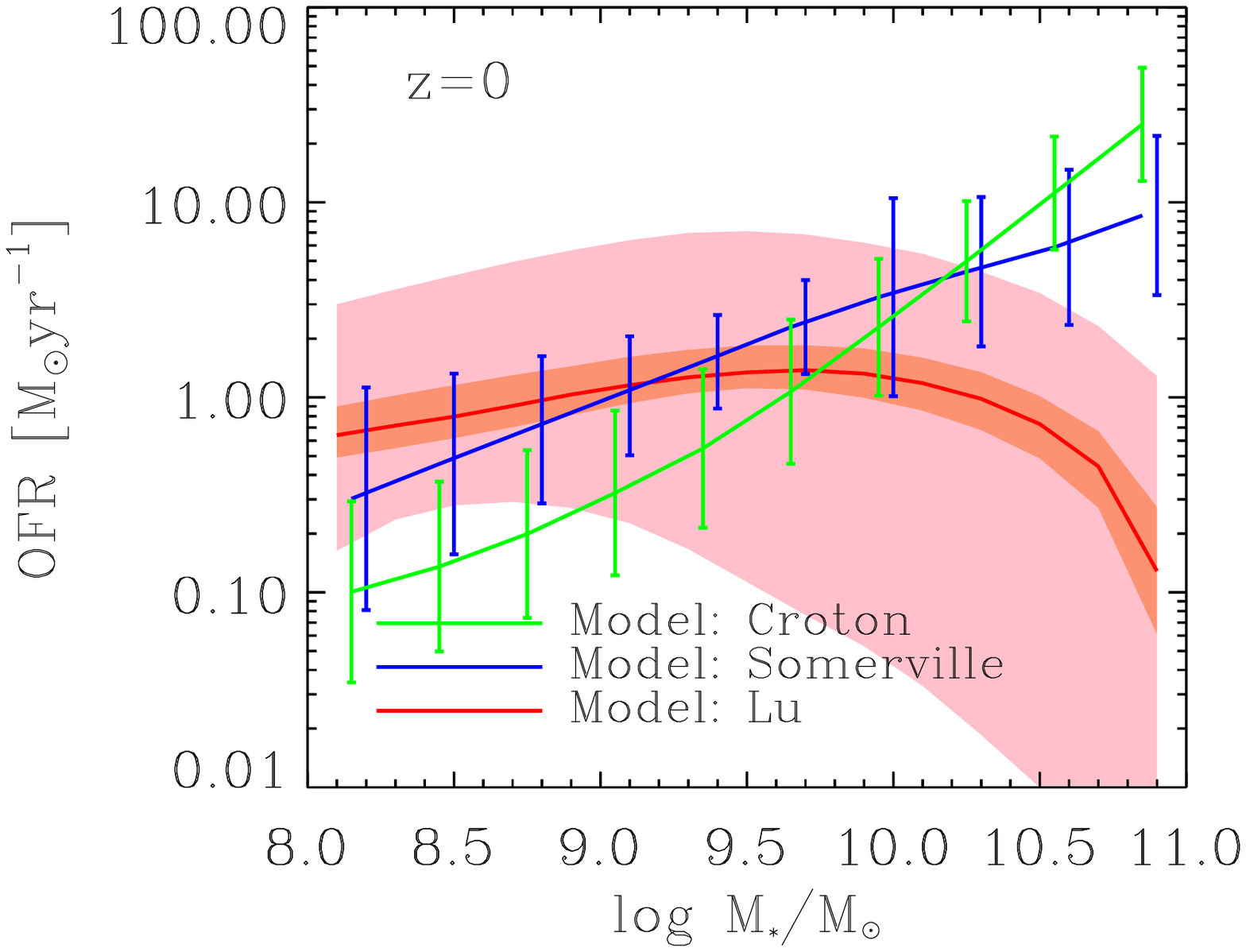} &
\includegraphics[width=0.45\textwidth]{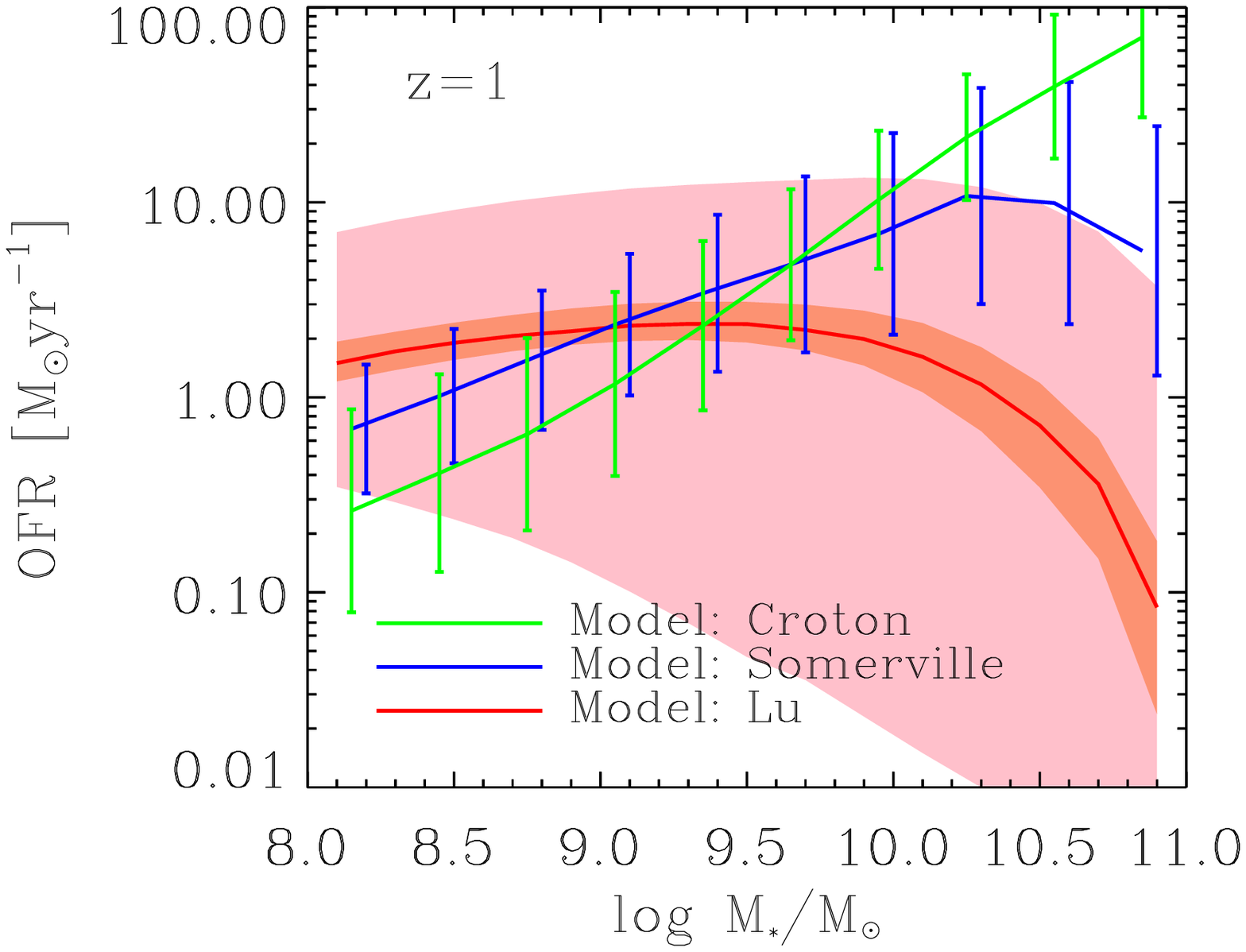} \\
\includegraphics[width=0.45\textwidth]{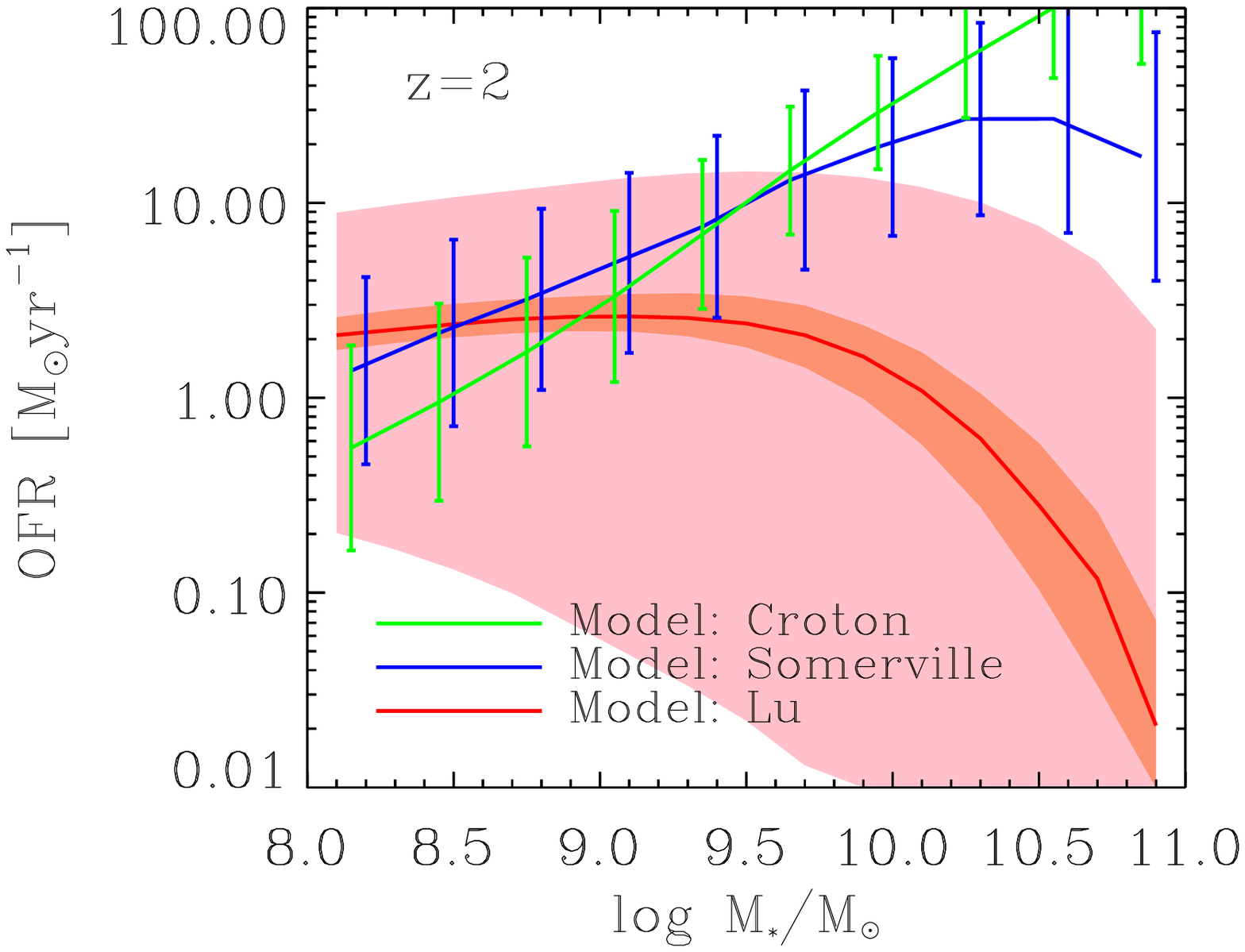} &
\includegraphics[width=0.45\textwidth]{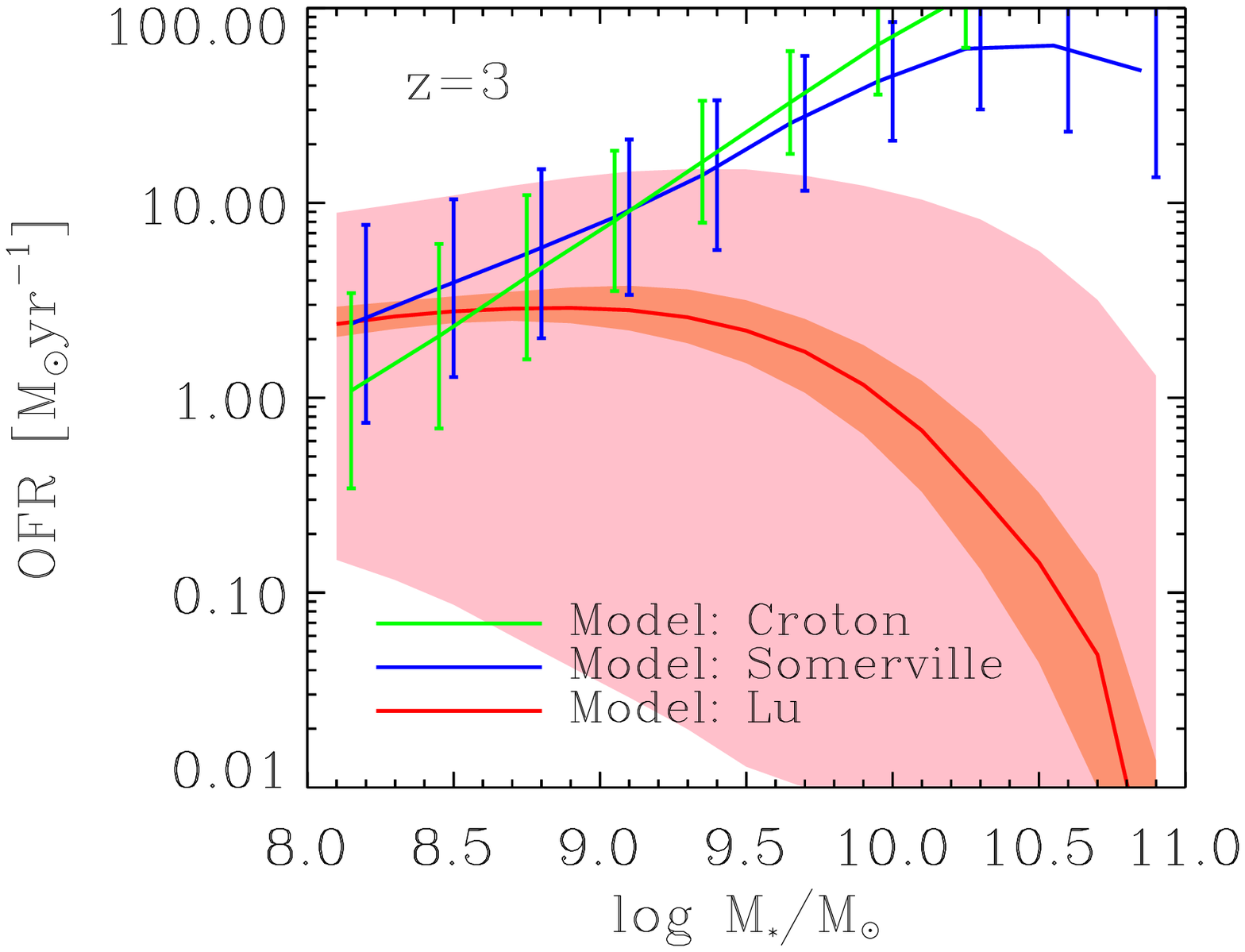}
\end{tabular}
\caption{Outflow rate of central galaxies as a function of stellar
  mass predicted by SAMs at $z=0, 1, 2,$ and 3.The green line denotes
  the prediction of the Croton model, the blue line denotes the
  prediction of the Somerville model, and the error bars on them show
  the 1-$\sigma$ scatter of the model galaxy samples. The dark and light
  red bands encompass 67\% and 95\% predictive posterior regions of
  the Lu model.  }
\label{fig:ofr}
\end{center}
\end{figure*}

In Figure \ref{fig:ofr} we plot the outflow rate (OFR) predicted by
the three models as a function of galaxy stellar mass at different
redshifts, $z=0, 1, 2$ and 3. The model predictions are very
different. For the Croton model, the outflow rate is basically
linearly proportional to the stellar mass, the reason being that the
assumed outflow rate is proportional to SFR, which is roughly
proportional to stellar mass. In this model, one finds that as the
star formation rate increases rapidly with redshift and stellar mass,
the outflow rate for high-z massive galaxies can reach more than
100 $\msun/$yr.

For the Somerville model, the outflow rate is similar to the Croton
model but increases with stellar mass more slowly. At the low-mass
end, the outflow rate is proportional to the stellar mass roughly as
$\propto M_*^{2/3}$. This can be understood because outflows in this
model are proportional to $\sim SFR/\vvir^{2}$, and the model is
tuned such that $SFR\propto M_*$ and $M_*\propto M_{\rm vir}^2\propto
\vvir^6$ for low mass galaxies. At the high-mass end, the outflow
rate starts to flatten out and even drop due the fact that star
formation is quenched.

The Lu model predicts a very different trend for how the outflow rate
scales with galaxy mass, and the predictive distribution is very
broad. For low-mass galaxies the outflow rate is almost constant, with
a weak stellar mass dependence. This is because the dominant mode of
the posterior has $OFR\propto SFR/\vvir^6$, resulting in
$M_*\propto M_{\rm vir}^2\propto \vvir^6$ and $SFR\propto M_*$ for
low-mass galaxies and an OFR that is nearly constant. When stellar
mass becomes large enough, the outflow rate decreases with increasing
stellar mass rapidly because of the strong halo circular velocity
dependence of the OFR. At higher redshifts, this characteristic OFR
decrease moves to lower stellar masses. Because the other two models
assume a much shallower circular velocity dependence for the OFR, they
have steeper slopes. The different predictions for the OFRs show how
they can sensitively depend on the model assumptions, and we expect
observations of galaxy OFRs and how they scale with galaxy mass could 
distinguish between such models.
Recent hydrodynamical simulations \citep{Puchwein2013, Dave2013} have
also found that a steeper circular velocity dependence for the
mass-loading factor tends to produce a shallower stellar mass function
in the low-mass end. It would be interesting to compare the SAM
predicted OFR with the simulation predictions. Our results also
suggest that accurate observational data for the faint-end slope of
the galaxy mass function and OFR are crucial to constraining feedback
models.

\section{Discussion and Conclusion}\label{sec:conclusion}

We have used three independently developed semi-analytic models
(SAMs), the Croton model, which is similar to the one presented in
\citet{Croton2006}, the Somerville model similar to the one presented
in \citet{Somerville2008}, and the Lu model updated from
\citet{Lu2011a}. Each model has been run on the same set of merger
trees extracted from the Bolshoi simulation \citep{Klypin2011}. After
carefully tuning the models against the stellar mass function of local
galaxies, we make predictions for the galaxy stellar and cold gas mass
evolution, star formation rate history, metallicities, and outflow
rates of the model galaxies. We compare the model predictions to 
understand the impact of the assumptions for star formation and
feedback implemented in those models.

Using hand tuning, we find parameter choices for the Croton and
Somerville models that fit the local stellar mass function reasonably
well after a large number of trials. Both models, however, still
overpredict the low mass number density somewhat, a common problem
with such techniques. For the Lu model, we have performed the
calibration using MCMC machinery, which allows us to explore the large
parameter space under observational data constraints in a systematic
way. The MCMC method identifies favored regions in the parameter
space, and randomly selects models from those regions to then make
predictions for the galaxy population. We have also noticed that,
without additional constraints, these favored regions in the Lu model
can be significantly offset from the values of the other SAMs.  In
particular, the posterior model for SN feedback outflows in the Lu
model suggests a very steep halo circular velocity dependence for the
outflow mass-loading factor, $OFR/SFR\propto \vvir^{-6}$. This
scaling is significantly steeper than what is normally assumed based
on energy conserving winds, $\propto \vvir^{-2}$, or momentum
conserving winds, $\propto \vvir^{-1}$.

The steep circular velocity dependence suggests that to explain the
shallow low-mass slope of the stellar mass function, galaxy formation
requires low-mass halos to have very efficient outflows that lead to
star formation suppression. Similar indications have been pointed out
by other studies. \citet{Mutch2013} employed a similar MCMC technique
to simultaneously reproduce the stellar mass function of galaxies at
$z=0$ and $z\approx 0.8$. The authors found that in order to fit the
data at both redshifts, their model had to assume extremely efficient
SN feedback in low-mass galaxies. \citet{Henriques2013} constrained
their model against both the stellar mass function and $B$- and
$K$-band luminosity functions over a larger redshift range, and found
that to achieve a good fit they needed to assume not only a large
mass-loading factor in low mass halos, but also that the timescale for
the ejected baryonic mass to reincorporate back into the host halo
needed to follow a particular halo mass and redshift dependence, 
which is different than what is assumed in the three models 
studied in this paper. It would be interesting to investigate if the extreme
circular velocity dependent outflow model used here in the Lu model
and new reincorporation model proposed by \citet{Henriques2013} are
essentially similar but just emphasize different processes. It would
also be interesting to study what observations can distinguish between
these two models.

When we take the locally constrained models and make predictions for
the galaxy properties at high redshifts we find that the variance of
the model predictions becomes significantly larger at higher
redshifts. However, the ``best fit'' versions of the Croton and the
Somerville models are usually encompassed by the 95\% range of the
posterior predictive distribution of the Lu model. This indicates
that, while the SAMs may adopt very different recipes for star
formation and feedback, when they are calibrated to the same data in
the local universe they agree with each other fairly well in terms of
predicting the stellar mass assembly histories of galaxies (at least
for the three SAMs considered here). The increasing differences
between the model predictions also suggest that more accurate
measurements at high-z could be used to discriminate between them and
to break some of the degeneracy between model parameters.

Comparing the model predictions with existing data we have found that
even though the SAMs are tuned to match only the local stellar mass
function, they generally produce star formation histories that are
qualitatively similar to those inferred observationally, but with some
discrepancies. In the SAMs considered here, low mass halos tend to
form stars more rapidly at high
redshift than what is inferred from observations, 
and their star formation rates are too low at low redshift compared with observations. 
When we look at the typical ages of stars predicted for local
galaxies, we find that low-mass model galaxies are systematically
older than those observational estimates. The predictions for the
stellar ages in this paper are very similar to those produced by other
SAMs \citep[e.g.][]{Fontanot2009,DeLucia2012a}, suggesting that this
issue is a general one, extending beyond just the three models we compare here.
For high-mass halos, the models have difficulty producing high
enough star formation rates at early times to match the data. The lack
of star formation in high-redshift high-mass halos results in 
a small underprediction of objects at
 the high-mass end of the stellar mass function at
these redshifts. However, we have found that when a reasonable
assumption about the uncertainty in estimating the observed stellar
masses is included, the model predictions are in 
reasonable agreement with the observations at the high-mass end.

The issues of overpredicting the star formation rate at high redshift
and underpredicting of the star formation rate at low redshift for
low-mass halos are closely related. 
All three models tend to produce rapid star formation in low-mass
halos at early times because gas cooling is efficient, and because
halos are denser, with higher circular velocities at a given mass,
making supernova feedback less effective
as it is currently parameterized.
Therefore, in the models, these halos have already formed a large
fraction of the expected present day stellar mass at early times.  
In this situation, to match the stellar mass function of local
galaxies, the models have to suppress further star formation. They
thus tend to predict rather low star formation rates locally for
low-mass systems, which is apparently
inconsistent with observations,
which show that low mass systems tend to still be star forming.
\citet{Weinmann2012} have reported similar behavior in several SAMs as
well as several sets of hydrodynamic simulations, and argued that the
problem is general, and may be connected with the widely used
``sub-grid'' recipes for star formation and stellar feedback.

The predictions for gas phase metallicity as a function of stellar
mass show a large discrepancy between the three models. For example,
the 95\% posterior predictive distribution of the Lu model no longer
encompasses the two other models. Moreover, the evolutionary trends of
the metallicity relations are all very different. The Lu model, which
fits the low-mass end of the local stellar mass function better than
other models, predicts the steepest metallicity-stellar mass
relation. It indicates that the model efficiently ejects
metal-enriched mass via strong feedback. The Croton model, which
retains the re-heated gas within the halo, predicts highest metallicity
for low-mass galaxies.

We have found that the differences between the model predictions
primarily stem from the different parameterizations of star formation
feedback, which is invoked to suppress star formation in low-mass
halos. This feedback is effectively modeled as outflow of the disk
gas. The outflow not only limits the cold baryon mass that is able to
fuel future star formation, but also blows out the metal-enriched ISM
in a galaxy. We have found that to match the faint-end slope of the
stellar mass function, the outflow rate is required to be as high as
about $1\, \msun$ yr$^{-1}$ for low-mass galaxies, but such a high
outflow rate results in metallicity-stellar mass relations that are
much steeper than current observational estimates.  The tension
between the match of the stellar mass function and the mismatch of the
metallicity-stellar mass relation strongly suggests that star
formation feedback is not properly modeled in these current models of
galaxy formation.

Among the observables we have compared, we have found that
metallicities and outflow rates show the largest discrepancies between
models. The fraction of galaxies that are quiescent is also highly
discrepant. Not only does each model predict a different trend for
those quantities as functions of stellar mass, but the variation
within one model family is large. This indicates that the
observational data for those quantities have strong power for further
constraining the uncertainties of galaxy formation models.  The
CANDELS survey, which has just finished taking data, will provide a
unique testbed for these models.  In particular, the survey will allow
a self consistent and comprehensive study of the star formation rates,
stellar masses, metallicities, and AGN fractions of galaxies from $z
\sim 1$ back to the reionization epoch, and the environmental
dependences of these properties.  This study provides theoretical
context for these data, and indicates that matching tighter
constraints on, for example, the connection between stellar mass and
halo mass at different cosmic epochs, the evolution of the fraction of
quiescent and star forming galaxies and the joint evolution of
metallicity and stellar mass, will provide insight into the physics of
galaxy formation.

\vspace{0.3in} 

DC wishes to acknowledge receipt of a QEII Fellowship by the
Australian Research Council.  
LP, CM and JP have been supported by the 
STScI CANDELS grant HST-GO-12060.12-A and by NSF-AST-1010033.
PB received support from an HST Theory
grant (program number HSTAR- 12159.01-A) as well as a Giacconi
Fellowship, both provided through grants from the Space Telescope
Science Institute, which is operated by the Association of
Universities for Research in Astronomy, Incorporated, under NASA
contract NAS5-26555. DCK is supported by NSF grant AST-08-08133 and
HST grant HST GO-12060.

\bibliographystyle{apj}
\bibliography{candels_sam}

\end{document}